\documentclass[11pt,letterpaper]{article}
\usepackage[margin=1in]{geometry}

\usepackage[T1]{fontenc}
\usepackage{textcomp}
\usepackage{amsmath,amsfonts,bm}
\usepackage{amssymb} 
\usepackage{hyperref}
\usepackage{wrapfig}
\usepackage{tabularx}
\usepackage{graphicx}
\usepackage{epstopdf}
\usepackage{subcaption}
\usepackage{algorithm}
\usepackage{algorithmic}
\usepackage{xcolor,url}
\usepackage[numbers]{natbib} 
\usepackage{bbm}
\usepackage{enumitem}

\usepackage{stfloats}
\usepackage{tikz}

\usetikzlibrary{fit}

\usepackage{amsthm}
\renewenvironment{proof}{{\sffamily\bfseries Proof. }}{\qed}
\newtheorem*{theoremnonum}{Theorem}

\newtheorem{lem}{Lemma}
\newtheorem{prop}{Proposition}

\theoremstyle{definition}
\newtheorem{de}{Definition}

\newtheorem{remark}{Remark}

\newcommand{\GW}{Galton-Watson }

\usepackage{thmtools}
\usepackage{thm-restate}
\usepackage{enumitem}
\graphicspath{ {./images/} }

\newcommand{\eps}{\epsilon}

\def\erdos{Erd\H os}
\def\renyi{R\'enyi }
\def\E{{\mathsf E}}
\def\P{{\mathsf P}}
\def\cg{{\mathcal{G}}}

\def\dist{{\mathrm{dist}}}
\def\ca{{\mathcal{A}}}
\def\cb{{\mathcal{B}}}
\def\cc{{\mathcal{C}}}
\def\cd{{\mathcal{D}}}
\def\ce{{\mathcal{E}}}
\def\cf{{\mathcal{F}}}
\def\ci{{\mathcal{I}}}

\def\ch{{\mathcal{H}}}
\def\cl{{\mathcal{L}}}
\def\cn{{\mathcal{N}}}
\def\cs{{\mathcal{S}}}
\def\cw{{\mathcal{W}}}
\def\ct{{\mathcal{T}}}
\def\cx{{\mathcal{X}}}

\def\cz{{\mathcal{Z}}}

\def\cv{{\mathcal{V}}}
\def\cp{{\mathcal{P}}}
\def\bp{{\mathbb{P}}}
\def\bq{{\mathbb{Q}}}
\def\deg{{\mathrm{deg}}}

\def\be{{\mathsf{E}}}
\def\gw{{\mathrm{GW}}}
\def\indi{{\mathbbm{1}}}
\def\poi{{\mathrm{Poi}}}
\def\sto{{\stackrel{\mathrm{sto.}}{\le}}}

\def\bin{{\mathrm{Binom}}}

\def\dtv{{d_\mathrm{TV}}}

\def\sfch{{\mathsf{Ch}}}

\def\dist{\mathrm{dist}}

\def\excl{{\mathsf{EXC}}}
\def\com{{\mathsf{COM}}}

\newcommand{\td}[1]{\tilde{#1}}
\newcommand{\wdtd}[1]{\widetilde{#1}}
\newcommand{\cond}{\mathchoice{\,\vert\,}{\mspace{2mu}\vert\mspace{2mu}}{\vert}{\vert}}

\allowdisplaybreaks

\title{\fontsize{18}{24}\selectfont Near-Quadratic Error-Free Graph Matching above Otter's Threshold via Approximate Local Tree Correlation Testing}

\author{
 Jiale Cheng\\
 University of Michigan\\
 \texttt{jlcheng@umich.edu}
 \and
 Ziao Wang\\
 University of Michigan\\
 \texttt{ziaow@umich.edu}
 \and
 Lei Ying\\
 University of Michigan\\
 \texttt{leiying@umich.edu}
}
\date{}

\begin{document}

\maketitle

\begin{abstract}
This paper presents a near-quadratic-time algorithm that correctly matches almost all nodes of two correlated \erdos--\renyi graphs without any incorrect matches. Each graph has average degree $\lambda$ and its edges are subsampled from a common parent \erdos--\renyi graph with $n$ nodes, with subsampling probability $s$. The nodes of one graph are then relabeled by an unknown permutation. The algorithm is built on a test for correlated versus independent  Galton--Watson trees with approximated likelihood ratios. For every constant $s>\sqrt{C_{\mathrm{Otter}}}$, where $C_{\mathrm{Otter}}\approx0.338$ is Otter's tree-counting constant, and every $\lambda=\omega(1)$ with $\lambda=(\log n)^{O(1)}$, the algorithm outputs a partial matching of size $n-o(n)$ that contains no incorrect matches with high probability. If, in addition, $\lambda s\ge(1+\eps)\log n$ for a constant $\eps>0$, a seeded completion step recovers the permutation exactly. In both cases, the time complexity of the algorithm is $n^{2+o(1)}$ with high probability, whereas previous polynomial-time algorithms with the same guarantees either have a time-complexity bound whose exponent goes to $\infty$ as $s$ approaches $\sqrt{C_{\mathrm{Otter}}}$ from above, or apply only when $s$ is close to $1$.
\end{abstract}

\section{Introduction}\label{sec:intro}

In applications such as social network de-anonymization and biological network alignment, two graphs are structurally similar even when their node identities are anonymized, hidden, or inconsistently labeled.  The \emph{graph matching problem} asks to recover the correspondence between the nodes of two such graphs. This problem arises in social network de-anonymization~\cite{Nar-Shm-deanonymizing2009,Nitish14Reconciliation}, the alignment of protein--protein interaction networks across species~\cite{singh2007pairwise}, and the matching of relational patterns in computer vision~\cite{Minsu12Pattern} and natural language processing~\cite{Hag-Ng-robust2005}. A standard formulation of this problem seeks a bijection between the two node sets that maximizes edge overlap. This optimization problem is a special case of the quadratic assignment problem~\cite{Burkard1998} and is computationally intractable in the worst case~\cite{QAPNPApproxi}. Recent studies have therefore focused on graph matching under random graph models, aiming to design computationally efficient algorithms with provable statistical guarantees.

This paper investigates the graph matching problem under the correlated \erdos--\renyi graph pair model $\mathrm{CER}(n,\lambda,s)$~\cite{Pedram11GA}, where $\lambda\in(0,ns]$ is the average-degree parameter and $s\in(0,1]$ is the correlation parameter. In this model, two graphs $G$ and $G'$ on the node set $[n]$ are obtained from the same parent \erdos--\renyi graph with edge probability $\frac{\lambda}{ns}$ by retaining every edge independently with probability $s$, so that each of them is marginally an \erdos--\renyi graph with edge probability $\lambda/n$. The graph $\td G$ is obtained by relabeling the nodes of $G'$ according to a uniformly random permutation $\pi$ of $[n]$. Given only the graph pair $(G,\td G)$, the goal is to recover $\pi$ accurately and efficiently.

Since sparse \erdos--\renyi graphs are locally tree-like, a natural approach is to compare the local neighborhoods of two nodes up to some depth $d$. For a node $u$ and its counterpart $\td u=\pi(u)$, the two local neighborhoods are close in distribution to a correlated pair of Poisson Galton--Watson trees with offspring mean $\lambda$,  whose intersection has offspring mean $\lambda s$. For most other pairs of nodes, the two local neighborhoods are close in distribution to an independent pair of such trees \cite{Luca20Tree,Hofstad24RGCN2}. Algorithms based on this approach~\cite{Luca24LRT,Luca24Stat,Cheng26RBAlign} test whether the local neighborhoods of two nodes form a correlated or an independent pair. The tests in~\cite{Luca24LRT,Cheng26RBAlign} are based on the exact computation of the likelihood ratio between the correlated and the independent tree-pair distributions. 
With high probability, the algorithm in~\cite{Cheng26RBAlign} has near-quadratic complexity when $\lambda=(\log n)^{\alpha+o(1)}$ for a constant $\alpha\in(0,1)$. When $\lambda=\Omega(\log n)$, the same analysis gives only a superpolynomial bound, because the complexity of computing each likelihood ratio using the method in \cite{Luca24LRT,Cheng26RBAlign} grows factorially with the number of children.
However,  $\lambda=\Omega(\log n)$ is necessary to achieve exact recovery of $\pi$ information-theoretically~\cite{Yihong22Threshold,Daniel16GA,Daniel18GA}. We therefore ask whether an algorithm based on the likelihood-ratio test can recover $\pi$ accurately and efficiently in this regime. The following informal theorem answers this question for every constant $s>\sqrt{C_{\mathrm{Otter}}}$ and every $\lambda$ at most polylogarithmic in $n$, where $C_{\mathrm{Otter}}\approx0.338$ is Otter's tree-counting constant~\cite{Richard48TreeCount}.

\begin{theoremnonum}[Informal statement of Theorems~\ref{th:TBAlign} and~\ref{th:TBAlign-runtime}]
Consider $(G,\td G)\sim\mathrm{CER}(n,\lambda,s)$, where $\lambda=\omega(1)$, $\lambda=(\log n)^{O(1)}$, and $s\in(\sqrt{C_{\mathrm{Otter}}},1]$ is a constant. Then our algorithm \texttt{TBAlign} proposed in Section~\ref{sec:tbalign} satisfies the following guarantees.
\begin{itemize}
    \item \textbf{Error-free almost-exact recovery.} With high probability, \texttt{TBAlign} outputs a partial matching $\pi'$ of size $n-o(n)$ with no incorrect matches.
    \item \textbf{Exact recovery.} If, in addition, $\lambda s\ge(1+\eps)\log n$ for a constant $\eps>0$, then with high probability, \texttt{TBAlign} can further extend $\pi'$ to a complete matching $\hat\pi$ that equals $\pi$ with a seeded completion step.
\end{itemize}
With high probability, the time complexity of \texttt{TBAlign} is $n^{2+o(1)}$.
\end{theoremnonum}

\subsection{Related Work}\label{sec:comparison}

Information-theoretic limits for graph matching have been studied under four recovery criteria. Partial recovery requires correctly matching a positive fraction of the nodes~\cite{Yihong22Threshold,Luca21Partial,Jian23Partial}; almost-exact recovery requires a complete matching with $o(n)$ mismatches~\cite{Yihong22Threshold}; error-free almost-exact recovery requires a partial matching of size $n-o(n)$ with no incorrect matches~\cite{Daniel19PartialAlign}; and exact recovery requires recovering the entire permutation $\pi$ correctly~\cite{Daniel16GA,Daniel18GA,Yihong22Threshold}.

Another line of work develops efficient algorithms that exploit structural features of the two graphs, such as canonical labels~\cite{Osman19CanonicalLabel}, degree profiles~\cite{Jian21DegProfile}, spectral signatures~\cite{Fan23Spectral1,Fan23Spectral2}, subgraph counts~\cite{Cheng26Chandelier}, and tree-based statistics of local neighborhoods~\cite{Luca20Tree,Luca24LRT,Cheng26RBAlign}. In the dense regime $\lambda=n^{\Omega(1)}$, the algorithm in~\cite{ding2026polynomial} achieves exact recovery in polynomial time for every fixed $s\in(0,1)$. In the sparse regime $\lambda=n^{o(1)}$, existing results point to a statistical--computational gap associated with the threshold $s=\sqrt{C_{\mathrm{Otter}}}$. Below this threshold, low-degree hardness results suggest that no polynomial-time algorithm achieves even partial recovery~\cite{Jian23LowDegree,li2025algorithmic}. Above it, polynomial-time algorithms achieve different recovery guarantees depending on different $\lambda$ regimes. We next review these algorithms by recovery criterion and compare their guarantees with ours.

\begin{figure}[t]
    \centering
    \begin{minipage}{0.49\linewidth}
        \centering
        \includegraphics[width=\linewidth]{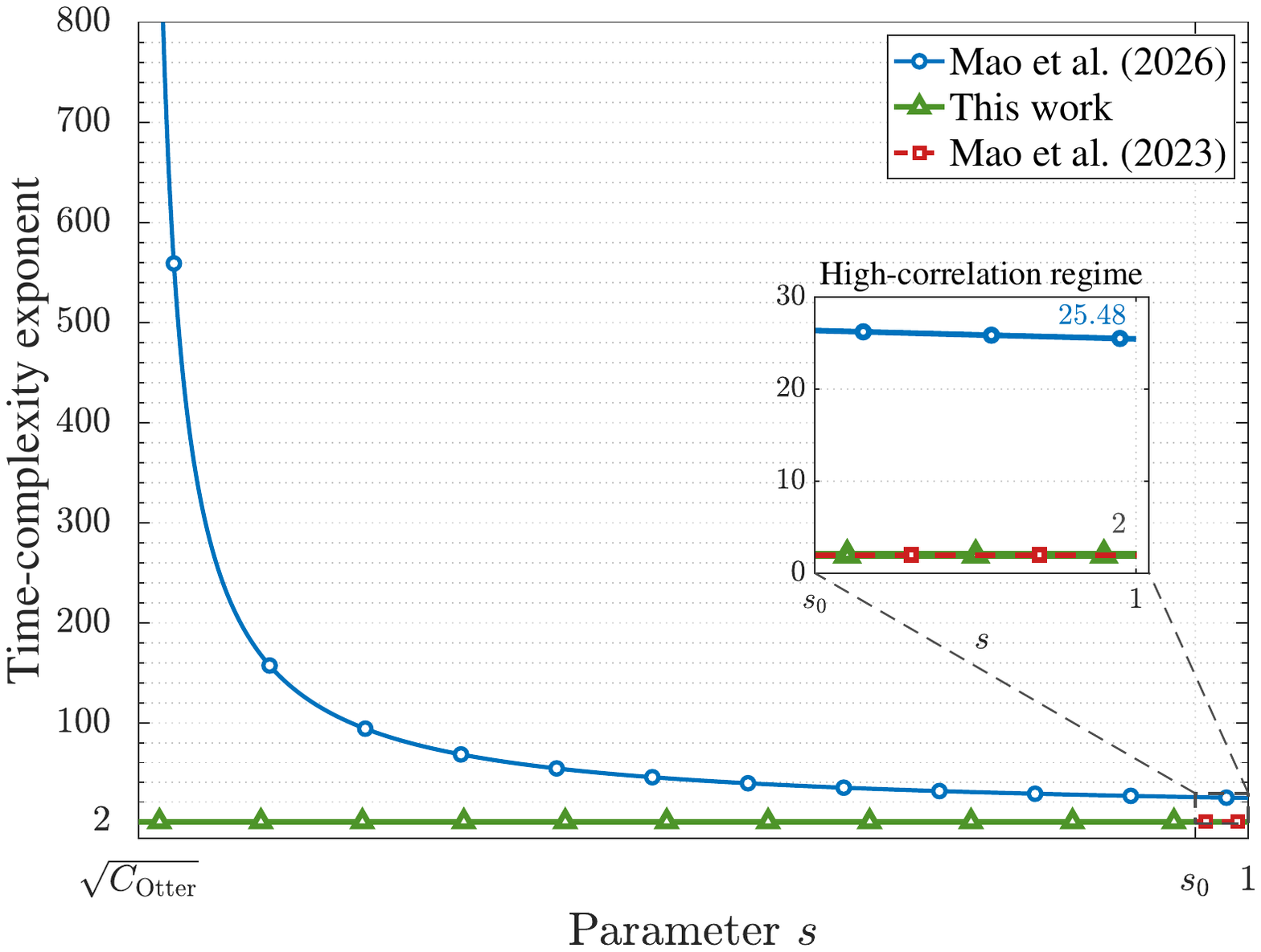}\\[2pt]
        (a)
    \end{minipage}\hfill
    \begin{minipage}{0.49\linewidth}
        \centering
        \includegraphics[width=\linewidth]{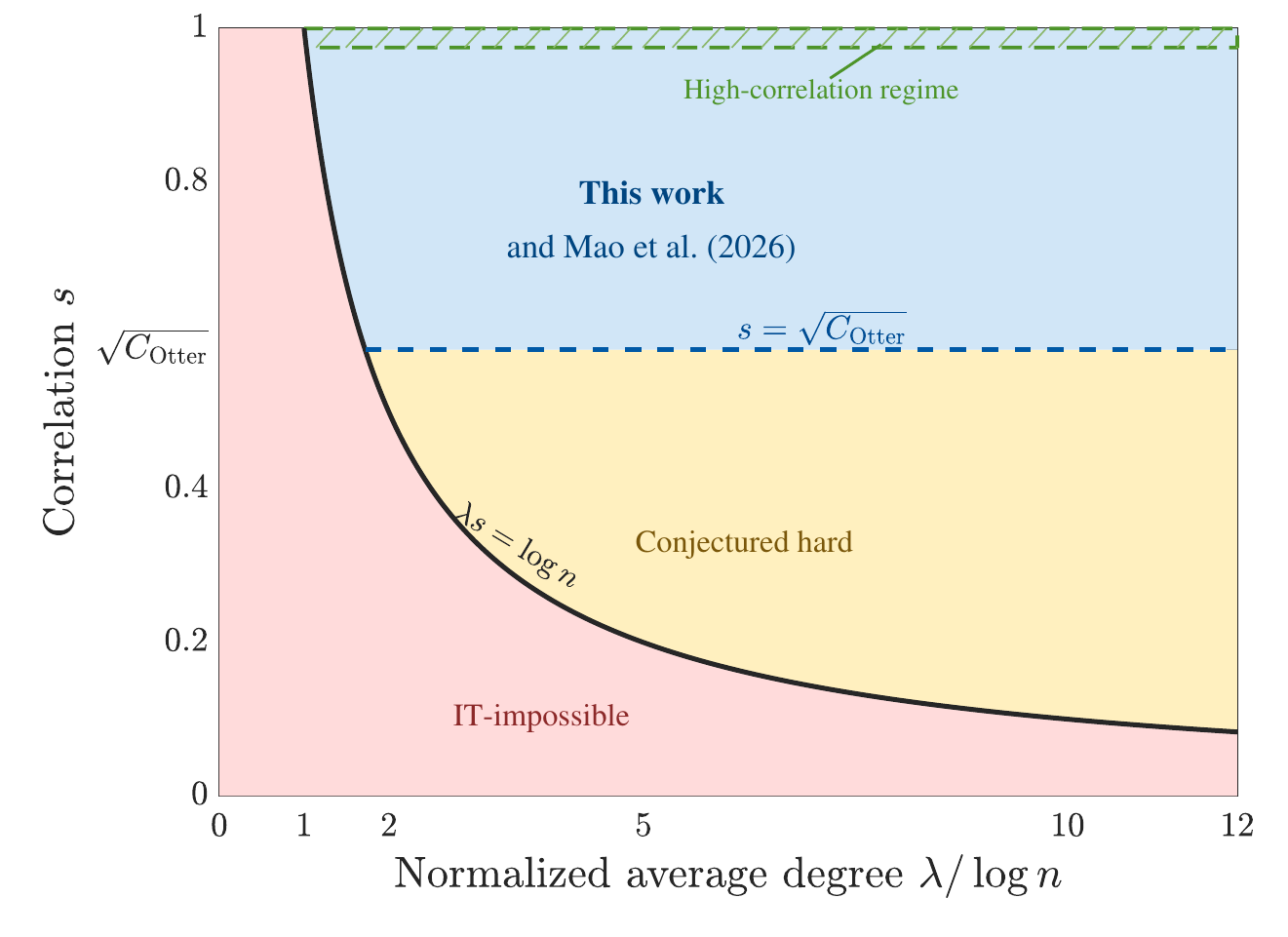}\\[2pt]
        (b)
    \end{minipage}
    \caption{Polynomial-time algorithms for exact recovery. (a)~Time-complexity exponents. The vertical axis displays the exponent $c$ in a time-complexity bound of the form $n^{c+o(1)}$. (b)~Exact-recovery regions. The labels Mao et al.~(2026) and Mao et al.~(2023) refer to~\cite{Cheng26Chandelier} and~\cite{Cheng23ConstantCorrelation}, respectively, and the high-correlation band in~(b) represents the algorithms in~\cite{Jian21DegProfile,Fan23Spectral2,mao21a-pmlr,Cheng23ConstantCorrelation}.}
    \label{fig:exact-comparison}
\end{figure}

\paragraph*{Partial recovery.} For a large enough constant $\lambda=\Theta(1)$, partial recovery can be achieved by counting a family of rooted trees called chandeliers~\cite{Cheng26Chandelier} or by tree correlation testing \cite{Luca20Tree,Luca24LRT}. 
The likelihood-ratio test used in~\cite{Luca24LRT} has vanishing type-I error and nonvanishing power for a large constant $\lambda$ when $s>\sqrt{C_{\mathrm{Otter}}}$, whereas no test has these properties for any $\lambda$ when $s\le\sqrt{C_{\mathrm{Otter}}}$~\cite{Luca24Stat}.
This result has been extended to asymmetric tree pairs in~\cite{maier2025}. The algorithm in our paper tests tree correlation using an approximate likelihood ratio, instead of the exact value, and it also require the condition $s>\sqrt{C_{\mathrm{Otter}}}$.

\paragraph*{Almost-exact recovery.} For $\lambda=\omega(1)$ and every constant $s>\sqrt{C_{\mathrm{Otter}}}$, the chandelier-counting algorithm in~\cite{Cheng26Chandelier} achieves error-free almost-exact recovery.\footnote{The condition in~\cite{Cheng26Chandelier} is stated for the edge correlation $\rho$ between the two graphs, and $\rho=s(1+o(1))$ when $\lambda=o(n)$.} The proved bound on its time complexity is $n^{c(s)}$, where the exponent $c(s)$ goes to infinity as $s$ approaches $\sqrt{C_{\mathrm{Otter}}}$ from above. In contrast, the algorithm in~\cite{Cheng26RBAlign} achieves almost-exact recovery for the same range of $s$ with time complexity $n^{2+o(1)}$, provided that $\lambda=(\log n)^{\alpha+o(1)}$ for a constant $\alpha\in(0,1)$. However, its output may contain $o(n)$ incorrect matches. Our algorithm achieves error-free almost-exact recovery and has $n^{2+o(1)}$ time complexity with high probability, provided that $\lambda=\omega(1)$ and $\lambda=(\log n)^{O(1)}$. 

\paragraph*{Exact recovery.} When $\lambda=n^{o(1)}$ and $\lambda s\ge(1+\eps)\log n$ for some constant $\eps>0$, there are known polynomial-time algorithms for exact recovery. The algorithm in~\cite{Cheng26Chandelier} achieves exact recovery with a time-complexity bound of $O(n^{c(s)}+n\lambda^2)$. The algorithms in~\cite{Jian21DegProfile,Fan23Spectral2,mao21a-pmlr} require $1-s$ to vanish as $n$ grows. The algorithm in~\cite{Cheng23ConstantCorrelation} has time complexity $n^{2+o(1)}$ in expectation, but it requires $s>s_0$ for an unspecified constant $s_0$ close to $1$.
In contrast, our algorithm achieves exact recovery with time complexity $n^{2+o(1)}$ for every $s>\sqrt{C_\mathrm{Otter}}$. Our guarantee, however, requires the more restrictive condition $\lambda=(\log n)^{O(1)}$, rather than $\lambda=n^{o(1)}$.
Figure~\ref{fig:exact-comparison} illustrates these comparisons. In Figure~\ref{fig:exact-comparison}(a), our algorithm keeps the time-complexity exponent $2+o(1)$ for every $s>\sqrt{C_{\mathrm{Otter}}}$ compared with the algorithms in~\cite{Cheng26Chandelier} and~\cite{Cheng23ConstantCorrelation}.  The magnified inset marks the exponents of the three time-complexity bounds at $s=1$ within the high-correlation regime $s>s_0$, where $s_0$ is placed schematically. Figure~\ref{fig:exact-comparison}(b) shows the exact-recovery regions of the algorithms we discussed when $\lambda=\Theta(\log n)$. The curve $\lambda s=\log n$ in this panel is the information-theoretic threshold for exact recovery~\cite{Yihong22Threshold}, and the conjectured hard region reflects the low-degree evidence in~\cite{Jian23LowDegree,li2025algorithmic}.

\paragraph*{Related models and extensions.}
Graph matching has also been studied beyond the $\mathrm{CER}(n,\lambda,s)$ model, including in correlated stochastic block models~\cite{racz2021correlated,Shuwen24SBMChandellier,Joonhyuk23SBM}, correlated geometric graph models~\cite{Haoyu22Geo}, and attributed graph models~\cite{Ning24AGA,Ziao24AGA,Ziao25AGA2,huang2026attributed}. Other extensions consider matching multiple graphs~\cite{ameen2025aligning} or partially observed graphs~\cite{wang2026,shiu2025information}. These works explore a broader range of graph-matching settings, while our comparison above focuses on the computational guarantees for exact recovery in the correlated Erdős--Rényi model.

\subsection{Technical Contributions} \label{sec:overview}
Our algorithm closely follows the graph matching algorithm in~\cite{Cheng26RBAlign}, which builds on the local-tree correlation testing developed in ~\cite{Luca24LRT}. Both algorithms identify matching vertices by comparing local trees rooted at their neighbors. Our stronger recovery guarantees and applicability to larger average degrees come from some novel changes to both the algorithm and its analysis. We next discuss these contributions at the algorithmic and analytical levels.
\subsubsection{Algorithmic contributions}
We make two main modifications to the algorithm in~\cite{Cheng26RBAlign}.
\paragraph{From 3-dangling to 5-dangling.}
The \emph{3-dangling-tree test}, introduced in~\cite{Luca24LRT} and subsequently adopted in~\cite{Cheng26RBAlign}, accepts a candidate pair of vertices only after identifying three pairs of their neighbors whose downstream local trees are classified as correlated. The three pairs use distinct neighbors on each side, and each downstream tree is rooted at the corresponding neighbor with the candidate vertex removed. Using this criterion, the algorithm in~\cite{Cheng26RBAlign} achieves almost-exact recovery. With high probability, it gives a complete matching with $o(n)$ mismatches.

We replace this criterion with a \emph{5-dangling-tree test}, which requires five such neighbor pairs. Under the conditions of our main theorem, this modification yields a partial matching of size $n-o(n)$ but without any incorrect matches with high probability. This error-free guarantee also allows us to use the matched pairs as seeds for a completion procedure based on the seeded graph matching algorithm in~\cite{Cheng26Chandelier}. Specifically, when $\lambda s\ge(1+\eps)\log n$ for a constant $\eps>0$, the completion procedure matches all the remaining vertices correctly with high probability, thereby achieving exact recovery.

\paragraph{Approximate likelihood-ratio testing.}
The algorithms in~\cite{Luca24LRT,Cheng26RBAlign} compute local-tree likelihood ratios using the recursive formula introduced in~\cite{Luca24LRT}. This recursion expresses the likelihood ratio for a pair of rooted trees in terms of the likelihood ratios for pairs of subtrees rooted at their children. Each update sums weighted products of these subtree likelihood ratios over all partial matchings between the two sets of children. The direct evaluation of this sum has factorial complexity in the number of children: when both nodes have $c$
children, the sum contains at least $c!$ terms. Since a typical node in the ER graph has about $\lambda$ children, this computation incurs a cost $e^{\Omega(\lambda\log\lambda)}$, which is superpolynomial in $n$ when $\lambda=\Omega(\log n)$. This computational bottleneck obstructs the extension of the approach in~\cite{Cheng26RBAlign} to the degree regime needed for exact recovery.

To overcome this difficulty, we express the matching sums in each recursive update through permanents of nonnegative matrices. 
We then approximate these permanents using the fully polynomial randomized approximation scheme (FPRAS) in~\cite{Jerrum04Permanent}, and use the resulting approximate likelihood ratios for tree correlation testing. With our choice of parameters, the computation takes $n^{o(1)}$ time per
tested local-tree pair, allowing the graph matching
algorithm to retain $n^{2+o(1)}$ time complexity with
high probability throughout the degree range considered
in this paper.

\subsubsection{Analytical contributions}
\paragraph{Analyzing the approximate likelihood-ratio test.}
The analysis of tree correlation tests in~\cite{Cheng26RBAlign} proceeds in an indirect manner. It constructs an auxiliary test that is analytically more tractable but computationally infeasible, establishes bounds on its type-I error and power, and then invokes Neyman--Pearson optimality to transfer these guarantees to an exact likelihood-ratio test. This argument establishes the existence of a suitable threshold without providing its value. To avoid computing this threshold, the rank-based algorithm in~\cite{Cheng26RBAlign} processes local-tree pairs in decreasing order of their likelihood ratios and stops after a prescribed number of vertex pairs have been matched. With this stopping rule, the algorithm achieves almost-exact recovery with high probability, without needing to compute the threshold.

This approach does not extend to our approximate likelihood-ratio test, which has no corresponding
Neyman--Pearson optimality guarantee. We show that the approximate likelihood ratio lies within a constant factor of the exact value with high probability. This guarantee reduces the analysis of the approximate test to controlling the exact test at thresholds shifted by constant factors. Specifically, controlling the approximate test's type-I error requires a bound for the exact test at a lower threshold, whereas controlling its power requires a bound at a higher threshold. These shifts are determined by the approximation guarantee. The existence of a suitable threshold from the Neyman--Pearson argument does not directly provide such control.

Therefore, we instead choose an explicit threshold and account for the constant-factor shifts caused by approximation, as well as its failure probability. This analysis gives a larger type-I error bound and a slightly weaker power guarantee than those of the auxiliary test, but the resulting bounds remain sufficient for our graph matching guarantees with the given threshold. After estabshing the guarantees of the the threshold-based matching algorithm, the corresponding rank-based matching algorithm provides the same gaurantees by following the same analysis in  \cite{Cheng26RBAlign}.  

\paragraph{Why 5-dangling gives an error-free output.} For the analysis, consider the union of the two graphs after aligning their vertices according to the true correspondence. We first show that, with high probability, every pair of local trees that passes the correlation test overlaps in this union graph. This follows by comparing non-overlapping local trees with independent Poisson Galton--Watson trees and combining the test's small type-I error with a union bound over all tested pairs.

On this event, consider an incorrectly matched pair of vertices. Each selected pair of their neighbors whose downstream local trees pass the correlation test yields a short path between the two vertices in the union graph. Since the criterion requires distinct neighbors on each side, these paths can be chosen so that no path arises from more than two of the selected neighbor pairs. Thus, the three neighbor pairs required by the 3-dangling criterion give at least two distinct short paths, whose union contains a cycle. With high probability, only $o(n)$ vertices have a cycle in the relevant local neighborhood. This argument therefore rules out wrong matches for almost all vertices, but does not exclude every wrong match.

The five neighbor pairs required by the 5-dangling criterion instead give at least three distinct short paths. Their union is a connected subgraph containing at least two cycles and $O(d)$ vertices, where $d$ is the depth of the local trees. A first-moment argument based on standard subgraph counting~\cite[Lemma~5.1]{frieze2015introduction} shows that, in our regime, no such subgraph occurs anywhere in the union graph with high probability.
Consequently, no incorrect vertex pair can satisfy the 5-dangling criterion, achieving error-free almost-exact recovery.

\section{Preliminaries}\label{sec:model-results}

\subsection{Notation}

Given a simple graph $H$, we use $\cv(H)$ and $\ce(H)$ to denote its node and edge sets, respectively. The distance between two nodes in $H$ is the length of the shortest path connecting them. A tree is a connected, acyclic simple graph, and a rooted tree is a tree with a distinguished node called the root. The depth of a node in a rooted tree is its distance from the root, and the depth of the tree is the maximum depth among all its nodes. We write $[n]:=\{1,\ldots,n\}$ and $[a:b]:=\{i\in\mathbb{Z}:a\le i\le b\}$ for integers $a$ and $b$.

For a correlated \erdos--\renyi graph pair $(G,\td G)$, we use the shorthand $\cv:=\cv(G)$, $\ce:=\ce(G)$, $\td\cv:=\cv(\td G)$, and $\td\ce:=\ce(\td G)$; both node sets are identified with $[n]$. Variables without tildes, such as $u$ and $v$, refer to nodes in $G$, whereas variables with tildes, such as $\td u$ and $\td v$, refer to nodes in $\td G$. For any $u\in\cv$, let $\cn(u)$ denote its neighborhood in $G$; analogously, let $\td\cn(\td u)$ denote the neighborhood of $\td u\in\td\cv$ in $\td G$.

We use $\P(\cdot)$ and $\E[\cdot]$ to denote probability and expectation, respectively, appending a subscript when the underlying distribution must be specified. We employ standard Landau's asymptotic notation, including $O(\cdot)$, $\Omega(\cdot)$, $\Theta(\cdot)$, $o(\cdot)$, $\omega(\cdot)$, and $\sim$, and we use $\td O(\cdot)$ to suppress logarithmic factors. Unless otherwise specified, all asymptotic statements assume $n\to\infty$. We say that an event holds with high probability if its probability is $1-o(1)$. Finally, \emph{computational complexity} or \emph{time complexity} is measured by the total number of elementary operations executed, assuming a uniform cost model where each arithmetic operation incurs a unit cost.

\subsection{Performance Criteria}
Given an observed graph pair $(G,\td G)$, a bijection $\hat\pi:\cv\to\td\cv$ is called a \emph{complete matching}, and an injective map $\pi'$ from a subset of $\cv$ to $\td\cv$ is called a \emph{partial matching}; nodes outside this subset are left \emph{unmatched}. A pair $(u,\td v)$ in a complete or partial matching is called a \emph{wrong match} if $\td v\neq\td u$. We consider the following four recovery criteria.
\begin{itemize}
    \item \emph{Partial recovery.} A complete matching $\hat\pi$ achieves partial recovery if there exists a constant $c>0$ such that $\bigl|\{u\in[n]:\hat\pi(u)=\td u\}\bigr|\ge cn$.
    \item \emph{Almost-exact recovery.} A complete matching $\hat\pi$ achieves almost-exact recovery if $\bigl|\{u\in[n]:\hat\pi(u)=\td u\}\bigr|=n-o(n)$. Such a matching matches every node and may contain $o(n)$ wrong matches.
    \item \emph{Error-free almost-exact recovery.} A partial matching $\pi'$ achieves error-free almost-exact recovery if $\bigl|\{u\in[n]:u\text{ is matched by }\pi'\}\bigr|=n-o(n)$ and $\pi'(u)=\td u$ for every matched node $u$. Such a matching may leave $o(n)$ nodes unmatched but contains no wrong match.
    \item \emph{Exact recovery.} A complete matching $\hat\pi$ achieves exact recovery if $\hat\pi(u)=\td u$ for every $u\in[n]$.
\end{itemize}
An algorithm achieves one of these criteria if its output satisfies the corresponding property with probability $1-o(1)$. The guarantees in our paper concern the last two criteria.

\subsection{Poisson Galton--Watson Tree-Pair Models}

\paragraph*{Poisson Galton--Watson tree.} Let $\lambda>0$. A Poisson Galton--Watson tree with offspring mean $\lambda$ is generated from a single root by letting every node independently produce a $\poi(\lambda)$ number of children. We order the children of each node and label them recursively by their positions among their siblings, starting with the first label at the root. Let $\ch$ denote the resulting rooted tree, let $\cx$ be its support, and let $\cx_k$ be the set of trees in $\cx$ with depth at most $k$. For an integer $k\ge0$, we write $\ch_k$ for the depth-$k$ truncation of $\ch$.

The likelihood-based tests depend only on rooted tree structure rather than the auxiliary order labels. For a labeled rooted tree $h$, let $[h]$ be the unlabeled rooted tree obtained by removing its labels while preserving its root and edges. We write $\cz:=\{[h]:h\in\cx\}$ and $\cz_k:=\{[h]:h\in\cx_k\}$ for the corresponding collections of rooted unlabeled trees.

\paragraph*{Independent tree pair.} Under the independent tree-pair distribution $\bq$, the trees $\ch$ and $\td\ch$ are sampled independently, and each is a Poisson Galton--Watson tree with offspring mean $\lambda$.

\paragraph*{Correlated tree pair.} Under the correlated tree-pair distribution $\bp$, first sample an intersection tree $\ch^*$ that is a Poisson Galton--Watson tree with offspring mean $\lambda s$. Starting from the same $\ch^*$, construct $\ch$ and $\td\ch$ through two conditionally independent augmentations. In each augmentation, at every node of $\ch^*$, add an independent $\poi(\lambda(1-s))$ number of exclusive children, and attach an independent Poisson Galton--Watson tree with offspring mean $\lambda$ below each exclusive child. Consequently, $\ch$ and $\td\ch$ share the intersection tree $\ch^*$ while having conditionally independent exclusive branches. Marginally, both resulting trees $\ch$ and $\td\ch$ are identically distributed as Poisson Galton--Watson trees with offspring mean $\lambda$.

\paragraph*{Likelihood ratio.} For $(t,\td t)\in\cz_k\times\cz_k$, define the likelihood ratio between the correlated and independent tree-pair distributions by
\begin{equation}
L_k(t,\td t):=\frac{\bp([\ch_k]=t,[\td\ch_k]=\td t)}{\bq([\ch_k]=t,[\td\ch_k]=\td t)}.
\label{eq:LHR}
\end{equation}
Large values of $L_k(t,\td t)$ provide evidence that the two observed rooted trees arise from the correlated rather than the independent model.

\section{Tree Correlation Based Graph Matching}\label{sec:tbalign-section}

This section presents our graph matching algorithm, which is based on the approximate likelihood-ratio test in Section~\ref{sec:test}. In Section~\ref{sec:tbalign}, we describe the threshold-based alignment algorithm \texttt{TBAlign}. It first constructs a partial matching from approximate likelihood-ratio tests on local trees, and then extends this partial matching to a complete matching by seeded completion. In Section~\ref{sec:tbalign-guarantees}, we state the recovery and time-complexity guarantees of \texttt{TBAlign} and sketch their proofs; the full proofs are deferred to the appendix.

\subsection{Threshold-based Alignment Algorithm: \texttt{TBAlign}}
\label{sec:tbalign}

\begin{de}[Local tree $\ct_{u\setminus v,k}^{\tau}$]
\label{de:local_tree_new}
Let $k\ge 0$ be a nonnegative integer and $\tau\ge 1$. For two adjacent nodes $u,v\in\cv$, the depth-zero local tree $\ct_{u\setminus v,0}^{\tau}$ is the singleton rooted at $u$. For $k\ge1$, let $\ct_{u\setminus v,k}^{\tau}$ be the subgraph of $G$ rooted at $u$, induced by the nodes at distance at most $k$ from $u$ in $G\setminus v$, provided that the depth-$k$ neighborhood of $u$ in $G$ is a tree and every node $y$ in this full neighborhood satisfies $\deg_G(y)\le\tau$. Otherwise, $\ct_{u\setminus v,k}^{\tau}$ is the empty tree without any nodes.
\end{de}

The local tree $\td\ct_{\td u\setminus\td v,k}^{\tau}$ is defined analogously for the graph $\td G$. For every nonnegative integer $k$, let $\cc_k^{\tau}$ and $\td\cc_k^{\tau}$ denote the indexed collections of all depth-$k$ local trees in $G$ and $\td G$, respectively. Local trees arising from distinct ordered pairs of adjacent nodes are treated as distinct elements, so each edge $(u,v)\in\ce$ yields two local trees $\ct_{u\setminus v,k}^{\tau}$ and $\ct_{v\setminus u,k}^{\tau}$. Therefore, $|\cc_k^{\tau}|=2|\ce|$ and $|\td\cc_k^{\tau}|=2|\td\ce|$.

To simplify notation throughout the algorithm, we omit the depth subscript $d$ and the maximum degree superscript $\tau$, writing $\ct_{w\setminus u}$, $\td\ct_{\td x\setminus\td v}$, $\cc$, and $\td\cc$ in place of $\ct^{\tau}_{w\setminus u,d}$, $\td\ct^{\tau}_{\td x\setminus\td v,d}$, $\cc_d^{\tau}$, and $\td\cc_d^{\tau}$. The algorithm proceeds in the following steps.

\begin{itemize}
\item {\bf Step 1: Local tree evaluation.} Set the following global parameters:
    \begin{align}
        d=\left\lceil(\log n)^{1/3}\right\rceil,\ \tau&=\left\lceil\max\left\{\sqrt{\log n},\lambda\left(1+\frac{1}{\log\log n}\right)\right\}\right\rceil,\ 
        \xi=n^{-11},\  \theta^*=e^{\frac{1}{16}}n^{\frac{\log\log n}{2}}.
        \label{eq:TBAlign-parameters}
    \end{align}
    Given the input graphs $G$ and $\td G$, construct the local tree collections $\cc$ and $\td\cc$ using parameters $d$ and $\tau$, and initialize a $2|\ce|\times2|\td\ce|$ zero matrix $\widehat{\mathbf L}$. For each pair of nonempty local trees $(\ct_{w\setminus u},\td\ct_{\td x\setminus\td v})\in\cc\times\td\cc$, compute the matrix entry
    \begin{align}
        \widehat{\mathbf L}(\ct_{w\setminus u},\td\ct_{\td x\setminus\td v})
        =\texttt{ApproxLR}\bigl([\ct_{w\setminus u}],[\td\ct_{\td x\setminus\td v}],\lambda,s,\xi,d\bigr),
        \label{eq:TBAlign-approximate-scores}
    \end{align}
    where \texttt{ApproxLR} is the likelihood-ratio approximation algorithm defined in Algorithm~\ref{algo:approximate-LR}. If either local tree is empty, the corresponding entry in $\widehat{\mathbf L}$ remains zero.

	\item {\bf Step 2: 5-dangling-tree test on witness graphs.} Initialize an $n\times n$ zero matrix $\mathbf M_{\mathrm{TB}}$. For each node pair $(u,\td v)\in\cv\times\td\cv$, construct the weighted bipartite witness graph $\cw(u,\td v)$ between $\cn(u)$ and $\td\cn(\td v)$, where each edge $(w,\td x)$ is assigned the weight $\widehat{\mathbf L}(\ct_{w\setminus u},\td\ct_{\td x\setminus\td v})$. Given the threshold $\theta^*$, apply the 5-dangling-tree test in Section~\ref{sec:overview} to $(u,\td v)$ by computing
    \begin{align}
        \mathbf M_{\mathrm{TB}}(u,\td v)
        :=\texttt{GreedyMatching}(\cw(u,\td v),\theta^*,5)\in\{0,1\}.
        \label{eq:TBAlign-witness-test}
    \end{align}
    Here, the subroutine \texttt{GreedyMatching} starts with an empty matching and examines the edges of $\cw(u,\td v)$ in an arbitrary order. It adds an edge if it shares no endpoint with any previously added edge and has weight strictly greater than $\theta^*$. It outputs $1$ as soon as the matching reaches size $5$, and outputs $0$ if it exhausts all edges without reaching this size.
    
   	\item {\bf Step 3: Partial mapping construction.} Define the set of matched nodes $\cv' \subseteq \cv$ by
    \begin{align}
        \cv':=\{u\in\cv:\exists\td v\in\td\cv \text{ such that } \mathbf M_{\mathrm{TB}}(u,\td v)=1\}.
        \label{eq:eq_matched_sets}
    \end{align} 
    Construct a partial mapping $\pi': \cv' \to \td\cv$ by assigning each $u \in \cv'$ to an arbitrary node $\td v \in \td\cv$ that satisfies $\mathbf M_{\mathrm{TB}}(u,\td v)=1$.

    \item {\bf Step 4: Seeded completion.} Initialize a mapping $\hat\pi\gets\pi'$. For each pair $(u,\td v)$ of currently unmatched nodes, let $W(u,\td v)$ denote the number of nodes $i$ already matched by $\hat\pi$ such that $i$ is adjacent to $u$ in $G$ and $\hat\pi(i)$ is adjacent to $\td v$ in $\td G$, i.e.,
    \begin{align}
        W(u,\td v):=\bigl|\{i\in\cv:i\text{ is matched in }\hat\pi,u\in\cn(i),\td v\in\td\cn(\hat\pi(i))\}\bigr|.
        \label{eq:def_common_seed}
    \end{align}
    We complete the mapping $\hat\pi:\cv\to\td\cv$ by iteratively adding the unmatched pairs. At each iteration, we set $\hat\pi(u)=\td v$ for a currently unmatched pair $(u,\td v)$ that maximizes $W(u,\td v)$, breaking ties arbitrarily. We then update $W$ for all remaining unmatched pairs, and repeat this process until $\hat\pi$ is defined on $\cv$.
\end{itemize}

\begin{algorithm}[htbp]
\caption{\texttt{TBAlign}: Threshold-based alignment algorithm}
\label{algo:TBAlign}
\begin{algorithmic}[1]
\REQUIRE Two graphs $G$ and $\td G$, average degree $\lambda$, and correlation $s$.
\ENSURE A partial mapping $\pi'$ and a complete mapping $\hat\pi$ defined on $\cv$.
\item[] \COMMENT{Step 1: Local tree evaluation}
\STATE Set $d$, $\tau$, $\xi$, and $\theta^*$ via~\eqref{eq:TBAlign-parameters}.
\STATE Construct local tree collections $\cc$ and $\td\cc$, and initialize a $2|\ce|\times 2|\td\ce|$ zero matrix $\widehat{\mathbf L}$.
\FOR{each pair $(\ct_{w\setminus u},\td\ct_{\td x\setminus\td v})\in\cc\times\td\cc$}
    \IF{$\ct_{w\setminus u}$ and $\td\ct_{\td x\setminus\td v}$ are both nonempty}
        \STATE $\widehat{\mathbf L}(\ct_{w\setminus u},\td\ct_{\td x\setminus\td v}) \gets \texttt{ApproxLR}\bigl([\ct_{w\setminus u}],[\td\ct_{\td x\setminus\td v}],\lambda,s,\xi,d\bigr)$.
    \ENDIF
\ENDFOR
\item[] \COMMENT{Step 2: 5-dangling-tree test on witness graphs}
\STATE Initialize $\mathbf M_{\mathrm{TB}}\gets\mathbf 0_{n\times n}$.
\FOR{each pair $(u,\td v)\in\cv\times\td\cv$}
    \STATE Construct the weighted bipartite witness graph $\cw(u,\td v)$.
    \STATE $\mathbf M_{\mathrm{TB}}(u,\td v)\gets\texttt{GreedyMatching}(\cw(u,\td v),\theta^*,5)$.
\ENDFOR
\item[] \COMMENT{Step 3: Partial mapping construction}
\STATE Initialize a partial mapping $\pi':\cv'\to\td\cv$, where $\cv'$ is given by~\eqref{eq:eq_matched_sets}.
\FOR{each $u\in\cv'$}
    \STATE Choose an arbitrary $\td v\in\td\cv$ satisfying $\mathbf M_{\mathrm{TB}}(u,\td v)=1$.
    \STATE $\pi'(u)\gets\td v$.
\ENDFOR
\item[] \COMMENT{Step 4: Seeded completion}
\STATE Initialize a mapping $\hat\pi\gets\pi'$.
\STATE Compute $W(u,\td v)$ for all currently unmatched pairs $(u,\td v)$ via~\eqref{eq:def_common_seed}.
\WHILE{there exist unmatched nodes in $\cv$}
    \STATE Choose an unmatched pair $(u^*,\td v^*)$ that maximizes $W(u,\td v)$ under $\hat\pi$, breaking ties arbitrarily.
    \STATE $\hat\pi(u^*)\gets\td v^*$.
    \STATE Update $W(u,\td v)$ for all remaining unmatched pairs.
\ENDWHILE
\RETURN $(\pi',\hat\pi)$.
\end{algorithmic}
\end{algorithm}

\subsection{Recovery and Computational Guarantees}
\label{sec:tbalign-guarantees}

The following two theorems establish the recovery and computational guarantees of \texttt{TBAlign}.

\begin{restatable}[Recovery guarantees]{theo}{TBAlignRecovery}
\label{th:TBAlign}
Consider $(G,\td G)\sim\mathrm{CER}(n,\lambda,s)$, where $\lambda=\omega(1)$, $\lambda=(\log n)^{O(1)}$, and $s\in(\sqrt{C_{\mathrm{Otter}}},1]$ is a constant independent of $n$.
Then, with probability $1-o(1)$, the partial mapping $\pi'$ output by Algorithm~\ref{algo:TBAlign} achieves error-free almost-exact recovery. More precisely,
\begin{enumerate}
	
    \item $\pi'$ contains no wrong match: $\pi'(u)=\pi(u)$ for every node $u$ matched under $\pi'$;
    \item Almost all nodes are matched:
    \begin{align}
        \bigl|\{u\in\cv:u\text{ is matched under }\pi'\}\bigr|\ge n(1-o(1)).
        \label{eq:TBAlign-recovery}
    \end{align}
\end{enumerate}
Moreover, if $\lambda s\ge(1+\eps)\log n$ for some arbitrarily small constant $\eps>0$, then the complete mapping $\hat\pi$ achieves exact recovery, i.e., $\hat\pi=\pi$, with probability $1-o(1)$.
\end{restatable}

\paragraph*{Proof sketch of Theorem~\ref{th:TBAlign}.} We outline the main ideas of the proof here, and the complete proof is given in Appendix~\ref{sec:5dang}. To establish the error-free almost-exact recovery of $\pi'$, it suffices to show that the matrix $\mathbf M_{\mathrm{TB}}$ computed in Step~2 has the following two properties with high probability:
\begin{itemize}
	\item[(i)] $\mathbf M_{\mathrm{TB}}(u,\td v)=0$ for every pair $(u,\td v)$ with $\td v\neq\td u$;
	\item[(ii)] $\mathbf M_{\mathrm{TB}}(u,\td u)=1$ for all but $o(n)$ nodes $u$.
\end{itemize}

By~(i), Step~3 assigns every node $u\in\cv'$ to $\td u$, so $\pi'$ contains no wrong match. By~(ii), $\pi'$ matches $n-o(n)$ nodes. For simplicity of presentation, we sketch the proofs of (i) and (ii) for $\tau=\infty$, in which case no degree cutoff is applied.
To extend the result to the cutoff $\tau$, we show that the cutoff only removes nodes from witness graphs, which creates no wrong match by the monotonicity of greedy matching under node deletion. Moreover, the cutoff changes only $o(n)$ of the witness graphs $\cw(u,\td u)$ with high probability, as shown in Lemma~\ref{lem:TBAlignBadNodes}.

{\bf Property~(i).} We follow the analysis of the 5-dangling-tree test in Section~\ref{sec:overview}. In Step~2, every edge $(w,\td x)$ added by \texttt{GreedyMatching} has weight greater than $\theta^*$, that is, its local trees pass the correlation test $\indi\{\widehat{\mathbf L}(\ct_{w\setminus u},\td\ct_{\td x\setminus\td v})>\theta^*\}$.
We define $\ca_1$ as the event that every pair of local trees that passes the correlation test shares a node in the union graph of $G$ and $\td G$, and $\ca_2$ as the event that the union graph contains no connected subgraph with at most $6d+6$ nodes and two cycles. To bound the probability of $\ca_1$, notice that two local trees that share no node are nearly independent, so their probability of passing the correlation test can be bounded using its type-I error. This error is small enough for a union bound over all pairs of local trees, and Lemma~\ref{lem:new_A1} shows that $\ca_1$ holds with high probability. For $\ca_2$, we use a first-moment bound on the number of small connected subgraphs with two cycles. The analysis in Lemma~\ref{lem:subgraph} shows that $\ca_2$ fails with probability $n^{-1+o(1)}$.

Finally, under $\ca_1\cap\ca_2$, assume by contradiction that a pair $(u,\td v)$ with $\td v\neq\td u$ passes the 5-dangling-tree test. Then \texttt{GreedyMatching} adds five edges of $\cw(u,\td v)$, and by $\ca_1$, the two downstream local trees of each edge share a node in the union graph. Each edge thus yields a path of length at most $2d+2$ from $u$ to $\td v$ through the shared node. Since the five edges share no endpoint, these five paths include at least three distinct ones, and their union is a connected subgraph with at most $6d+6$ nodes and two cycles, which contradicts $\ca_2$. Hence $\ca_1\cap\ca_2$ implies property~(i), and the detailed proof is given in Lemma~\ref{lem:new-sufficient}.

{\bf Property~(ii).}  Intuitively, for a fixed node $u$, the nodes $u$ and $\td u$ have about $\lambda s$ common neighbors, and the two downstream local trees of each common neighbor are correlated and pass the correlation test with high probability. This implies that $(u,\td u)$ has many neighbor pairs whose local trees pass the correlation test. We define $\ca_{3,u}$ as the event that $\mathbf M_{\mathrm{TB}}(u,\td u)=1$. 

To bound the probability of $\ca_{3,u}$, we only consider the neighbor pairs $(w,\td w)$ formed by the common neighbors $w$ of $u$ and $\td u$. We first couple the neighborhoods of $u$ and $\td u$ up to depth $d+1$ with a correlated tree pair $(\ch_{d+1},\td\ch_{d+1})$ drawn from $\bp$, and the coupling succeeds with probability $1-o(1)$. Under this coupling, the number of common neighbors follows $\poi(\lambda s)$. For each common neighbor $w$, the local trees $\ct_{w\setminus u}$ and $\td\ct_{\td w\setminus\td u}$ form a correlated tree pair of depth $d$, so they pass the correlation test with high probability by Theorem~\ref{thm:LRT}. The local tree pairs of different common neighbors are independent and are tested by separate invocations of \texttt{ApproxLR}. By Poisson thinning, the number of common neighbors whose local trees pass the correlation test is at least $9$ with probability $1-o(1)$. Since $9$ such neighbor pairs force \texttt{GreedyMatching} to output~$1$ in whatever order it examines the edges, $\ca_{3,u}$ holds with high probability, and the detailed proof is given in Lemma~\ref{lem:new-correct}. Finally, the expected number of nodes $u$ for which $\ca_{3,u}$ fails is $o(n)$, and Markov's inequality shows that property~(ii) holds with high probability.

{\bf Exact recovery.} Suppose that $\lambda s\ge(1+\eps)\log n$. Since $\pi'$ contains no wrong match, its $n-o(n)$ matched nodes serve as correct seeds for Step~4. We adapt the seeded graph matching method in~\cite{Cheng26Chandelier} to our greedy rule, which adds an unmatched pair $(u,\td v)$ maximizing $W(u,\td v)$ at each iteration of Step~4. We show that this rule yields $\hat\pi=\pi$ with high probability, and the detailed proof is given in Lemma~\ref{lem:greedy-seeded-completion} in Appendix~\ref{appd:seeded}.

\begin{restatable}[Computational complexity]{theo}{TBAlignRuntime}
\label{th:TBAlign-runtime}
Under the assumptions of Theorem~\ref{th:TBAlign}, for every realization of the input graphs $(G,\td G)$ and every execution of \texttt{TBAlign}, the computational complexity is
\begin{align}
    O\left(n^2+\bigl(|\ce|+|\td\ce|+|\ce|\,|\td\ce|\bigr)n^{o(1)}\right).
    \label{eq:TBAlign-deterministic-complexity}
\end{align}
Moreover, with probability $1-o(e^{-n})$, this complexity is bounded by $n^{2+o(1)}$.
\end{restatable}

The proof of Theorem~\ref{th:TBAlign-runtime} bounds the computational complexity of each step of Algorithm~\ref{algo:TBAlign}. The degree cutoff $\tau$ ensures that every node of a nonempty local tree has at most $\tau$ children, so each invocation of \texttt{ApproxLR} has computational complexity $n^{o(1)}$. The proof details are given in Appendix~\ref{appd:runtime}.

\begin{remark}[Rank-based implementation]
\label{rem:rank-based}
Following~\cite{Cheng26RBAlign}, Step~2 of \texttt{TBAlign} can be implemented without specifying $\theta^*$. We process the entries of $\widehat{\mathbf L}$ in non-increasing order, applying the same greedy 5-dangling-tree criterion, until $\lceil n-n/\lambda\rceil$ nodes are matched or all entries are examined. The same comparison argument as in~\cite[Lemma~1]{Cheng26RBAlign} shows that, with high probability, this procedure stops before examining any entry at most $\theta^*$. Thus, with the same seeded completion step, this implementation retains the recovery guarantees of \texttt{TBAlign}. Sorting costs $O(|\ce||\td\ce|\log n)$, so the overall time complexity remains $n^{2+o(1)}$ with high probability.
\end{remark}

\section{Approximate Likelihood-Ratio Testing for \GW Trees}\label{sec:test}

This section develops an approximate likelihood-ratio test to distinguish correlated \GW trees from independent ones. The following theorem gives a test $\Psi$ with an explicit threshold, type-I error $O(n^{-11})$, and power $1-O(\lambda^{-1})$.

\begin{restatable}{theo}{RankLRT}
\label{thm:LRT}
Let $s\in(\sqrt{C_{\mathrm{Otter}}},1]$ be a constant independent of $n$, and suppose that $\lambda=\omega(1)$ with $\lambda=(\log n)^{O(1)}$. Set
\[
    d=\left\lceil(\log n)^{1/3}\right\rceil,
    \qquad
    \theta^*=e^{1/16}n^{(\log\log n)/2},
    \qquad
    \xi=n^{-11}.
\]
Then the algorithm \texttt{ApproxLR} defined in Algorithm~\ref{algo:approximate-LR} outputs a nonnegative estimate $\widehat L_d(t,\td t)$ for any input $(t,\td t)\in\cz_d\times\cz_d$ and parameters $\lambda,s,\xi,d$. Moreover, the test
\begin{align}
    \Psi(t,\td t)=\indi\left\{\widehat L_d(t,\td t)>\theta^*\right\}
    \label{eq:rank-LRT}
\end{align}
satisfies
\begin{align}
    \bq\bigl(\Psi([\ch_d],[\td\ch_d])=1\bigr)
    &\le n^{-(\log\log n)/2}+n^{-11},
    \label{eq:rank-LRT-typeI}\\
    \bp\bigl(\Psi([\ch_d],[\td\ch_d])=1\bigr)
    &\ge 1-\frac{7}{\lambda}-e^{1/8}n^{-(\log\log n)/2}-n^{-11}
    \label{eq:rank-LRT-power}
\end{align}
for all sufficiently large $n$.
\end{restatable}

The remainder of this section is organized as follows. In Section~\ref{sec:subset-dp-calculation}, we reduce likelihood-ratio computation to a sequence of permanent computations for nonnegative matrices and construct \texttt{ApproxLR} using the fully polynomial randomized approximation scheme (FPRAS) developed by Jerrum, Sinclair, and Vigoda~\cite{Jerrum04Permanent}, and bound its computational complexity. In Section~\ref{sec:tree-test-statistical}, we sketch the proof of Theorem~\ref{thm:LRT}. The analysis of \texttt{ApproxLR} is given in Appendix~\ref{appd:approxlr}, and the full proof of Theorem~\ref{thm:LRT} is given in Appendix~\ref{appd:lrt}.

\subsection{Efficient Approximation of Local-Tree Likelihood Ratios}
\label{sec:subset-dp-calculation}

This subsection presents the \texttt{ApproxLR} algorithm used to compute the estimate $\widehat L_d$ in Theorem~\ref{thm:LRT}. We express the likelihood-ratio recursion in terms of nonnegative matrix permanents and then define \texttt{ApproxLR}. 

For finite sets $A$ and $B$, let $\mathrm{Inj}(A,B)$ denote the set of injective mappings from $A$ to $B$. Let $c$ and $\td c$ be non-negative integers, and let $\mathbf Z=(Z_{ij})_{i\in[1:c],\,j\in[1:\td c]}$ be a nonnegative $c\times\td c$ matrix. For any integer $q\in[0:\min\{c,\td c\}]$, we define
\begin{align}
    S_q(\mathbf Z):=\frac{1}{q!}\sum_{\substack{\sigma\in\mathrm{Inj}([1:q],[1:c])\\\td\sigma\in\mathrm{Inj}([1:q],[1:\td c])}}\prod_{a=1}^{q}Z_{\sigma(a),\td\sigma(a)}.
    \label{eq:approximate-L-matching-sum}
\end{align}
By convention, we set $S_0(\mathbf Z)=1$. The following lemma expresses the likelihood-ratio recursion from~\cite[Lemma~2.1]{Luca24LRT} in terms of $S_q$.

\begin{restatable}[Likelihood-ratio recursion]{lem}{LRRecursion}
\label{lem:approximate-L-recursion}
Let $k\ge1$, $\lambda>0$, and $s\in(0,1]$. Fix $(t,\td t)\in\cz_k\times\cz_k$, and let $t_1, \ldots, t_c$ and $\td t_1, \ldots, \td t_{\td c}$ denote the unlabeled subtrees rooted at the children of $t$ and $\td t$, respectively. Define the matrix $\mathbf Z=(Z_{ij})$ by setting
\[
    Z_{ij}:=\frac{s}{\lambda}L_{k-1}(t_i,\td t_j),\qquad (i,j)\in[1:c]\times[1:\td c].
\]
Then, under the convention $0^0=1$, we have
\begin{align}
    L_k(t,\td t)=e^{\lambda s}\sum_{q=0}^{\min\{c,\td c\}}(1-s)^{c+\td c-2q}S_q(\mathbf Z).
    \label{eq:approximate-L-exact-recursion}
\end{align}
Furthermore, for the base case at depth zero, $L_0(t,\td t)=1$ for the unique pair in $\cz_0\times\cz_0$.
\end{restatable}

As discussed in Section~\ref{sec:overview}, evaluating the sums $S_q(\mathbf Z)$ in this recursion directly has time complexity superpolynomial in $n$ when $\lambda$ is at least of order $\log n$. To avoid this cost, we express $S_q(\mathbf Z)$ as the permanent of a nonnegative matrix, which can be approximated by the FPRAS. For an $m\times m$ matrix $\mathbf A$, define the permanent of $\mathbf A$ by
\[
    \operatorname{per}(\mathbf A):=\sum_{\sigma\in\mathrm{Inj}([1:m],[1:m])}\prod_{i=1}^{m}A_{i,\sigma(i)}.
\]
Since every injection from $[1:m]$ to itself is a bijection, this sum is taken over all permutations of $[1:m]$. By convention, we set $\operatorname{per}(\mathbf A)=1$ for $m=0$. The following lemma expresses $S_q(\mathbf Z)$ as the permanent of an augmented matrix.

\begin{restatable}[{\cite[Theorem~2.1]{Friedland06Matchings}}]{lem}{PermanentIdentity}\label{lem:approximate-L-matching-permanent}
For a nonnegative $c\times\td c$ matrix $\mathbf Z$ and an integer $q\in[0:\min\{c,\td c\}]$, define the $(c+\td c-q)\times (c+\td c-q)$ matrix
\begin{align}
    \mathbf A_q(\mathbf Z):=
    \begin{pmatrix}
        \mathbf Z & \mathbf 1_{c\times(c-q)}\\
        \mathbf 1_{(\td c-q)\times\td c} & \mathbf 0_{(\td c-q)\times(c-q)}
    \end{pmatrix},
    \label{eq:approximate-L-augmented-matrix}
\end{align}
where $\mathbf 1$ and $\mathbf 0$ denote matrices of all ones and all zeros whose dimensions are specified by their subscripts. Then we have
\begin{align}
    S_q(\mathbf Z)=\frac{\operatorname{per}(\mathbf A_q(\mathbf Z))}{(c-q)!(\td c-q)!}.
    \label{eq:approximate-L-permanent-identity}
\end{align}
\end{restatable}

We defer the proofs of Lemmas~\ref{lem:approximate-L-recursion} and~\ref{lem:approximate-L-matching-permanent} to Appendix~\ref{appd:approxlr-lemmas}.
Throughout this subsection, we use $\operatorname{Poly}$ to denote a polynomial with fixed nonnegative coefficients and degree independent of the problem parameters. We use the following FPRAS for nonnegative matrix permanents developed by Jerrum, Sinclair, and Vigoda~\cite{Jerrum04Permanent}.

%Throughout this subsection, we use $\operatorname{Poly}$ to denote a polynomial with fixed nonnegative coefficients and degree independent of the problem parameters. We rely on the following permanent approximation algorithm by Jerrum, Sinclair, and Vigoda~\cite{Jerrum04Permanent}.

\begin{lem}[{FPRAS Permanent Approximation~\cite[Theorem~1.1]{Jerrum04Permanent},\cite[Corollary~2]{Chen26FastPermanent}}]
\label{lem:approximate-L-permanent-FPRAS}
Let $\mathbf A$ be a nonnegative $m\times m$ matrix, and let $\delta, \eta \in (0,\frac12)$. There exists a randomized algorithm $\texttt{ApproxPer}(\mathbf A,\delta,\eta)$ that outputs a nonnegative random variable satisfying
\begin{align}
    \P\left((1-\delta)\operatorname{per}(\mathbf A)\le\texttt{ApproxPer}(\mathbf A,\delta,\eta)\le(1+\delta)\operatorname{per}(\mathbf A)\right)\ge1-\eta,
    \label{eq:approximate-L-permanent-guarantee}
\end{align}
where the probability is taken over the random choice of $\texttt{ApproxPer}$. Its time complexity is bounded by $\operatorname{Poly}(m,\delta^{-1},\log(\eta^{-1}))$.
\end{lem}

\begin{remark}[Independent executions]
\label{rem:independent-executions}
The algorithm in Lemma~\ref{lem:approximate-L-permanent-FPRAS} can be viewed as a deterministic function $\texttt{ApproxPer}(\mathbf A,\delta,\eta;R)$, where $R$ is a sequence of independent unbiased random bits~\cite[Section~2.1]{Jerrum04Permanent}. These bits supply all randomness used by the algorithm, including the random choices in its Markov chain simulations; fixing them therefore makes the algorithm deterministic. Throughout this paper, each execution uses a fresh sequence of random bits, independent of its input matrix and all randomness used by preceding executions. Consequently, for independent random matrices $\mathbf A,\mathbf B$ and fixed $\delta,\eta$, the outputs $\texttt{ApproxPer}(\mathbf A,\delta,\eta;R_1)$ and $\texttt{ApproxPer}(\mathbf B,\delta,\eta;R_2)$ are independent, since they are deterministic functions of the independent pairs $(\mathbf A,R_1)$ and $(\mathbf B,R_2)$, respectively.
\end{remark}

We now define the algorithm $\texttt{ApproxLR}$ for a fixed pair of trees $(t,\td t)\in\cz_d\times\cz_d$ with parameters $\lambda>0$, $s\in(0,1]$, $d\ge 0$ and an error tolerance $\xi\in(0,1/2)$. Let $\rho$ and $\td\rho$ denote their respective roots. For $j\in[0:d]$, let $\cs_j(t)$ and $\cs_j(\td t)$ denote the sets of vertices at depth $j$ in $t$ and $\td t$, respectively. For each $h\in[0:d]$ and vertices $u\in\cs_{d-h}(t)$ and $\td v\in\cs_{d-h}(\td t)$, let $t_u$ and $\td t_{\td v}$ denote the subtrees in $\cz_h$ rooted at $u$ and $\td v$, respectively. Let $\sfch_t(u)$ denote the set of children of $u$ in $t$, and define $\sfch_{\td t}(\td v)$ analogously. The algorithm proceeds in the following steps.

\begin{itemize}

\item {\bf Step 1: Choose the approximation parameters.} Compute the maximum degree $D$ among nodes in the trees $t$ and $\td t$
    \begin{equation}
        D:=\max\left\{2,\max_{u\in\cv(t)}|\sfch_t(u)|,\max_{\td v\in\cv(\td t)}|\sfch_{\td t}(\td v)|\right\},
        \label{eq:approximate-L-input-degree}
    \end{equation}
    and set
    \begin{align}
    \delta:=\frac{1}{64D^d},\qquad \eta:=\frac{\xi}{(D+1)\sum_{j=0}^{d}D^{2j}}.
    \label{eq:approximate-L-parameters}
	\end{align}
    These parameters are computed once and remain fixed globally; they are not recomputed for individual subtrees.
    
    \item {\bf Step 2: Initialize at remaining depth $0$.} For every pair $(u,\td v)\in\cs_d(t)\times\cs_d(\td t)$, set
    \begin{equation}
        \widehat\ell(u,\td v)\gets1.
        \label{eq:approximate-L-initialization}
    \end{equation}
    If either $\cs_d(t)$ or $\cs_d(\td t)$ is empty, no initialization occurs at this depth.

    \item {\bf Step 3: Recursive Calculation.} For $h=1,\ldots,d$ and every pair $(u,\td v)\in\cs_{d-h}(t)\times\cs_{d-h}(\td t)$, compute $\widehat\ell(u,\td v)$ as follows. Enumerate the child sets as $\sfch_t(u)=\{y_1,\ldots,y_c\}$ and $\sfch_{\td t}(\td v)=\{\td y_1,\ldots,\td y_{\td c}\}$, and construct the $c\times\td c$ matrix $\widehat{\mathbf Z}(u,\td v)$ by setting
    \begin{align}
        \widehat{\mathbf Z}(u,\td v)_{ij}:=\frac{s}{\lambda}\widehat\ell(y_i,\td y_j),\qquad (i,j)\in[1:c]\times[1:\td c].
        \label{eq:approximate-L-approximate-entries}
    \end{align}
    Set $\widehat S_0(u,\td v)=1$. For each $q\in[1:\min\{c,\td c\}]$, construct matrix $\mathbf A_q(\widehat{\mathbf Z}(u,\td v))$ according to~\eqref{eq:approximate-L-augmented-matrix}, and compute
    \begin{align}
        \widehat S_q(u,\td v):=\frac{\texttt{ApproxPer}(\mathbf A_q(\widehat{\mathbf Z}(u,\td v)),\delta,\eta)}{(c-q)!(\td c-q)!}.
        \label{eq:approximate-L-approximate-matching-sum}
    \end{align}
    Here, $\texttt{ApproxPer}$ is the randomized algorithm in Lemma~\ref{lem:approximate-L-permanent-FPRAS}. Finally, set
    \begin{align}
        \widehat\ell(u,\td v)\gets e^{\lambda s}\sum_{q=0}^{\min\{c,\td c\}}(1-s)^{c+\td c-2q}\widehat S_q(u,\td v),
        \label{eq:approximate-L-update}
    \end{align}
    where we maintain the convention $0^0=1$.
    \item {\bf Step 4: Return the root entry.} Output $\widehat \ell(\rho,\td\rho)$ for roots $(\rho,\td\rho)$:
    \begin{align}
        \widehat L_d(t,\td t):=\widehat\ell(\rho,\td\rho).
        \label{eq:approximate-L-output}
    \end{align}
\end{itemize}

The procedures are summarized in Algorithm~\ref{algo:approximate-LR}. When $d=0$, it returns $1$ by Step~2. For a positive depth, a pair of vertices with no children has value $e^{\lambda s}$ by Step~3, which gives the same likelihood-ratio as the recursion~\eqref{eq:approximate-L-exact-recursion}. 
Following Remark~\ref{rem:independent-executions}, for fixed $\lambda,s,d,\xi$, the output of the $i$-th execution of \texttt{ApproxLR} is a deterministic function of its input tree pair $(T_i,\td T_i)$ and the collection $\mathcal R_i$ of independent random-bit sequences supplying all randomness used by its calls to \texttt{ApproxPer}. These collections $\mathcal R_i$ are mutually independent across executions and jointly independent of all input tree pairs. Consequently, mutually independent input tree pairs $(T_i,\td T_i)$ yield mutually independent outputs, since these outputs are deterministic functions of the mutually independent triples $(T_i,\td T_i,\mathcal R_i)$.
The following lemma shows that the computational complexity of \texttt{ApproxLR} is polynomial in $D^d$ and $\log(\xi^{-1})$. Its proof is presented in Appendix~\ref{sec:approximate-L-complexity}.

\begin{restatable}[Computational complexity of \texttt{ApproxLR}]{lem}{ApproxLRComplexity}
\label{lem:approximate-L-complexity}
Let $\lambda>0$, $s\in(0,1]$, $\xi\in(0,1/2)$, and let $d\ge1$ be an integer. For a fixed pair $(t,\td t)\in\cz_d\times\cz_d$, let $D$ be the parameter defined in~\eqref{eq:approximate-L-input-degree}. The computational complexity of Algorithm~\ref{algo:approximate-LR} with parameters $\lambda, s, \xi$, and $d$ is
\begin{align}
    (D+1)\sum_{j=0}^{d}D^{2j}\operatorname{Poly}\left(D,D^d,\log(\xi^{-1})+(d+1)\log D\right).
    \label{eq:approximate-L-complexity-general}
\end{align}
In particular, if $\xi=n^{-11}$ and $D^d=n^{o(1)}$, then the computational complexity is $n^{o(1)}$.
\end{restatable}

\begin{algorithm}[htbp]
\caption{\texttt{ApproxLR}: likelihood-ratio approximation for one tree pair}
\label{algo:approximate-LR}
\begin{algorithmic}[1]
\REQUIRE $(t,\td t)\in\cz_d\times\cz_d$, $\lambda>0$, $s\in(0,1]$, $\xi\in(0,1/2)$, integer $d\ge0$. 
\ENSURE The nonnegative approximation $\widehat L_d(t,\td t)$.
\STATE Compute $D$, $\delta$, and $\eta$ by~\eqref{eq:approximate-L-input-degree} and~\eqref{eq:approximate-L-parameters}.
\STATE For every pair $(u,\td v)\in\cs_d(t)\times\cs_d(\td t)$, set $\widehat\ell(u,\td v)\gets1$ by~\eqref{eq:approximate-L-initialization}.
\FOR{$h=1,\ldots,d$}
    \FOR{each $(u,\td v)\in\cs_{d-h}(t)\times\cs_{d-h}(\td t)$}
        \STATE Enumerate the children of $u$ and $\td v$ as $\{y_1,\ldots,y_c\}$ and $\{\td y_1,\ldots,\td y_{\td c}\}$, respectively.
        \STATE Form $\widehat{\mathbf Z}(u,\td v)$ by~\eqref{eq:approximate-L-approximate-entries}.
        \STATE Set $\widehat S_0(u,\td v)\gets1$.
        \FOR{$q=1,\ldots,\min\{c,\td c\}$}
            \STATE Form $\mathbf A_q(\widehat{\mathbf Z}(u,\td v))$ by~\eqref{eq:approximate-L-augmented-matrix}.
            \STATE Compute $\widehat S_q(u,\td v)$ by~\eqref{eq:approximate-L-approximate-matching-sum}.
        \ENDFOR
        \STATE Compute $\widehat\ell(u,\td v)$ by~\eqref{eq:approximate-L-update}.
    \ENDFOR
\ENDFOR
\RETURN $\widehat L_d(t,\td t):=\widehat\ell(\rho,\td\rho)$.
\end{algorithmic}
\end{algorithm}

\subsection{Proof sketch of Theorem~\ref{thm:LRT}.}
\label{sec:tree-test-statistical}

We outline the main ideas of the proof here, and the complete proof is given in Appendix~\ref{appd:lrt}. The proof is based on a lower bound of the power in a likelihood-ratio test at every threshold $\theta$ in terms of the errors of another test. Let $\Phi:\cz_d\times\cz_d\to\{0,1\}$ be a test on pairs of rooted unlabeled trees of depth at most $d$, and let $\theta>0$ be a real number. Since $L_d$ is the ratio of $\bp$ to $\bq$, every tree pair $(t,\td t)$ with $L_d(t,\td t)\le\theta$ has
\begin{align}
    \bp([\ch_d]=t,[\td \ch_d]=\td t)\le\theta\,\bq([\ch_d]=t,[\td \ch_d]=\td t).\label{eq:change_measure}
\end{align}
The power of the likelihood ratio test $\mathbbm{1}\{L_d([\ch_d],[\td\ch_d])\le\theta\}$ can be bounded by applying~\eqref{eq:change_measure} and summing over the tree pairs accepted by $\Phi$, i.e.,
\begin{align}
	\bp\bigl(L_d([\ch_d],[\td\ch_d])\le\theta\bigr)
    &\le\bp\bigl(\Phi([\ch_d],[\td\ch_d])=0\bigr)+\bp\bigl(L_d([\ch_d],[\td\ch_d])\le\theta,\Phi([\ch_d],[\td\ch_d])=1 \bigr)\nonumber\\
    &\le\bp\bigl(\Phi([\ch_d],[\td\ch_d])=0\bigr)+\theta\, \bq\bigl(\Phi([\ch_d],[\td\ch_d])=1\bigr).
    \label{eq:aux-LR-power}
\end{align}

Inequality~\eqref{eq:aux-LR-power} shows that the likelihood-ratio test $\indi\{L_d>\theta\}$ has power close to $1$ if $\Phi$ has power close to $1$ and type-I error $o(\theta^{-1})$. We next construct such a test $\Phi$ and reduce the error events of $\Psi$ to events of $L_d$, by analyzing the approximation error of $\widehat L_d$.

{\bf Auxiliary test.} Following~\cite{Cheng26RBAlign}, we construct such an auxiliary test $\Phi$ based on a collection $\cf_{d-l}$ of tree pairs of depth $d-l$, where $l=\lceil2\log\log n\rceil$. Given a tree pair $(t,\td t)$ of depth $d$, we form a bipartite graph between the nodes in generation $l$ of $t$ and those of $\td t$. A node of $t$ and a node of $\td t$ are adjacent in the bipartite graph if the pair of subtrees rooted at them lies in $\cf_{d-l}$. The test $\Phi$ accepts $(t,\td t)$ if this bipartite graph has a matching of size at least $(\lambda s)^l/(2e)$. We show that $\Phi$ has type-I error at most $n^{-\log\log n}$ and power at least $1-\frac7\lambda$. By~\eqref{eq:aux-LR-power}, these two bounds imply that the likelihood-ratio test $\indi\{L_d>\theta\}$ has power at least $1-\frac7\lambda-\theta n^{-\log\log n}$ for every $\theta>0$. The construction of $\Phi$ and the detailed proof of these two bounds are given in Appendix~\ref{sec:aux_test}.

{\bf Approximation of $L_d$ by $\widehat L_d$.} The power bound above holds for the likelihood-ratio test $\indi\{L_d>\theta\}$, whereas the test $\Psi$ in Theorem~\ref{thm:LRT} thresholds the estimate $\widehat L_d$ of $L_d$. To relate the two tests, we first show that $\widehat L_d$ is close to $L_d$. \texttt{ApproxLR} computes $\widehat L_d$ from the leaves to the root through the recursion in Lemma~\ref{lem:approximate-L-recursion}, and uses the FPRAS at every level to approximate the permanents with a relative error $\delta$ and error probability at most $\eta$. Both the approximation error and the failure probability accumulate over the levels. We show that the choice $\delta=1/(64D^d)$ in Step~1 compensates for this error accumulation and keeps $\widehat L_d$ within a factor $e^{\pm1/16}$ of $L_d$. The failure probabilities add up over all calls, whose number is polynomial in $D^d$, and the choice of $\eta$ keeps their sum at most $\xi=n^{-11}$. Hence, conditioned on any input pair, $\widehat L_d$ lies between $e^{-1/16}L_d$ and $e^{1/16}L_d$ with probability at least $1-n^{-11}$. The detailed proof is given in Appendix~\ref{sec:approximate-L-guarantees}.

{\bf Type-I error and power of $\Psi$.} Since $\widehat L_d$ is close to $L_d$, we can reduce the error events of $\Psi$ to events of $L_d$. Specifically, when $\widehat L_d$ lies between $e^{-1/16}L_d$ and $e^{1/16}L_d$, the event $\Psi=1$ implies $L_d>e^{-1/16}\theta^*$, and the event $\Psi=0$ implies $L_d\le e^{1/16}\theta^*$. It remains to bound the probabilities of these two events of $L_d$ under $\bq$ and $\bp$, respectively.

 Throughout the analysis, $L_d$ and $\Phi$ are evaluated at the random tree pair $([\ch_d],[\td\ch_d])$. For the type-I error, notice that $L_d$ has expectation one under $\bq$ because it is the likelihood ratio of $\bp$ to $\bq$. Markov's inequality then gives $\bq(L_d>e^{-1/16}\theta^*)\le e^{1/16}/\theta^*$. For the power, we apply~\eqref{eq:aux-LR-power} with $\theta=e^{1/16}\theta^*$ and obtain
\[
    \bp\bigl(L_d\le e^{1/16}\theta^*\bigr)\le\bp(\Phi=0)+e^{1/16}\theta^*\,\bq(\Phi=1)\le\frac{7}{\lambda}+e^{1/16}\theta^*n^{-\log\log n},
\]
where the last inequality follows from the two bounds on $\Phi$. Since the type-I error bound decreases in $\theta^*$ while the last term of the display increases in $\theta^*$, we choose $\theta^*=e^{1/16}n^{(\log\log n)/2}$ to make both of order $n^{-(\log\log n)/2}$. Finally, adding the failure probability $n^{-11}$ of the approximation to both bounds yields~\eqref{eq:rank-LRT-typeI} and~\eqref{eq:rank-LRT-power}. The detailed proof of both error bounds is given in Appendix~\ref{appd:pf-lrt}.

\section{Conclusion}\label{sec:conclusion}

We proposed a near-quadratic-time algorithm for graph matching in the correlated \erdos--\renyi graph pair model $\mathrm{CER}(n,\lambda,s)$. For every constant $s>\sqrt{C_{\mathrm{Otter}}}$ and every $\lambda=\omega(1)$ with $\lambda=(\log n)^{O(1)}$, our algorithm achieves error-free almost-exact recovery with high probability, and it achieves exact recovery with high probability if in addition $\lambda s\ge(1+\eps)\log n$ for a constant $\eps>0$. In both cases, its time complexity is $n^{2+o(1)}$ with high probability.  
Our results on tree correlation test are also of independent interest. We proposed the randomized algorithm \texttt{ApproxLR}, which approximates the likelihood ratio of every pair of trees within a constant factor with high probability; for trees of depth $d$ whose nodes have at most $D$ children, its time complexity is polynomial in $D^d$. Based on this approximation, we proposed a test with an explicit threshold, which has type-I error $O(n^{-11})$ and power $1-O(1/\lambda)$ for every diverging $\lambda$ with $\lambda=(\log n)^{O(1)}$.

The analysis of our algorithm separates the properties of the graphs from those of the correlation test. The 5-dangling-tree test excludes every wrong match as long as every pair of local trees that passes the correlation test shares a node in the union graph and the union graph contains no small connected subgraph with two cycles. In $\mathrm{CER}(n,\lambda,s)$, both conditions hold with high probability. We therefore expect our approach to extend to other correlated graph models in which the local neighborhoods are close in distribution to pairs of correlated trees, such as correlated stochastic block models~\cite{racz2021correlated,Shuwen24SBMChandellier,Joonhyuk23SBM} and attributed graph models~\cite{Ning24AGA,Ziao24AGA,Ziao25AGA2,huang2026attributed}.

\section*{Funding Acknowledgements}
{This work was supported in part by U.S. National Science Foundation (NSF) under grants 2207548 and 2324769; and AFOSR grant FA9550-24-1-0002. The work of Ziao Wang was also supported in part by the  Natural Sciences and Engineering
Research Council of Canada (NSERC) under grant CPRA-609642-2026. }
\section*{AI Methodology Statement}
During the preparation of this paper, we used ChatGPT to assist in understanding results in cited papers, checking our proofs for potential errors, and improving the clarity of the writing. The main ideas and arguments underlying the proofs were developed by the authors. We take full responsibility for the content of the paper and the correctness of its results.
\bibliographystyle{IEEEtranN}
\bibliography{main3_private.bib}

\newpage

\appendix

\section{Proof of Theorem~\ref{th:TBAlign}}\label{sec:5dang}
In this appendix, we present the detailed proof of Theorem~\ref{th:TBAlign}, which is restated below.
\TBAlignRecovery*

Appendices~\ref{appd:5dang-infty} and~\ref{appd:5dang-tau} analyze the matrix $\mathbf M_{\mathrm{TB}}$ computed in Step~2 of Algorithm~\ref{algo:TBAlign}, first without the degree cutoff and then with the cutoff $\tau$ in~\eqref{eq:TBAlign-parameters}. The lemmas used in this analysis are proved in Appendix~\ref{appd:5dang-lemmas}. Appendix~\ref{appd:pf-TBAlign} combines this analysis with the seeded completion in Appendix~\ref{appd:seeded} to prove Theorem~\ref{th:TBAlign}.

\subsection{Step 1. Analysis with $\tau=\infty$}\label{appd:5dang-infty}
Let ${\bf M}_\mathrm{TB}^\infty$ denote the matching matrix under \texttt{TBAlign} with $\tau=\infty$,
and $\cw^\infty(u,\td v)$ denote the witness graph for the vertex pair $(u,\td v)$ under $\tau=\infty$.
Let $\bar{G}$ denote the union graph of $G$ and $\td G$. We use the bar notation to distinguish the notations associated with $\bar G$. More specifically, $\bar\cv:=[n]$ and $\bar\ce:=\ce\cup\{(i,j):(\pi(i),\pi(j))\in\td\ce\}$ are the node set and the edge set of $\bar G$, respectively. Moreover, $u$, $\tilde u$, and $\bar u$ represent the {\em same} node but in different graphs $G$, $\tilde G$, and $\bar G$, respectively, and $\bar\cv(H)$ denotes the set of nodes of $\bar G$ that represent the nodes of a subgraph $H$ of $G$ or $\td G$. Recall that $\Psi([\ct],[\td\ct])=\indi\{\widehat{\mathbf L}(\ct,\td\ct)>\theta^*\}$ for a pair of local trees $(\ct,\td\ct)$, where $\widehat{\mathbf L}$ is defined in Step~1 of Algorithm~\ref{algo:TBAlign}.
We define
    \begin{align}
\ca_{1}:=\Bigl\{&
\nexists({\ct}^{\infty}_{w\setminus u},
{\td\ct}^{\infty}_{\td x\setminus \td v})
\in{\cc}^{\infty}_d\times {\td\cc}_d^{\infty}\text{ such that }
\Psi([{\ct}^{\infty}_{w\setminus u}],
[{\td\ct}^{\infty}_{\td x\setminus \td v}])=1
\text{ and } \notag\\
&\quad \bar{\cv}({\ct}^{\infty}_{w\setminus u})
\cap \bar{\cv}({\td\ct}^{\infty}_{\td x\setminus \td v})=\emptyset \Bigr\},\label{eq:new-A1}
\end{align}
and $\ca_2$ as the event that there exists no subgraph $\bar S$ of $\bar{G}$ that satisfies the following two properties:
\begin{enumerate}
    \item $|\cv(\bar S)|<|\ce (\bar S)|$;
    \item $|\cv(\bar S)|\le 6d+6$.
\end{enumerate}

Lemma~\ref{lem:new-sufficient} states that the event $\ca_1\cap\ca_2$ implies that \texttt{TBAlign} does not output any wrong matches. Lemmas~\ref{lem:new_A1} and~\ref{lem:subgraph} state that these two events both happen with high probability. These three lemmas together imply
\begin{equation}
    \P({\bf M}_\mathrm{TB}^\infty(u,\td v)=0,\text{for all pairs }(u,\td v)\text{ such that }\td v\neq \td u)=1- \td O(n^{-1/3}).
    \label{eq:infty-correct}
\end{equation}

For a vertex $u$ in $G$, define
\begin{align}
    \ca_{3,u}&:=\left\{\texttt{GreedyMatching}({\cw}^{\infty}(u,\td u),\theta^*,5)=1\right\}.
    \label{eq:new-A2}
\end{align}
This is the event that \texttt{TBAlign} correctly matches $u$ with $\td u$, and Lemma~\ref{lem:new-correct} states that this event happens with high probability.
Let $X^\infty:=\sum_{u\in \cv}\mathbbm{1}\{{\bf M}_\mathrm{TB}^\infty(u,\td u)=1\}$. It follows by Lemma~\ref{lem:new-correct} that 
\begin{equation}
    \E[n-X^\infty] = O\left(n\exp\left(-\frac{\lambda s}{2}\right)+\sqrt{n}\right).
\end{equation}
By Markov's inequality, 
\begin{align*}
    \P\left(n-X^\infty\ge n\exp\left(-\frac{\lambda s}{4}\right)+n^{3/4}\right)= O\left( \exp\left(-\frac{\lambda s}{4}\right)+n^{-1/4}\right),
\end{align*}
which further implies 
\begin{equation}
    \P\left(X^\infty\le n-n\exp\left(-\frac{\lambda s}{4}\right)-n^{3/4}\right)= O\left( \exp\left(-\frac{\lambda s}{4}\right)+n^{-1/4}\right).\label{eq:x-infty}
\end{equation}

\subsection{Step 2. Analysis with the Degree Cutoff $\tau$}\label{appd:5dang-tau}
We now reduce $\tau$ to the actual value given in~\eqref{eq:TBAlign-parameters}. Recall that ${\bf M}_\mathrm{TB}$ is the matching matrix computed in Step~2 of Algorithm~\ref{algo:TBAlign}. We couple Algorithm~\ref{algo:TBAlign} with $\tau=\infty$ and with the actual $\tau$ as follows. For each pair of nonempty local trees that is the same for $\tau=\infty$ and for the actual $\tau$, the two executions of \texttt{ApproxLR} that compute its entries of $\widehat{\mathbf L}$ use the same collection of random bits, so the two entries are equal by Remark~\ref{rem:independent-executions}. We also let \texttt{GreedyMatching} examine the edges of ${\cw}^\infty(u,\td v)$ and ${\cw}(u,\td v)$ in the same order. We define $\cb$ (resp. $\td\cb$) to be the set of vertices in $G$ (resp. $\td G$) whose $(d+1)$-hop neighborhood contains a vertex with degree larger than $\tau$, i.e.,
\begin{align*}
    \cb
    &:=
    \left\{
    u\in {\cv}:
    \exists v\in {\cv}
    \text{ such that }
    \dist(u,v)\le d+1
    \text{ and }
    \deg(v)> \tau
    \right\},
    \\
    \td\cb
    &:=
    \left\{
    \td u\in \td{\cv}:
    \exists \td v\in \td{\cv}
    \text{ such that }
    \dist(\td u,\td v)\le d+1
    \text{ and }
    \deg(\td v)> \tau
    \right\}.
\end{align*}
We note that
\begin{itemize}
    \item Given that $u\notin {\cb},$ a local tree ${\ct}_{w\setminus u},$ where $w\in{\cn}(u)$, remains the same for $\tau=\infty$ and $\tau=\max\{\sqrt{\log n},\lambda(1+1/\log\log n)\}$ because (i) no vertex in the $(d+1)$-hop neighborhood of $u$ has degree larger than $\max\{\sqrt{\log n},\lambda(1+1/\log\log n)\},$ and (ii) any vertex in the tree ${\ct}_{w\setminus u}$ is at most $(d+1)$ hops away from $u$. The same conclusion holds for a vertex $\td{v}\not\in \td \cb.$
Therefore, we have 
\begin{align}
    {\bf M}_\mathrm{TB}(u,\td v)= {\bf M}_\mathrm{TB}^\infty(u,\td v),&
    \qquad
    \forall (u,\td v)\in {\cb}^c\times \td{\cb}^c.\label{eq:tau-equal}
\end{align}

\item On the other hand, if $u\in {\cb},$ ${\ct}_{w\setminus u}$ may become an empty tree if it includes a vertex with degree larger than $\tau$, which would result in $\widehat{\bf L}({\ct}_{w\setminus u}, \td{\ct}_{\tilde{x}\setminus \tilde{v}})=0$ for every $\td x\in\td\cn(\td v)$. In this case, the witness graph ${\cw}(u,\td v)$ will be changed by removing node $w$ and all connected edges. By the vertex-deletion monotonicity of greedy matching under a fixed query order \cite[Lemma~4.2 and the proof of Fact~3.1]{Derakhshan26GreedyMatching}, deleting witness vertices while preserving the relative order of the remaining edges cannot increase the cardinality of the greedy matching. Applying this fact successively to all witness vertices removed by the truncation, we have 
${\bf M}_\mathrm{TB}(u, \tilde{v})=0$ if ${\bf M}_\mathrm{TB}^\infty(u, \tilde{v})=0.$ In other words,  
\begin{align}
    {\bf M}_\mathrm{TB}(u,\td v)\leq {\bf M}_\mathrm{TB}^\infty(u,\td v),&
    \qquad
    \hbox{if } u\in {\cb} \hbox{ or } \tilde{v}\in\td{\cb}.\label{eq:tau-le}
\end{align}
\end{itemize}
Let $X:=\sum_{u\in\cv}\mathbbm{1}\{{\bf M}_\mathrm{TB}(u,\td u)=1\}$. It follows by~\eqref{eq:tau-equal} that $X\ge X^\infty-|\cb|-|\td\cb|$ almost surely. By Lemma \ref{lem:TBAlignBadNodes} in Appendix~\ref{appd:bad-nodes}, we have
\begin{equation}
\label{eq:bad-nodes}
    \P\left(|\cb|+|\td\cb|  >
    2n\exp\left( -(\log n )^{2/5} \right)\right)=O\left(\exp\left( -(\log n )^{2/5} \right)\right).
\end{equation}
It then follows by~\eqref{eq:x-infty} and~\eqref{eq:bad-nodes} that
\begin{align}
    &\P\left(X> n-n\exp\left(-\frac{\lambda s}{4}\right)-n^{3/4}-2n\exp\left( -(\log n )^{2/5} \right)\right)\nonumber\\
    &=1-O\left(\exp\left(-\frac{\lambda s}{4}\right)+\exp\left( -(\log n )^{2/5} \right)\right).\label{eq:tau-num}
\end{align}
On the other hand,~\eqref{eq:tau-equal} and~\eqref{eq:tau-le} together imply that ${\bf M}_\mathrm{TB}(u,\td v)\leq {\bf M}_\mathrm{TB}^\infty(u,\td v)$ for all $(u,\td v)$. It then follows by~\eqref{eq:infty-correct} that
\begin{equation}
    \P({\bf M}_\mathrm{TB}(u,\td v)=0,\text{for all pairs }(u,\td v)\text{ such that }\td v\neq \td u)=1- \td O(n^{-1/3}).
    \label{eq:tau-correct}
\end{equation}

\subsection{Analysis of the Events $\ca_1$, $\ca_2$, and $\ca_{3,u}$}\label{appd:5dang-lemmas}\begin{lem}
\label{lem:new-sufficient}
    The event $\ca_{1}\cap\ca_2$ implies that \texttt{TBAlign} does not match any wrong pair of vertices, i.e., ${{\bf M}^\infty_\mathrm{TB}(u,\td v)}=0$, for all pairs $(u,\td v)$ such that $\td v\neq \td u$.
\end{lem}

\begin{proof}
    We prove using contradiction that when event $\ca_{1}\cap \ca_2$ occurs, ${\bf M}_\mathrm{TB}^\infty(u,\td v)=0$ for all pairs $(u,\td v)$ such that $\td v\neq \td u$. Suppose that there exists some $\td v\neq \td u$ such that
\begin{align}
    {\bf M}_\mathrm{TB}^\infty(u,\td v)=\texttt{GreedyMatching}({\cw}^{\infty}(u,\td v),\theta^*,5)=1.
    \nonumber
\end{align}
Then there exist {\em at least} five distinct neighbors $u_1,u_2,u_3,u_4,u_5$ of $u$ and five distinct neighbors $\td v_1,\td v_2,\td v_3,\td v_4,\td v_5$ of $\td v$ such that
\begin{align}
    \Psi([{\ct}^{\infty}_{u_i\setminus u}],[{\td\ct}^{\infty}_{\td v_i\setminus \td v}])
     = 1,
    \qquad
    i=1,2,3,4,5.
    \label{eq:wrong-five-witnesses}
\end{align}
Conditioned on $\ca_{1}$,  the local tree pair $({\ct}^{\infty}_{u_i\setminus u},{\td\ct}^{\infty}_{\td v_i\setminus \td v})$ has non-empty intersection in $\bar G$. In other words, there exists
\begin{align}
    \bar{x}_i\in
    \bar{\cv}({\ct}^{\infty}_{u_i\setminus u})
    \cap
    \bar{\cv}({\td\ct}^{\infty}_{\td v_i\setminus \td v}).
    \nonumber
\end{align}
By definition of the local trees, we have the following statements:
\begin{enumerate}
    \item there is a path $(y_i^{(0)},y_i^{(1)},\ldots,y_i^{(t)})$ in $G$ from $u$ to $x_i$ with $y_i^{(0)}=u$, $y_i^{(1)}=u_i$, $y_i^{(t)}=x_i$, and $t\le d+1$,
    \item there is a path $(\td z_i^{(0)},\td z_i^{(1)},\ldots,\td z_i^{(\td t)})$ in $\td G$ from $\td v$ to $\td x_i$ with $\td z_i^{(0)}=\td v$, $\td z_i^{(1)}=\td v_i$, $\td z_i^{(\td t)}=\td x_i$, and $\td t\le d+1$.
\end{enumerate}
 Passing to the union graph, we obtain two paths
\begin{align}
    (\bar y_i^{(0)},\bar y_i^{(1)},\ldots,\bar y_i^{(t)})
    \qquad\text{and}\qquad
    (\bar z_i^{(0)},\bar z_i^{(1)},\ldots,\bar z_i^{(\td t)}),
    \nonumber
\end{align}
where $\bar y_i^{(0)}=\bar u$, $\bar y_i^{(1)}=\bar u_i$, $\bar z_i^{(0)}=\bar v$, $\bar z_i^{(1)}=\bar v_i$, and $\bar y_i^{(t)}=\bar z_i^{(\td t)}=\bar x_i$.

We now claim that there exists a path $\bar{\cp}_i=(\bar p_i^{(0)},\bar p_i^{(1)},\ldots,\bar p_i^{(\bar t)})$
from $\bar u$ to $\bar v$ in $\bar G$, with $\bar t\le 2(d+1)$, such that either $\bar p_i^{(1)}=\bar u_i$, or $\bar p_i^{(\bar t-1)}=\bar v_i$. We consider three cases. 
\begin{itemize}
\item {\bf{Case (i):}} $\bar v=\bar y_i^{(\iota)}$ for some $\iota\in\{0,1,\ldots,t\}$. Since $\bar v\neq \bar u$, we have $\iota\neq 0$. We can construct the desired path $\bar{\cp}_i :=(\bar y_i^{(0)},\bar y_i^{(1)},\ldots,\bar y_i^{(\iota)})$. Since $\bar y_i^{(0)}=\bar u$, $\bar y_i^{(1)}=\bar u_i$ and $\bar y_i^{(\iota)}=\bar v$, $\bar{\cp}_i$
is a path from $\bar u$ to $\bar v$ of length $\bar t=\iota\le d+1$, and $\bar p_i^{(1)}=\bar y_i^{(1)}=\bar u_i$.

\item {\bf{Case (ii):}} $\bar u=\bar z_i^{(\iota)}$ for some $\iota\in\{0,1,\ldots,\td t\}$. Since $\bar v\neq \bar u$, we have $\iota\neq 0$. We can construct the desired path $\bar{\cp}_i:=(\bar z_i^{(\iota)},\bar z_i^{(\iota-1)},\ldots,\bar z_i^{(1)},\bar z_i^{(0)})$. Since $\bar z_i^{(\iota)}=\bar u$, $\td z_i^{(1)}=\td v_i$ and $\td z_i^{(0)}=\td v$, $\bar{\cp}_i$
is a path from $\bar u$ to $\bar v$ of length $\bar t=\iota\le d+1$, and $\bar p_i^{(\bar t-1)}=\bar z_i^{(1)}=\bar v_i$.

\item {\bf{Case (iii):}} $\bar v\neq \bar y_i^{(\iota)}$ for all $\iota\in\{0,1,\ldots,t\}$ and $\bar u\neq \bar z_i^{(\iota)}$ for all $\iota\in\{0,1,\ldots,\td t\}$. Since $\bar y_i^{(t)}=\bar z_i^{(\td t)}=\bar x_i$, the two paths intersect in $\bar G$. Let $\bar w_i$ be the first vertex on the path $(\bar y_i^{(0)},\bar y_i^{(1)},\ldots,\bar y_i^{(t)})$ that also lies on $(\bar z_i^{(0)},\bar z_i^{(1)},\ldots,\bar z_i^{(\td t)})$. Then the initial segment from $\bar u$ to $\bar w_i$ along the $\bar y$-path and the initial segment from $\bar v$ to $\bar w_i$ along the $\bar z$-path are disjoint except at $\bar w_i$. Hence we construct the desired path
\begin{align}
    \bar{\cp}_i:=(\bar y_i^{(0)},\bar y_i^{(1)},\ldots,\bar w_i,\ldots,\bar z_i^{(1)},\bar z_i^{(0)}).
    \nonumber
\end{align}
Since $\bar y_i^{(0)}=\bar u$ and $\td z_i^{(0)}=\td v$, $\bar{\cp}_i$ is a path from $\bar u$ to $\bar v$ in $\bar G$. Its length satisfies $\bar t\le t+\td t\le 2(d+1)$, and it has both $\bar p_i^{(1)}=\bar y_i^{(1)}=\bar u_i$ and $\bar p_i^{(\bar t-1)}=\bar z_i^{(1)}=\bar v_i$.
\end{itemize}
Some illustrating examples for all the $3$ cases of constructing path $\bar{\mathcal{P}}_i$ are shown in Figure \ref{fig:3cases-new}.

\begin{figure}
    \centering 
        \includegraphics[width=0.6\textwidth]{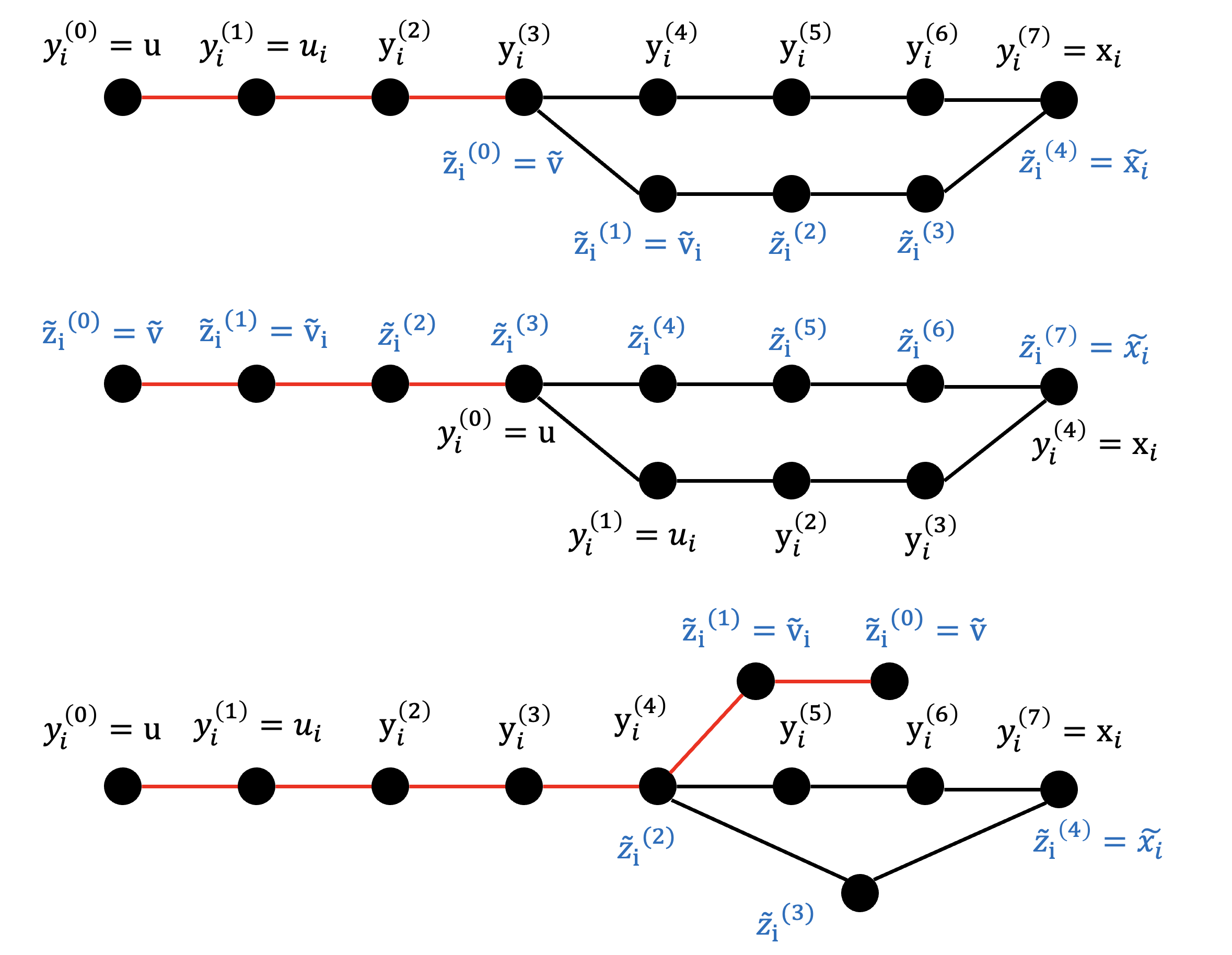}
        \caption{Examples of constructing path $\bar{\mathcal{P}}_i$ in Case (i) (top), Case (ii) (middle) and Case (iii) (bottom). In each case, the red edges denotes path $\bar{\mathcal{P}}_i$. The black notations next to each node represents the nodes of path $(y_i^{(0)},y_i^{(1)},\ldots,y_i^{(t)})$ in $G$, while the blue notations next to each node represents the nodes of path $(\td z_i^{(0)},\td z_i^{(1)},\ldots,\td z_i^{(\td t)})$ in $\td G$.} 
        \label{fig:3cases-new}
\end{figure}

Applying this construction for $i=1,2,3,4,5$, we obtain five paths $\bar{\cp}_1,\ldots,\bar{\cp}_5$ from $\bar u$ to $\bar v$ in $\bar G$, each of length at most $2(d+1)$. Since $\bar{\cp}_i$ has to include either $\bar u_i$ or $\bar v_i$, by the pigeonhole principle, one of the following two statements must hold true:
\begin{enumerate}
    \item There exist three paths of length at most $2(d+1)$ from $\bar u$ to $\bar v$ such that their first vertex after $\bar u$ is all distinct.
    \item There exist three paths of length at most $2(d+1)$ from $\bar u$ to $\bar v$ such that their last vertex before reaching $\bar v$ is all distinct.
\end{enumerate}

Let us assume without loss of generality that the first statement holds. Denote the union graph of these three paths by $H$, and the three distinct vertices by $\bar r_1$, $\bar r_2$ and $\bar r_3$ respectively.
Because each of these paths has at most $2d+1$ vertices other than $\bar u$ and $\bar v$, we have $|\cv(H)|\le 3(2d+1)+2\le 6(d+1)$.

We claim that $|\cv(H)|<|\ce(H)|$. To see this, let $H'$ denote the graph obtained by removing vertex $\bar u$ and all its incident edges from $H$. Notice that $\bar u$ is only connected to the three vertices $\bar r_1$, $\bar r_2$ and $\bar r_3$ in $H$. Therefore, we have $|\cv(H')|=|\cv(H)|-1$ and $|\ce(H')|=|\ce(H)|-3$. Since $H'$ is a connected graph, we also have $|\ce(H')|\ge |\cv(H')|-1$. These together imply that $|\ce(H)|\ge |\cv(H)|+1$, which proves the claim.
Now we know that $H$ is a subgraph of $\bar G$ that satisfies both $|\cv(H)|<|\ce(H)|$ and $|\cv(H)|\le 6d+6$. This contradicts with event $\ca_2$, which completes the proof.
\end{proof}

\begin{lem}
\label{lem:new_A1}
Under the assumptions in Theorem~\ref{th:TBAlign}, we have
\begin{align}
    \P(\ca^c_{1})
    =
    \td O\bigl(n^{-1/3}\bigr).
    \nonumber
\end{align}
\end{lem}
The proof of this lemma largely follows that of Lemma~2 in~\cite{Cheng26RBAlign}, which assumes that $\lambda=(\log n)^{\alpha+o(1)}$ for some $\alpha\in(0,1)$. Here we extend the result to the broader regime $\lambda=\omega(1)$ and $\lambda=(\log n)^{O(1)}$. The key arguments remain unchanged, and we present the proof here for completeness.

\begin{proof}
    Recall the definition of local tree $\ct_{w\setminus u}^\infty$ for a pair of adjacent vertices $w$ and $u$ in $G$.  In the following proof, we slightly extend this definition. 
    For a vertex $w\in\cv$, not necessarily adjacent to $u$ in $G$, we define $\ct^\infty_{w\setminus u}$ in the same way as before: it is the $d$-hop neighborhood of $w$ in the graph $G\setminus u$ if the $d$-hop neighborhood of $w$ in $G$ is a tree; otherwise, $\ct^\infty_{w\setminus u}$ is defined to be the empty tree. We extend the definition of $\td\ct^\infty_{\td x\setminus\td v}$ to the case of non-adjacent $(\td x,\td v)$ in the similar way. 
    
    With this extended definition, we define $\ca_{1}'$ as the event that there exist $(u,w)\in \cv \times \cv$ and $(\td v,\td x)\in \td \cv\times \td\cv$ such that
    \begin{enumerate}
        \item $\Psi([{\ct}^{\infty}_{w\setminus u}],[{\td\ct}^{\infty}_{\td x\setminus \td v}])=1$;

        \item $\bar\cv(\ct^\infty_{w\setminus u})\cap \bar\cv(\td\ct^\infty_{\td x\setminus \td v})=\emptyset$. 
    \end{enumerate}
    By definition, we have $\ca_{1}^c\subseteq \ca_{1}'$, so to prove the lemma, it suffices to prove
    $\P(\ca_{1}')=\td O(n^{-1/3}).$

    We also define an auxiliary event 
    \[
    \cd:=\{|\cn(i,d)|< 8e^2\lambda^d\log n,\  \forall i\in{\cv}\}\bigcap\{|\td\cn(\td j,d)|< 8e^2\lambda^d\log n,\ \forall \tilde j \in \td{\cv}\},
    \]
    where $\cn(i,d)$ (resp. $\td\cn(\td j,d)$) denotes the set of vertices within distance $d$ from $i$ in $G$ (resp. $\td j$ in $\td G$). It follows by Lemma~\ref{lem:neighbor_size} in Appendix~\ref{appd:neighbor-size} and the union bound that
    \begin{equation}
    \label{eq:cd}
    \P(\cd^c)=\td O(n^{-1/3}).
    \end{equation}
    Then we have
    \begin{align}
        \P(\ca_{1}')&\le \P(\ca_{1}'\cap \cd)+\P(\cd^c)\nonumber\\
        &\le \P(\cd^c)+\sum_{\substack{
    (u,w)\in\cv\times\cv\\
    (\td v,\td x)\in \td\cv\times\td\cv
    }}\P(\{\Psi([{\ct}^{\infty}_{w\setminus u}],[{\td\ct}^{\infty}_{\td x\setminus \td v}])=1\}\cap \{\bar\cv(\ct^\infty_{w\setminus u})\cap \bar\cv(\td\ct^\infty_{\td x\setminus \td v})=\emptyset\}\cap\cd)\label{eq:sum_pairs}
    \end{align}

    Now, we fix vertex pairs $(u,w)\in\cv\times\cv$ and $(\td v,\td x)\in \td\cv\times\td\cv$. We omit the subscript and superscript from notations $\ct^\infty_{w\setminus u}$ and $\td\ct^\infty_{\td x\setminus \td v}$, and simply write $\ct$ and $\td\ct$. Recall that $\cz_d$ denote the set of all rooted unlabeled trees with depth up to $d$. We then have
    \begin{align}
        &\P(\{\Psi([{\ct}],[{\td\ct}])=1\}\cap \{\bar\cv(\ct)\cap \bar\cv(\td\ct)=\emptyset\}\cap\cd)\nonumber\\
        ={}& \sum_{\substack{
        (t,\td t)\in \cz_d\times\cz_d
        }} \P\big(\{[{\ct}]=t\}\cap\{[{\td\ct}]=\td t\}\cap \{\bar\cv(\ct)\cap \bar\cv(\td\ct)=\emptyset\}\cap\cd\big)\P(\Psi(t,\td t)=1)\nonumber\\
        \stackrel{(a)}{\le} {} & \sum_{\substack{
    (t,\td t)\in \cz_d\times\cz_d:\\
    |\cv(t)|\le 8e^2\lambda^d\log n\\
    |\cv(\td t)|\le 8e^2\lambda^d\log n 
    }} \P\big(\{[{\ct}]=t\}\cap\{[{\td\ct}]=\td t\}\cap \{\bar\cv(\ct)\cap \bar\cv(\td\ct)=\emptyset\}\big)\P(\Psi(t,\td t)=1)\label{eq:expend_t}
    \end{align}
    where (a) follows because both $\ct$ and $\td\ct$ have at most $8e^2\lambda^d\log n$ vertices under event $\cd$. Here, $\P(\Psi(t,\td t)=1)$ denotes the probability that $\Psi$ outputs $1$ on the fixed input $(t,\td t)$. The first equality holds because the random bits used by \texttt{ApproxPer} in computing $\Psi([\ct],[\td\ct])$ are independent of $(G,\td G,\pi)$.

    Recall that $\bq(\ch,\td\ch)$ is the independent tree-pair distribution.
    By Lemma~\ref{lem:tree_pmf} in Appendix~\ref{appd:tree-pmf}, we know that for any pair of unlabeled trees $(t,\td t)\in \cz_d\times\cz_d$ satisfying $|\cv(t)|\le 8e^2\lambda^d\log n$ and  $|\cv(\td t)|\le 8e^2\lambda^d\log n$, 
    \begin{align}
        &\frac{\P\big(\{[{\ct}]=t\}\cap\{[{\td\ct}]=\td t\}\cap \{\bar\cv(\ct)\cap \bar\cv(\td\ct)=\emptyset\}\big)}{\bq([\ch_d]=t,[\td\ch_d]=\td t)}\nonumber\\
        \le{} & \exp\left(\frac{\lambda}{n}\left((8e^2\lambda^d\log n+1)8e^2\lambda^d\log n+8e^2\lambda^d\log n(1+8e^2\lambda^d\log n+8e^2\lambda^d\log n)\right)\right)\nonumber\\
        \le{} & \exp\left(\frac{\lambda}{n}320e^4\lambda^{2d}(\log n)^2\right)=O(1),\label{eq:const_ratio}
    \end{align}
    where the last equality follows because $\lambda^{2d}=n^{o(1)}$. Then~\eqref{eq:expend_t} and~\eqref{eq:const_ratio} together imply that
    \begin{align}
        &\P(\{\Psi([{\ct}],[{\td\ct}])=1\}\cap \{\bar\cv(\ct)\cap \bar\cv(\td\ct)=\emptyset\}\cap\cd)\nonumber\\
        \le{}&  \exp\left(\frac{\lambda}{n}320e^4\lambda^{2d}(\log n)^2\right)\sum_{\substack{
    (t,\td t)\in \cz_d\times\cz_d\\
    |\cv(t)|\le 8e^2\lambda^d\log n\\
    |\cv(\td t)|\le 8e^2\lambda^d\log n 
    }}\bq([\ch_d]=t,[\td\ch_d]=\td t)\P(\Psi(t,\td t)=1)\nonumber\\
    \le{}&  \exp\left(\frac{\lambda}{n}320e^4\lambda^{2d}(\log n)^2\right)\bq(\Psi([\ch_d],[\td\ch_d])=1)\nonumber\\
    \le{}& \exp\left(\frac{\lambda}{n}320e^4\lambda^{2d}(\log n)^2\right)\left(n^{-(\log\log n)/2}+n^{-11}\right)=O(n^{-11}),\label{eq:fix_uwvx}
    \end{align}
    where the second inequality holds by $\bq(\Psi([\ch_d],[\td\ch_d])=1)=\sum_{(t,\td t)\in\cz_d\times\cz_d}\bq([\ch_d]=t,[\td\ch_d]=\td t)\P(\Psi(t,\td t)=1)$, and the last inequality follows by~\eqref{eq:rank-LRT-typeI} in Theorem~\ref{thm:LRT}. Notice that~\eqref{eq:fix_uwvx} holds for any $w$ and $(\td v,\td x)$. Substituting~\eqref{eq:fix_uwvx} and~\eqref{eq:cd} into~\eqref{eq:sum_pairs} yields
    \begin{equation}
        \P(\ca_{1}')=\td O(n^{-1/3})+O(n^{-7})=\td O(n^{-1/3}),
    \end{equation}
    which completes the proof.
\end{proof}

\begin{lem}\label{lem:subgraph}
Under the assumptions in Theorem~\ref{th:TBAlign}. We have $$\P(\ca_2^c)=n^{-1+o(1)}.$$
\end{lem}

\begin{proof}
    Let $H$ be a fixed unlabeled graph that satisfies $|\cv(H)|\le 6d+6$ and $|\cv(H)|<|\ce(H)|$. Let $X_H$ denote the number of subgraphs in $\bar G$ that are isomorphic to $H$. Recall that $\bar G$ is an \erdos--\renyi random graph with edge probability $\frac{\lambda(2-s)}{n}$. It follows by Lemma 5.1 in~\cite{frieze2015introduction} that
    \begin{align}
        \E[X_H]&=\binom{n}{|\cv(H)|}\frac{|\cv(H)|!}{\mathrm{aut}(H)}\left(\frac{\lambda(2-s)}{n}\right)^{|\ce(H)|}\nonumber\\
        &\le n^{|\cv(H)|}\left(\frac{\lambda(2-s)}{n}\right)^{|\ce(H)|}\nonumber\\
        &\le n^{|\cv(H)|}\left(\frac{\lambda(2-s)}{n}\right)^{|\cv(H)|+1}\nonumber\\
        &\le \frac{(\lambda(2-s))^{|\cv(H)|+1}}{n}=n^{-1+o(1)}.\label{eq:subgraph-exp}
    \end{align}
    Here $\mathrm{aut}(H)$ is the automorphism number of $H$, which is at least $1$, and the last equality follows because $(\lambda(2-s))^{6d+7}=n^{o(1)}$ under the assumption that $d= \lceil(\log n)^{1/3}\rceil$. Moreover, the number of isomorphism classes one can form with at most $6d+6$ vertices is upper-bounded by $2^{(6d+6)^2}=n^{o(1)}$, since $(6d+6)^2=O((\log n)^{2/3})$. Let $X$ denote the number of subgraphs $\bar S$ of $\bar G$ that satisfies $|\cv(\bar S)|\le 6d+6$ and $|\ce(\bar S)|>|\cv(\bar S)|$. It follows by~\eqref{eq:subgraph-exp} that
    \begin{align*}
        \E[X]=n^{o(1)}\E[X_H]=n^{-1+o(1)}.
    \end{align*}
    By Markov's inequality,
    \[
    \P(\ca_2^c)=\P(X\ge 1)=n^{-1+o(1)},
    \]
    which completes the proof.
\end{proof}

\begin{lem}\label{lem:new-correct}
Under the assumptions in Theorem~\ref{th:TBAlign}. We have $$\P(\ca_{3,u}^c)=O\left(n^{-1/2}+\exp\left(-\frac{\lambda s}{2}\right)\right).$$
\end{lem}

\begin{proof}
    Define $\ct$ as the local tree rooted at $u$ in $G$ such that $\ct$ is the $(d+1)$-hop-neighborhood of $u$ in $G$ if it is a tree; otherwise, $\ct$ is an empty tree. We similarly define $\td\ct$ as the local tree rooted at $\td u$ in $\td G$.
    By Lemma~\ref{lem:tree_tv_new}, we can couple $(\ct,\td\ct)$ and a pair of correlated Poisson Galton--Watson trees $(\ch,\td\ch)\sim\bp$ so that
    \begin{equation}
        \label{eq:tree_couple_5dang}
        \P([\ct]=[\ch_{d+1}],[\td\ct]=[\td\ch_{d+1}])
        =1-O(n^{-1/2}),
    \end{equation}
    where $\ch_{d+1}$ is the truncation of $\ch$ up to depth $d+1$, and $\td\ch_{d+1}$ is the truncation of $\td\ch$ up to depth $d+1$. Let
    \[
        \mathsf{Coup}:=\{[\ct]=[\ch_{d+1}],\,
              [\td\ct]=[\td\ch_{d+1}]\}.
    \]

    We now switch to study the output of the subroutine \texttt{GreedyMatching} when applied to the correlated tree pair $\ch_{d+1}$ and $\td\ch_{d+1}$. More precisely, let $r$ and $\td r$ denote the root nodes of $\ch_{d+1}$ and $\td\ch_{d+1}$, respectively. Define $\sfch_r$ as the set of children of $r$ in $\ch_{d+1}$, and for each $i\in\sfch_r$, define $\ch_{d+1,i}$ to be the subtree rooted at $i$ in $\ch_{d+1}$. We also define $\wdtd\sfch_{\td r}$ and $\td\ch_{d+1,\td j}$ for each $\td j\in\wdtd\sfch_{\td r}$ similarly. The witness graph $\cw^\infty(r,\td r)$ is defined as the bipartite graph between vertex sets $\sfch_r$ and $\wdtd\sfch_{\td r}$, with the weight of edge $(i,\td j)$ being $\widehat L_d([\ch_{d+1,i}],[\td\ch_{d+1,\td j}])$.

    We extend the coupling above to include the randomness used by \texttt{ApproxLR}. First generate mutually independent collections of random-bit sequences for the evaluations defining $\cw^\infty(u,\td u)$, independently of the graphs and the coupled trees. On $\mathsf{Coup}$, fix root-preserving vertex relabelings, depending only on the coupled trees, that map $\ch_{d+1}$ to $\ct$ and $\td\ch_{d+1}$ to $\td\ct$. With these relabelings, each edge of $\cw^\infty(r,\td r)$ corresponds to an edge of $\cw^\infty(u,\td u)$ with the same pair of unlabeled subtrees. We supply the \texttt{ApproxLR} evaluations on corresponding edges with the same collections of random-bit sequences. On $\mathsf{Coup}^c$, we use fresh independent collections of random-bit sequences to compute the weights of $\cw^\infty(r,\td r)$.

    Conditionally on the graphs and coupled trees, the collections used within each witness graph have their prescribed independent distributions. Thus this construction preserves the marginal law of each weighted witness graph. Moreover, on $\mathsf{Coup}$, corresponding evaluations have identical inputs, parameters, and random-bit sequences, so their outputs agree by Remark~\ref{rem:independent-executions}. We let the auxiliary \texttt{GreedyMatching} execution examine the edges of $\cw^\infty(r,\td r)$ in the order in which it examines the corresponding edges of $\cw^\infty(u,\td u)$ on $\mathsf{Coup}$, and use any fixed edge order on $\mathsf{Coup}^c$. Therefore, on $\mathsf{Coup}$, we have
    \[
        \{\texttt{GreedyMatching}
          (\cw^\infty(u,\td u),\theta^*,5)=1\}
        \Longleftrightarrow
        \{\texttt{GreedyMatching}
          (\cw^\infty(r,\td r),\theta^*,5)=1\}.
    \]

    We define the random variable
    \[
        X
        :=
        \sum_{i\in\sfch_r^*}
        \Psi\bigl([\ch_{d+1,i}],
                  [\td\ch_{d+1,\td i}]\bigr),
    \]
    where $\sfch_r^*$ denotes the set of children of the root in the intersection tree $\ch^*$. Here each test uses the same \texttt{ApproxLR} output already assigned to its corresponding witness edge. Equivalently, $X$ is the number of children $i$ of the root in $\ch^*$ for which
    \[
        \widehat L_d
        ([\ch_{d+1,i}],[\td\ch_{d+1,\td i}])>\theta^*.
    \]
    Throughout the proof, probabilities involving $\Psi$ include its internal random bits; in particular, we use $\bp$ also for the joint law of the GW trees and the independent algorithmic randomness.

    We claim that the event $X\ge9$ implies $\texttt{GreedyMatching} (\cw^\infty(r,\td r),\theta^*,5)=1$. This implication holds for every ordering of the edges. Under the event $X\ge9$, there exist nine non-overlapping vertex pairs $(i_1,\td i_1),\ldots,(i_9,\td i_9) \in\sfch_r\times\wdtd\sfch_{\td r}$ such that for each $k\in[9]$,
    \[
        \widehat L_d
        ([\ch_{d+1,i_k}],[\td\ch_{d+1,\td i_k}])
        >\theta^*.
    \]
    Towards a contradiction, suppose $\texttt{GreedyMatching} (\cw^\infty(r,\td r),\theta^*,5)=0$. Then, the matching found by the subroutine before it terminates has size less than or equal to $4$, which includes at most $4$ vertices in $\sfch_r$ and at most $4$ vertices in $\wdtd\sfch_{\td r}$. This implies that there must exist some $k^*\in[9]$ such that the edge $(i_{k^*},\td i_{k^*})$ does not conflict with any edge in the matching. Since this edge has weight greater than $\theta^*$, the subroutine could not have terminated without adding it, which leads to a contradiction.

    Notice that the number of children in the set $\sfch_r^*$ follows distribution $\poi(\lambda s)$. Conditionally on $|\sfch_r^*|$, the subtree pairs $([\ch_{d+1,i}],[\td\ch_{d+1,\td i}])$, $i\in\sfch_r^*$, are independent and identically distributed as $([\ch_d],[\td\ch_d])$ under $\bp$. Their test outputs use mutually independent collections of random-bit sequences, jointly independent of the trees. Hence, conditionally on $|\sfch_r^*|$, the events
    \[
        \Psi\bigl([\ch_{d+1,i}],
                  [\td\ch_{d+1,\td i}]\bigr)=1
    \]
    are mutually independent, each with probability $\bp(\Psi([\ch_d],[\td\ch_d])=1)$, which does not depend on $|\sfch_r^*|$. It then follows by Poisson thinning that
    \[
        X\sim
        \poi\left(
            \lambda s\,
            \bp(\Psi([\ch_d],[\td\ch_d])=1)
        \right).
    \]
    Therefore, we have
    \begin{align}
        &\P\bigl(
            \texttt{GreedyMatching}
            (\cw^\infty(r,\td r),\theta^*,5)=0
        \bigr)
        \le \P(X\le8)
        \nonumber\\
        &\qquad=
        \P\left(
            \poi\left(
                \lambda s\,
                \bp(\Psi([\ch_d],[\td\ch_d])=1)
            \right)\le8
        \right).
        \nonumber
    \end{align}
    Notice that for any $\mu\ge0$, we have
    \[
        \P(\poi(\mu)\le8)
        =e^{-\mu}\sum_{j=0}^{8}\frac{\mu^j}{j!}
        =O\left(e^{-\mu/2}\right).
    \]
    Then by~\eqref{eq:rank-LRT-power} in Theorem~\ref{thm:LRT} and the fact that $\lambda=\omega(1)$, we have
    \[
        \lambda s-8s
        \le
        \lambda s\,
        \bp(\Psi([\ch_d],[\td\ch_d])=1)
        \le
        \lambda s
    \]
    for all sufficiently large $n$, and hence
    \begin{align}
        &\P\bigl(
            \texttt{GreedyMatching}
            (\cw^\infty(r,\td r),\theta^*,5)=0
        \bigr)
        \nonumber\\
        &\qquad=
        O\left(
            \exp\left(-\frac{\lambda s}{2}+4s\right)
        \right)
        =
        O\left(
            \exp\left(-\frac{\lambda s}{2}\right)
        \right),
        \label{eq:tree_pass_5dang}
    \end{align}
    where the last equality uses $s\le1$.

    Finally, the equivalence on $\mathsf{Coup}$ gives
    \begin{align*}
        &\P\bigl(
            \texttt{GreedyMatching}
            (\cw^\infty(u,\td u),\theta^*,5)=0
        \bigr)\\
        &=\P\big(\mathsf{Coup}^c,\texttt{GreedyMatching}
            (\cw^\infty(u,\td u),\theta^*,5)=0\big)+\P\big(\mathsf{Coup},\texttt{GreedyMatching}
            (\cw^\infty(u,\td u),\theta^*,5)=0\big)\\
        &\le
        \P(\mathsf{Coup}^c)
        +
        \P\bigl(
            \texttt{GreedyMatching}
            (\cw^\infty(r,\td r),\theta^*,5)=0
        \bigr).
    \end{align*}
    By equations~\eqref{eq:tree_couple_5dang}
    and~\eqref{eq:tree_pass_5dang}, we therefore have
    \[
        \P\bigl(
            \texttt{GreedyMatching}
            (\cw^\infty(u,\td u),\theta^*,5)=1
        \bigr)
        =
        1-O\left(
            n^{-1/2}
            +\exp\left(-\frac{\lambda s}{2}\right)
        \right),
    \]
    which completes the proof.
\end{proof}

\subsection{Combining the Bounds}\label{appd:pf-TBAlign}
By~\eqref{eq:tau-num} and~\eqref{eq:tau-correct} in Appendix~\ref{appd:5dang-tau}, $\mathbf M_{\mathrm{TB}}(u,\td v)=0$ for all pairs $(u,\td v)$ with $\td v\neq\td u$ with high probability, and $\mathbf M_{\mathrm{TB}}(u,\td u)=1$ for all but $o(n)$ nodes $u$ with high probability. We now combine these two bounds with Lemma~\ref{lem:greedy-seeded-completion} on the seeded completion to prove Theorem~\ref{th:TBAlign}.

\begin{proof}
We first establish the error-free almost-exact recovery of $\pi'$ using~\eqref{eq:tau-num} and~\eqref{eq:tau-correct}, and then show the exact recovery of $\hat\pi$ using Lemma~\ref{lem:greedy-seeded-completion}.

We consider the event that $\mathbf M_{\mathrm{TB}}(u,\td v)=0$ for all pairs $(u,\td v)$ with $\td v\neq\td u$ and that $X>n-n\exp(-\lambda s/4)-n^{3/4}-2n\exp(-(\log n)^{2/5})$, where $X=\sum_{u\in\cv}\indi\{\mathbf M_{\mathrm{TB}}(u,\td u)=1\}$ is defined in Appendix~\ref{appd:5dang-tau}. The probabilities of these two conditions are bounded in~\eqref{eq:tau-correct} and~\eqref{eq:tau-num}, respectively. On this event, the first condition ensures that the only possible nonzero entry in row $u$ of the matrix $\mathbf M_{\mathrm{TB}}$ is located at $\td u=\pi(u)$. Consequently, for every $u\in\cv'$, Step~3 correctly assigns $\pi'(u)=\pi(u)$, implying that $\pi'$ contains no wrong match. By the definition of $\cv'$ in \eqref{eq:eq_matched_sets}, the number of matched nodes satisfies
\begin{align}
    |\cv'|
    &=\sum_{u\in\cv}\indi\{\mathbf M_{\mathrm{TB}}(u,\td u)=1\}\nonumber\\
    &\ge n-n\exp\left(-\frac{\lambda s}{4}\right)
    -2n\exp\left(-(\log n)^{2/5}\right)-n^{3/4}
    =n(1-o(1)),
    \label{eq:TBAlign-initial-size}
\end{align}
where the final equality uses $\lambda=\omega(1)$ and the fact that $s>0$ is constant. By the union bound of~\eqref{eq:tau-correct} and~\eqref{eq:tau-num}, we have
\begin{align}
    &\P\Bigl(\bigl\{\mathbf M_{\mathrm{TB}}(u,\td v)=0\text{ for all pairs }(u,\td v)\text{ such that }\td v\neq\td u\bigr\}\nonumber\\
    &\qquad\cap\Bigl\{X>n-n\exp\Bigl(-\frac{\lambda s}{4}\Bigr)-n^{3/4}-2n\exp\bigl(-(\log n)^{2/5}\bigr)\Bigr\}\Bigr)\nonumber\\
    \ge{}&1-\td O\bigl(n^{-1/3}\bigr)-O\left(\exp\left(-\frac{\lambda s}{4}\right)+\exp\left(-(\log n)^{2/5}\right)\right)\nonumber\\
    ={}&1-O\left(\exp\left(-\frac{\lambda s}{4}\right)
    +\exp\left(-(\log n)^{2/5}\right)\right)=1-o(1),\label{eq:partial_error_prob}
\end{align}
where the first equality holds because $\td O(n^{-1/3})=o(\exp(-(\log n)^{2/5}))$. This establishes the error-free almost-exact recovery guarantee in~\eqref{eq:TBAlign-recovery}.

It remains to prove the exact recovery guarantee when $\lambda s\ge(1+\eps)\log n$ for an arbitrarily small constant $\eps>0$. For all sufficiently large $n$, the matched size satisfies
\[
    n-n\exp\left(-\frac{\lambda s}{4}\right)
    -2n\exp\left(-(\log n)^{2/5}\right)-n^{3/4}
    \ge\left(1-\frac{\eps}{16}\right)n.
\]
Therefore, $\pi'$ contains at least $(1-\eps/16)n$ correct matches. Additionally, the assumption $\lambda=(\log n)^{O(1)}$ implies $\lambda=o(n)$, which yields
\[
    \frac{\lambda}{n}\le\frac12,\qquad \frac{ns-\lambda}{n-\lambda}=s-o(1)\ge\eps
\]
for all sufficiently large $n$ and $\eps<s$. These inequalities show that the conditions of Lemma~\ref{lem:greedy-seeded-completion} are satisfied. Since Step~4 of Algorithm~\ref{algo:TBAlign} precisely executes the seeded completion procedure described in Algorithm~\ref{algo:greedy-seeded-completion}, Lemma~\ref{lem:greedy-seeded-completion} ensures the exact recovery for every correct initial matching $\pi'$ in \texttt{TBAlign}. Combining the error probability $O(n^{-\eps/8})$ from Lemma~\ref{lem:greedy-seeded-completion} with~\eqref{eq:partial_error_prob} and applying a union bound, we have
\begin{align*}
    \P(\hat\pi=\pi)&\ge1-O\left(\exp\left(-\frac{\lambda s}{4}\right)
    +\exp\left(-(\log n)^{2/5}\right)\right)
    -O(n^{-\eps/8})\\
    &=1-O\left(\exp\left(-(\log n)^{2/5}\right)\right)=1-o(1),
\end{align*}
where the first equality follows from the asymptotic domination bounds
\[
    \exp\left(-\frac{\lambda s}{4}\right)
    \le n^{-(1+\eps)/4}
    =o\left(\exp\left(-(\log n)^{2/5}\right)\right),
    \qquad
    n^{-\eps/8}
    =o\left(\exp\left(-(\log n)^{2/5}\right)\right).
\]
This completes the proof of Theorem~\ref{th:TBAlign}.
\end{proof}

\section{Proof of Theorem~\ref{th:TBAlign-runtime}}\label{appd:runtime}
In this appendix, we present the detailed proof of Theorem~\ref{th:TBAlign-runtime}, which is restated below.

\TBAlignRuntime*

\begin{proof}
We sequentially bound the computational complexity of each step of Algorithm~\ref{algo:TBAlign}. Fix an arbitrary realization of the input graphs and an execution of \texttt{TBAlign}. By~\eqref{eq:TBAlign-parameters} in Step~1, the parameters satisfy
\begin{align}
    \log\tau=O(\log\log n),\qquad \tau^{2d}
    =\exp\left(O\left((\log n)^{1/3}\log\log n\right)\right)
    =n^{o(1)}.
    \label{eq:TBAlign-tree-size-complexity}
\end{align}
\noindent\underline{\textbf{Step 1: Local tree evaluation.}}
We first access the neighbor nodes and compute the degrees for all nodes in $G$ and $\td G$, which takes $O(n^2)$ time. To construct $\ct_{w\setminus u}$, we perform a breadth-first search starting from $w$ in both $G\setminus u$ and $G$ up to depth $d$. The search terminates early if any explored node has a degree exceeding $\tau$ in $G$, in which case the local tree evaluates to the empty tree. Each search explores at most $\sum_{j=0}^{d}\tau^j=O(\tau^d)$ nodes. Verifying whether the full depth-$d$ neighborhood in $G$ is a tree requires examining at most $O(\tau^{2d})$ node pairs. Thus, the computational complexity of constructing and testing $\ct_{w\setminus u}$ is bounded by $O(\tau^{2d})$. An identical bound holds for $\td\ct_{\td x\setminus\td v}$. Because $|\cc|=2|\ce|$ and $|\td\cc|=2|\td\ce|$, constructing and testing all local trees requires
\[
    O\left(n^2+(|\ce|+|\td\ce|)\tau^{2d}\right)
    =O\left(n^2+(|\ce|+|\td\ce|)n^{o(1)}\right)
\]
time, where we applied~\eqref{eq:TBAlign-tree-size-complexity}.

Furthermore, every nonempty local tree has at most $\tau$ children per node. Consequently, the maximum degree parameter $D$ computed in~\eqref{eq:approximate-L-input-degree} by \texttt{ApproxLR} satisfies $D\le\tau$ for all sufficiently large $n$, yielding $D^d\le\tau^d=n^{o(1)}$. By~\eqref{eq:approximate-L-complexity-general} in Lemma~\ref{lem:approximate-L-complexity}, each invocation of \texttt{ApproxLR} with failure parameter $\xi=n^{-11}$ executes in $n^{o(1)}$ time. Since there are at most $4|\ce|\,|\td\ce|$ such invocations, and initializing the matrix $\widehat{\mathbf L}$ takes $O(|\ce|\,|\td\ce|)$ time, the total computational complexity of Step~1 is bounded by
\[
    O\left(n^2+\bigl(|\ce|+|\td\ce|+|\ce|\,|\td\ce|\bigr)n^{o(1)}\right).
\]

\noindent\underline{\textbf{Step 2: 5-dangling-tree test on witness graphs.}}
Initializing the matrix $\mathbf M_{\mathrm{TB}}$ takes $O(n^2)$ time. Each pair in $\cc\times\td\cc$ contributes an edge to exactly one witness graph. Since $|\cc|=2|\ce|$ and $|\td\cc|=2|\td\ce|$, the total number of edges across all witness graphs is $4|\ce|\,|\td\ce|$. Moreover, the node sets $\cn(u)$ and $\td\cn(\td v)$ of each witness graph $\cw(u,\td v)$ are the neighbor sets computed in Step~1, so specifying them for all node pairs $(u,\td v)\in\cv\times\td\cv$ has complexity $O(n^2)$. Therefore, constructing all witness graphs requires $O(n^2+|\ce|\,|\td\ce|)$ time.

Since the \texttt{GreedyMatching} subroutine examines edges sequentially, it processes each witness edge at most once. For each examined edge, the algorithm checks if its weight exceeds $\theta^*$ and whether it shares an endpoint with any already-selected edge. Because the procedure terminates as soon as the matching reaches size five, at most four edges are held in the matching at any time, and such check has complexity $O(1)$. Consequently, executing the matching across all witness graphs has complexity $O\left(|\ce|\,|\td\ce|\right)$. The total computational complexity of Step~2 is thus

\[
    O\left(n^2+|\ce|\,|\td\ce|\right).
\]

\noindent\underline{\textbf{Step 3: Partial mapping construction.}}
We identify the node set $\cv'$ and select an arbitrary nonzero entry for each row indexed by $\cv'$ by scanning the rows of the $n\times n$ matrix $\mathbf M_{\mathrm{TB}}$, which has complexity $O(n^2)$.

\noindent\underline{\textbf{Step 4: Seeded completion.}}
By the complexity analysis provided in the proof of Lemma~\ref{lem:greedy-seeded-completion} in Appendix~\ref{appd:seeded}, the computational complexity of expanding the partial mapping $\pi'$ to the complete mapping $\hat\pi$ via seeded completion is bounded by
\[
    O\left(n^2+|\ce|\,|\td\ce|\right).
\]

Summing the bounds across all four steps yields the deterministic upper bound \eqref{eq:TBAlign-deterministic-complexity} for the overall computational complexity of \texttt{TBAlign}. 

Finally, each graph has marginal distribution $\mathrm{ER}(n,\lambda/n)$, and hence each edge count has distribution $\bin(\binom{n}{2},\lambda/n)$ with mean $(n-1)\lambda/2$. Applying~\eqref{eq:CB_upper2} in Lemma~\ref{lem:CB} with $\delta=1$ and taking a union bound gives
\[
    \P\left(|\ce|>n\lambda\ \text{or}\ |\td\ce|>n\lambda\right)
    \le 2\exp\left(-\frac{(n-1)\lambda}{6}\right)
    =o(e^{-n}),
\]
where the last equality uses $\lambda=\omega(1)$. On the complementary event, the deterministic bound~\eqref{eq:TBAlign-deterministic-complexity} is at most
\[
    O\left(n^2+(n\lambda+n^2\lambda^2)n^{o(1)}\right)
    =n^{2+o(1)},
\]
since $\lambda=(\log n)^{O(1)}$. This completes the proof of Theorem~\ref{th:TBAlign-runtime}.
\end{proof}

\section{Proof of Theorem~\ref{thm:LRT}}\label{appd:lrt}
In this appendix, we present the detailed proof of Theorem~\ref{thm:LRT}, which is restated below.

\RankLRT*

We prove the statistical guarantees for the test $\Psi$ in Theorem~\ref{thm:LRT} using an auxiliary correlation test $\Phi$. The following proposition extends~\cite[Proposition~2]{Cheng26RBAlign} to a wider range of $\lambda$ with the same choice of depth $d=\lceil(\log n)^{1/3}\rceil$ throughout this range.

\begin{restatable}{prop}{CorTest}
\label{prop:CorTest}
Let $s\in(\sqrt{C_{\mathrm{Otter}}},1]$ be a constant independent of $n$, and suppose that $\lambda=\omega(1)$ with $\lambda=(\log n)^{O(1)}$ and $d=\left\lceil(\log n)^{1/3}\, \right\rceil$.
Then there exists a test $\Phi:\cz_d\times \cz_d\to\{0,1\}$
that satisfies
\begin{align}
    {\bq}\bigl(\Phi([\ch_d],[\td\ch_d])=1\bigr) &\le n^{-\log\log n},
    \label{eq:CorTest-typeI} \\
    {\bp}\bigl(\Phi([\ch_d],[\td\ch_d])=1\bigr)
    &\ge 1-\frac{7}{\lambda}
    \label{eq:CorTest-power}
\end{align}
for all sufficiently large $n$.
\end{restatable}

We construct $\Phi$ and prove Proposition~\ref{prop:CorTest} in Appendix~\ref{sec:aux_test}. In the proof of Theorem~\ref{thm:LRT}, we combine Proposition~\ref{prop:CorTest} with the approximation result in Lemma~\ref{lem:LR-approximation} of Appendix~\ref{sec:approximate-L-guarantees} to establish the statistical guarantees for $\Psi$.  
The detailed proof of Theorem~\ref{thm:LRT} is given in Appendix~\ref{appd:pf-lrt}.

\subsection{Proof of Proposition~\ref{prop:CorTest}: An Auxiliary Test $\Phi$}\label{sec:aux_test}

Let $l=\lceil2\log\log n\rceil$. We construct the correlation test $\Phi$ satisfying guarantees~\eqref{eq:CorTest-typeI} and~\eqref{eq:CorTest-power} in Proposition~\ref{prop:CorTest} from an auxiliary set $\cf_{d-l}\subseteq \cz_{d-l}\times\cz_{d-l}$ given by the following lemma. A proof of this lemma for $\lambda=(\log n)^{\alpha+o(1)}$ with a constant $\alpha\in(0,1)$ is given in~\cite[Appendix~A-C]{Cheng26RBAlign}, and it carries over verbatim to $\lambda=\omega(1)$, $\lambda=(\log n)^{O(1)}$ and $d$ and $l$ given above.

\begin{lem}[{\cite[Appendix~A-C]{Cheng26RBAlign}}]
\label{lem:matching}
Let $s\in(\sqrt{C_{\mathrm{Otter}}},1]$ be a constant. Suppose that
\begin{align}
    \lambda=\omega(1),\qquad \lambda=(\log n)^{O(1)},
    \qquad
    d=\lceil(\log n)^{1/3}\rceil, \qquad l=\lceil2\log\log n\rceil. \nonumber
\end{align}
 Then there exist constants $C_3>0$ and $C_4>0$ independent of $n$, and a set $\cf_{d-l}\subseteq \cz_{d-l}\times\cz_{d-l}$, such that
\begin{align}
    \bq&\bigl(([\ch_{d-l}],[\td\ch_{d-l}])\in\cf_{d-l}\bigr)\leq C_3e^{-d+l}, \label{eq:L2_Type1}\\
    \bp&\bigl(([\ch_{d-l}],[\td\ch_{d-l}])\in\cf_{d-l}\bigr)\geq  1-C_4 \lambda^{-1}, \label{eq:L2_Power}
\end{align}
for all large enough $n$.
\end{lem}

\begin{figure}[htbp]
    \centering 
        \includegraphics[width=0.85\textwidth]{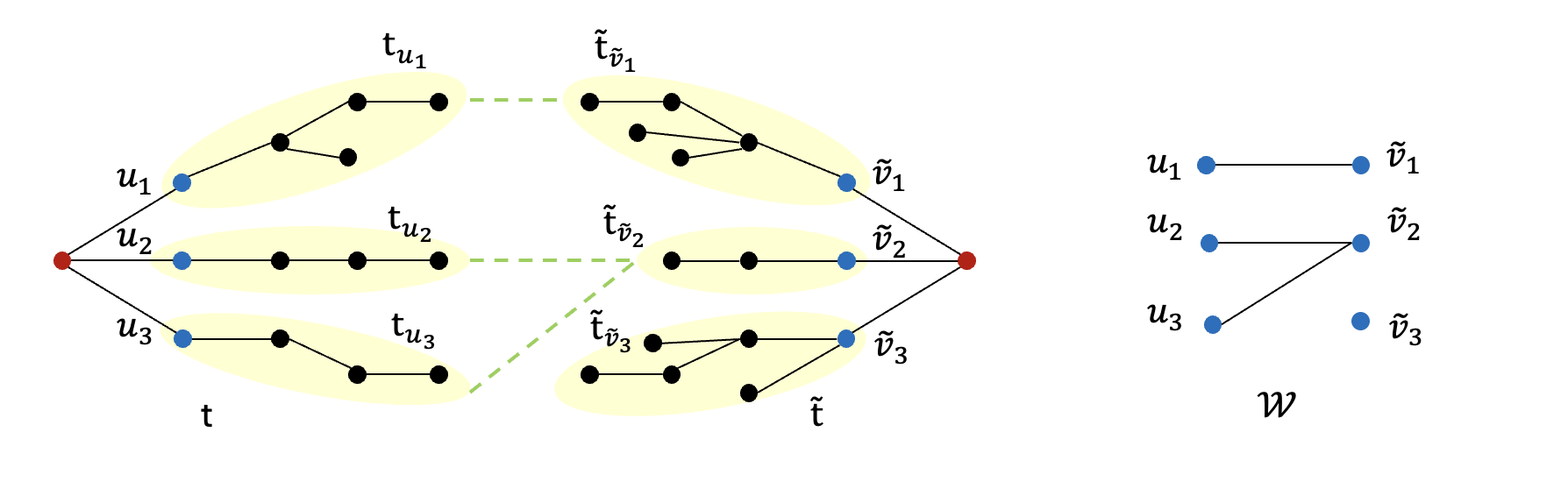}
        \caption{An example of constructing bipartite graph $\cw$ from trees $(t,\td t)\in\cz_4\times\cz_4$ when $l=1$. The red nodes in both tree $t$ and $\td t$ are roots, while the blue nodes in $t$ and $\td t$ denote the two parts in $\cw$. The green dashed line between two subtrees denotes that they appear in set $\cf_{3}$, which corresponds to the edges in the bipartite graph $\cw$.}
        \label{fig:gw-cortest-construction}
\end{figure}

We now utilize $\cf_{d-l}$ to define the test $\Phi$ via the maximum matching over a bipartite graph $\cw$ defined below. Given a pair of rooted unlabeled trees
$(t,\td t)\in\cz_d\times\cz_d$, we first assign arbitrary labels to the
vertices in the $l$-th generations of $t$ and $\td t$, and denote the resulting
sets by $\cs_l(t)$ and $\td\cs_l(\td t)$, respectively.
We then construct a bipartite graph $\cw$ between $\cs_{l}(t)$ and $\td\cs_{l}(\td t)$ such that an edge exists between $u\in \cs_l(t)$ and $\td v \in \td\cs_l(\td t)$ if $(t_u,\td t_{\td v})\in \cf_{d-l}, $ where $t_u$ and $\td t_{\td v}$ are the
unlabeled subtrees of $t$ and $\td t$ rooted at $u$ and $\td v$, respectively. An illustration of this construction is provided in Figure~\ref{fig:gw-cortest-construction}.

We now define the test $\Phi$ such that it declares that $(t,\td t)$ is drawn from the
correlated model $\bp$ when the maximum matching size of $\cw$ is at least
$\frac{1}{2e}(\lambda s)^l$. That is,
    \begin{align}
    \Phi(t,\td t)= \mathbbm{1}
    \left\{ \texttt{MaxMatching}\bigl(\cw\bigr) \ge \frac{1}{2e}(\lambda s)^{l} \right\},
    \label{eq:Phi_def}
\end{align}

We next prove Proposition~\ref{prop:CorTest} based on Lemma~\ref{lem:matching}.

\begin{proof}
We prove that $\Phi$ satisfies the type-I error bound~\eqref{eq:CorTest-typeI} and the power bound~\eqref{eq:CorTest-power} in the following two parts.

\underline{\textbf{Proof of \eqref{eq:CorTest-typeI}}.} 
Suppose $(\ch,\td\ch)\sim\bq$. Throughout the proof, let $\cs_l = \cs_l(\ch)$ and $\td\cs_l = \cs_l(\td\ch)$ denote the level-$l$ vertex sets of $\ch$ and $\td\ch$, respectively. For each $u\in\cs_l$ and $\td v\in\td\cs_l$, let
$\ch_{d,u}$ and $\td\ch_{d,\td v}$ denote the subtrees of
$\ch_d$ and $\td\ch_d$ rooted at $u$ and $\td v$, respectively. Recall from~\eqref{eq:Phi_def} that $\Phi$ outputs $1$ if and only if
\[
    \texttt{MaxMatching}(\cw)\ge \frac{1}{2e}(\lambda s)^l,
\]
where $l = \lceil 2\log\log n \rceil$ and $d = \lceil (\log n)^{1/3} \rceil$ in our setting.

We first record the conditional edge probabilities in $\cw$. Under $\bq$, the trees $\ch$ and $\td\ch$ are two independent \GW trees with $\mathrm{Poi}(\lambda)$ offspring. By the definition of $\ch$ and $\td\ch$, conditioned on $(\cs_l, \td\cs_l)$, the subtrees in the two collections
\[
    \{\ch_{d,u}:u\in\cs_l\}
    \quad\text{and}\quad
    \{\td\ch_{d,\td v}:\td v\in\td\cs_l\}
\]
are mutually independent and identically distributed Galton--Watson trees with $\mathrm{Poi}(\lambda)$ offspring truncated at depth $d-l$. By construction, $(u,\td v)$ is an edge of $\cw$ if and only if $([\ch_{d,u}],[\td\ch_{d,\td v}])\in\cf_{d-l}$, so the edge indicators associated with any fixed matching in $\cw$ are conditionally independent.
Moreover, by the definition of $\cw$ and the type-I bound~\eqref{eq:L2_Type1} in Lemma~\ref{lem:matching}, for every $u\in\cs_l$ and $\td v\in\td\cs_l$,
\begin{align}
    \bq\bigl((u,\td v)\in\cw\mid\cs_l,\td\cs_l\bigr)
    &=\bq\bigl(([\ch_{d-l}],[\td\ch_{d-l}])\in\cf_{d-l}\bigr)\nonumber\\
    &\le C_3e^{-d+l} \label{eq:edge-prob-rev}
\end{align}
for all sufficiently large $n$, where $C_3$ is a constant independent of $n$.

To bound the type-I error, we now condition on the event
\begin{align}
    \ca_4:=\Bigl\{\max\{|\cs_l|,|\td\cs_l|\}\le(e\lambda)^l\Bigr\}, \label{eq:def-A5}
\end{align}
on which both level-$l$ vertex sets have at most $(e\lambda)^l$ vertices, and split
\begin{align}
    \bq\bigl(\Phi([\ch_d],[\td\ch_d])=1\bigr)
    \le\bq\Bigl(\texttt{MaxMatching}(\cw)\ge\tfrac{1}{2e}(\lambda s)^l\,\Big|\,\ca_4\Bigr)
    +\bq(\ca_4^c). \label{eq:two_terms}
\end{align}
We bound the two terms on the right-hand side of~\eqref{eq:two_terms} in turn.

We first bound the conditional probability in~\eqref{eq:two_terms}. Since $\texttt{MaxMatching}(\cw)$ takes integer values, the event $\{\texttt{MaxMatching}(\cw)\ge\frac{1}{2e}(\lambda s)^l\}$ holds if and only if $\cw$ contains a matching of size $\bigl\lceil\frac{1}{2e}(\lambda s)^l\bigr\rceil$. For any fixed realizations of $(\cs_l,\td\cs_l)$ consistent with $\ca_4$ and any positive integer $k$, a matching of size $k$ between $\cs_l$ and $\td\cs_l$ is specified by choosing $k$ vertices on each side together with a bijection between them. Thus, the number of such matchings is at most

\[\binom{|\cs_l|}{k}\binom{|\td\cs_l|}{k}k!\le(|\cs_l|\,|\td\cs_l|)^k/k!\le \frac{(e\lambda)^{2lk}}{k!}\]

Therefore, conditioned on the event $\ca_4$ where $|\cs_l|,|\td\cs_l|\le(e\lambda)^l$, by taking a union bound over these matchings, we obtain
\begin{align}
    \bq\bigl(\cw\text{ has a matching of size }k\mid\cs_l,\td\cs_l\bigr)\le& \frac{(e\lambda)^{2lk}}{k!}(\bq\bigl((u,\td v)\in\cw\mid\cs_l,\td\cs_l\bigr))^k\nonumber\\
    \le &\frac{(e\lambda)^{2lk}}{k!}\bigl(C_3 e^{-d+l}\bigr)^{k}\nonumber\\
    \le &\Bigl(\frac{e\,C_3\,(e\lambda)^{2l}\,e^{-d+l}}{k}\Bigr)^{k}, \label{eq:typeI-cond-rev}
\end{align}
where the first inequality follows from the conditional independence of the edges in each matching, the second inequality holds by~\eqref{eq:edge-prob-rev}, and the last inequality uses $k!\ge(k/e)^k$. Applying~\eqref{eq:typeI-cond-rev} with $k=\bigl\lceil\frac{1}{2e}(\lambda s)^l\bigr\rceil\ge\frac{1}{2e}(\lambda s)^l$, the base on~\eqref{eq:typeI-cond-rev} satisfies
\begin{align}
    \frac{eC_3(e\lambda)^{2l}e^{-d+l}}{k}
    &\le \frac{eC_3(e\lambda)^{2l}e^{-d+l}}{\frac{1}{2e}(\lambda s)^l}\nonumber\\
    &=\exp\Bigl(\log(2e^2C_3)-d
    +l\bigl(3-\log s+\log\lambda\bigr)\Bigr). \label{eq:base-rev}
\end{align}
The assumption $\lambda=(\log n)^{O(1)}$ and the identity $l=\lceil2\log\log n\rceil$ imply that all terms in the exponent of~\eqref{eq:base-rev} other than $-d$ are $O((\log\log n)^2)$. Since $d=\lceil(\log n)^{1/3}\,\rceil$, this remainder is $o(d)$, so~\eqref{eq:base-rev} is at most $e^{-d/2}$ for all sufficiently large $n$. Using again that $k\ge\frac{1}{2e}(\lambda s)^l$, we have
\begin{align}
    \bq\Bigl(\texttt{MaxMatching}(\cw)\ge\tfrac{1}{2e}(\lambda s)^l\,\Big|\,\cs_l,\td\cs_l\Bigr)
    \le\bigl(e^{-d/2}\bigr)^{\frac{1}{2e}(\lambda s)^l}
    =\exp\Bigl(-\frac{(\lambda s)^l d}{4e}\Bigr) \label{eq:typeI-uniform-rev}
\end{align}
holds for every realization of $(\cs_l,\td\cs_l)$ consistent with $\ca_4$. 

Since $\lambda=\omega(1)$,  we have $\lambda s\ge e$ for all sufficiently large $n$, and $l\ge2\log\log n$ gives
\[
    \frac{(\lambda s)^l d}{4e}
    \ge \frac{1}{4e} (\log n)^{\frac13+2\log(\lambda s)}\ge \log 2+\log n\log\log n
\]
for all sufficiently large $n$. Therefore,~\eqref{eq:typeI-uniform-rev} indicates that conditional on $\ca_4$,
\begin{align}
    \bq\Bigl(\texttt{MaxMatching}(\cw)\ge\tfrac{1}{2e}(\lambda s)^l\,\Big|\,\ca_4\Bigr)\le\frac12 n^{-\log\log n}. \label{eq:first-term}
\end{align}

Second, we bound $\bq(\ca_4^c)$. Since $\lambda=\omega(1)$, we have $\lambda\ge2$ for all sufficiently large $n$. Thus the upper-tail bound~\eqref{eq:gen_LB} in Lemma~\ref{lem:gen-size-rev} of Appendix~\ref{appd:gen-size-rev} applies with offspring mean $\mu=\lambda$ to each marginal tree under $\bq$. Moreover, since $l\ge2\log\log n$, we have
\[
\frac14 e^l\ge \frac14 (\log n)^2 \ge \log n\log\log n+2\log 2+\frac12
\]
for all sufficiently large $n$. Applying~\eqref{eq:gen_LB}, we obtain
\begin{align}
    \bq(\ca_4^c)
    \le \P\bigl(|\cs_l|\ge(e\lambda)^l\bigr) +\P\bigl(|\td\cs_l|\ge(e\lambda)^l\bigr)
    \le 2\exp\Bigl( -\tfrac14 e^l+\tfrac12 \Bigr)
    \le \frac12 n^{-\log\log n}, \label{eq:ca5-prob}
\end{align}
for all sufficiently large $n$. Substituting~\eqref{eq:first-term} and~\eqref{eq:ca5-prob} into~\eqref{eq:two_terms} yields 
\[
\bq(\Phi([\ch_d],[\td\ch_d])=1)\le n^{-\log\log n}
\]
for all sufficiently large $n$, which completes the proof of~\eqref{eq:CorTest-typeI}.

\underline{\textbf{Part 2: Proof of \eqref{eq:CorTest-power}}.} 
Suppose $(\ch,\td\ch)\sim\bp$, and let $\ch^*$ be the intersection tree in the construction of the correlated trees. We write $\cs_l^*$ for the set of vertices in the $l$-th generation of $\ch^*$. We first exhibit a matching in $\cw$ indexed by true pairs $(u,\td u)$, and then control its size.

For each $u^*\in\cs_l^*$, let $u$ and $\td u$ denote the corresponding nodes in $\ch_d$ and $\td\ch_d$ at level $l$, respectively. If $([\ch_{d,u}],[\td\ch_{d,\td u}])\in\cf_{d-l}$, then $(u,\td u)\in \ce(\cw)$ by the definition of $\cw$. Therefore, the set of pairs $(u,\td u)$ such that $([\ch_{d,u}],[\td\ch_{d,\td u}])\in\cf_{d-l}$ forms a matching of $\cw$, and we have
\[
\texttt{MaxMatching}(\cw)\geq \sum_{u^*\in \cs^*_l}\mathbbm{1}\{([\ch_{d,u}],[\td\ch_{d,\td u}])\in\cf_{d-l}\}.
\]
Therefore,
\begin{align}
    \bp\left(
\texttt{MaxMatching}(\cw)
\le \frac{1}{2e}(\lambda s)^{l}
\right)
\le 
\bp\left(
\sum_{u^*\in \cs^*_l}\mathbbm{1}\{([\ch_{d,u}],[\td\ch_{d,\td u}])\in\cf_{d-l}\} \le \frac{1}{2e}(\lambda s)^{l}
\right).\label{eq:s3_bd1}
\end{align}
Next, we define a typical event
\begin{align*}
    \mathcal{A}_5 
&:= \Bigl\{ |\cs_l^*| \ge \frac{(\lambda s)^{l}}{e} \Bigr\}.
\end{align*}
The probability $\bp(\ca_5)$ can be bounded by Lemma~\ref{lem:gen-size-rev}. First, since $\ch^*$ is a Galton--Watson tree with offspring distribution $\mathrm{Poi}(\lambda s)$ and $\lambda s\ge 2$ for all sufficiently large $n$, Lemma~\ref{lem:gen-size-rev} applies for every $l\in\mathbb N$, and by~\eqref{eq:gen_UB},
\begin{align}
    \mathsf{P}(\ca_5)\geq  1-\frac{e^2}{(e-1)^2}\cdot\frac{1}{\lambda s-1}. \label{eq:ca4-prob}
\end{align}
Next, conditioned on $|\cs^*_l|$, the subtree pairs $\{([\ch_{d,u}],[\td\ch_{d,\td u}]):u^*\in\cs_l^*\}$ are independent and identically distributed with the correlated model $\bp$. Therefore,
\begin{align}
    \sum_{u^*\in \cs^*_l}\mathbbm{1}\{([\ch_{d,u}],[\td\ch_{d,\td u}])\in\cf_{d-l}\}\sim \mathrm{Bin}\left(
|\cs^*_l|,\, \bp\bigl(([\ch_{d-l}],[\td\ch_{d-l}])\in\cf_{d-l}\bigr)
\right). \nonumber
\end{align}

Conditioned on event $\ca_5$, we have $\lvert\cs^*_l\rvert\ge e^{-1}(\lambda s)^{l}$. Moreover, by~\eqref{eq:L2_Power} in Lemma~\ref{lem:matching}, we have $\bp\bigl(([\ch_{d-l}],[\td\ch_{d-l}])\in\cf_{d-l}\bigr)\geq  1-C_4 \lambda^{-1}$ for all sufficiently large $n$. Therefore,
\begin{align}
    \sum_{u^*\in \cs^*_l}\mathbbm{1}\{([\ch_{d,u}],[\td\ch_{d,\td u}])\in\cf_{d-l}\}\stackrel{\text{sto.}}{\geq}\mathrm{Bin}\left(\frac{1}{e}(\lambda s)^{l},\, 1-C_4\lambda^{-1}\right).\label{eq:s3_power_bd2}
\end{align}
By Chernoff bound \eqref{eq:CB_Lower} in Lemma \ref{lem:CB}, we have
\begin{align} 
&\bp\left(
\sum_{u^*\in \cs^*_l}\mathbbm{1}\{([\ch_{d,u}],[\td\ch_{d,\td u}])\in\cf_{d-l}\} \le \frac{1}{2e}(\lambda s)^{l}\,\middle|\, \ca_5
\right)\nonumber\\
\leq{} & \P\left( \textrm{Bin}\left(
\frac{1}{e}(\lambda s)^{l},\, 1-C_4\lambda^{-1}
\right)\le \frac{1}{2e}(\lambda s)^{l}
\right)\nonumber\\
\leq{} &
\exp\left(
-\frac{1}{8e}(\lambda s)^{l}
\cdot
\frac{(1-2C_4\lambda^{-1})^2}{1-C_4\lambda^{-1}}
\right),\label{s3_bd2}
\end{align}
Therefore, combining \eqref{eq:ca4-prob} and \eqref{s3_bd2} gives
\begin{align*}
    &\bp\left(
\texttt{MaxMatching}(\cw)
\geq \frac{1}{2e}(\lambda s)^{l}
\right) \\\geq{}& 1- \bp\left(
|\{u^*\in \cs^*_l: ([\ch_{d,u}],[\td\ch_{d,\td u}])\in\cf_{d-l}\}| \le \frac{1}{2e}(\lambda s)^{l}\,\middle|\, \ca_5
\right)- \bp(\ca_5^c)\\
\geq{} & 1- \exp\left(
-\frac{1}{8e}(\lambda s)^{l}
\cdot
\frac{(1-2C_4\lambda^{-1})^2}{1-C_4\lambda^{-1}}
\right)-\frac{e^2}{(e-1)^2}\cdot\frac{1}{\lambda s-1}\\
\geq{}& 1-\frac{7}{\lambda},
\end{align*}
for all sufficiently large $n$. In the last inequality, the exponential term is at most $1/\lambda$ since $(\lambda s)^l\ge e^l\ge(\log n)^2$ and $\lambda=(\log n)^{O(1)}$, and the last term is at most $6/\lambda$ since $s>\sqrt{C_{\mathrm{Otter}}}>1/2$. This completes the proof of \eqref{eq:CorTest-power}.
\end{proof}

\subsection{Error Analysis of $\Psi$}\label{appd:pf-lrt}
Proposition~\ref{prop:CorTest} bounds the type-I error and the power of the auxiliary test $\Phi$, and Lemma~\ref{lem:LR-approximation} shows that $\widehat L_d$ lies within a factor $e^{\pm1/16}$ of $L_d$ with probability at least $1-\xi$. We now use these results to prove the type-I error bound~\eqref{eq:rank-LRT-typeI} and the power bound~\eqref{eq:rank-LRT-power} of $\Psi$ in Theorem~\ref{thm:LRT}.

\begin{proof}
For each fixed tree pair $(t,\td t)\in\cz_d\times\cz_d$, define the event
\[
    \ca_6(t,\td t):=\left\{\widehat L_d(t,\td t)<e^{-1/16}L_d(t,\td t),\text{ or } {\widehat L_d}(t,\td t)> e^{1/16}L_d(t,\td t)\right\}.
\]
By the law of total probability,
\begin{align}
    \bq\bigl(\ca_6([\ch_d],[\td\ch_d])\bigr)
    &=\sum_{(t,\td t)\in\cz_d\times\cz_d}\bq\bigl([\ch_d]=t,[\td\ch_d]=\td t\bigr)\P\left(\ca_6(t,\td t)\right)\nonumber\\
    &\le n^{-11}\sum_{(t,\td t)\in\cz_d\times\cz_d}\bq\bigl([\ch_d]=t,[\td\ch_d]=\td t\bigr)=n^{-11},
    \label{eq:LRT-approximation-failure-Q}
\end{align}
where the equality holds because the random bits used by \texttt{ApproxLR} are independent of $([\ch_d],[\td\ch_d])$, and the inequality follows from Lemma~\ref{lem:LR-approximation} proved in Appendix~\ref{sec:approximate-L-guarantees}. Applying the same argument under $\bp$, we have
\begin{align}
    \bp\bigl(\ca_6([\ch_d],[\td\ch_d])\bigr)\le n^{-11}\sum_{(t,\td t)\in\cz_d\times\cz_d}\bp\bigl([\ch_d]=t,[\td\ch_d]=\td t\bigr)=n^{-11}.
    \label{eq:LRT-approximation-failure-P}
\end{align}

We first prove the type-I error bound~\eqref{eq:rank-LRT-typeI}. Since the likelihood-ratio $L_d([\ch_d],[\td\ch_d])$ is nonnegative and $\E_{\bq}[L_d([\ch_d],[\td\ch_d])]=1$, Markov's inequality gives
\begin{align}
    \bq\left(L_d([\ch_d],[\td\ch_d])>e^{-1/16}\theta^*\right)&\le\frac{e^{1/16}}{\theta^*}=n^{-(\log\log n)/2}, \label{eq:LRT-exact-upper-tail}
\end{align}
and by the union bound, the type-I error of $\Psi$ satisfies
\begin{align}
    \bq\bigl(\Psi([\ch_d],[\td\ch_d])=1\bigr)&\le\bq\bigl(\Psi([\ch_d],[\td\ch_d])=1,\ca_6([\ch_d],[\td\ch_d])^c\bigr)+\bq\bigl(\ca_6([\ch_d],[\td\ch_d])\bigr)\nonumber\\
    &\le\bq\left(L_d([\ch_d],[\td\ch_d])>e^{-1/16}\theta^*\right)+\bq\bigl(\ca_6([\ch_d],[\td\ch_d])\bigr)\nonumber\\
    &\le n^{-(\log\log n)/2}+n^{-11},
    \label{eq:rank_LRT_Type1_Proof}
\end{align}
where the first inequality follows by splitting according to whether $\ca_6([\ch_d],[\td\ch_d])$ occurs. The second inequality holds because for every $(t,\td t)\in\cz_d\times\cz_d$, if $\ca_6(t,\td t)^c$ occurs and $\Psi(t,\td t)=1$, then $L_d(t,\td t)\ge e^{-1/16}\widehat L_d(t,\td t)>e^{-1/16}\theta^*$. The last inequality follows from~\eqref{eq:LRT-exact-upper-tail} and~\eqref{eq:LRT-approximation-failure-Q}. This proves~\eqref{eq:rank-LRT-typeI}.

We next prove the power bound~\eqref{eq:rank-LRT-power}. We have
\begin{align}
    \bp\bigl(\Psi([\ch_d],[\td\ch_d])=1\bigr)&=1-\bp\bigl(\Psi([\ch_d],[\td\ch_d])=0\bigr)\nonumber\\
    &\stackrel{\text{(a)}}{\ge}1-\bp\bigl(\Psi([\ch_d],[\td\ch_d])=0,\ca_6([\ch_d],[\td\ch_d])^c\bigr)-\bp\bigl(\ca_6([\ch_d],[\td\ch_d])\bigr)\nonumber\\
    &\stackrel{\text{(b)}}{\ge}1-\bp\left(L_d([\ch_d],[\td\ch_d])\le e^{1/16}\theta^*\right)-\bp\bigl(\ca_6([\ch_d],[\td\ch_d])\bigr)\nonumber\\
    &\stackrel{\text{(c)}}{\ge}1-\bp\left(L_d([\ch_d],[\td\ch_d])\le e^{1/16}\theta^*\right)-n^{-11},
    \label{eq:LRT-power-reduction}
\end{align}
where (a) follows by splitting according to whether $\ca_6([\ch_d],[\td\ch_d])$ occurs, (b) holds since for every $(t,\td t)\in\cz_d\times\cz_d$, if $\ca_6(t,\td t)^c$ occurs and $\Psi(t,\td t)=0$, then $L_d(t,\td t)\le e^{1/16}\widehat L_d(t,\td t)\le e^{1/16}\theta^*$. (c) follows from~\eqref{eq:LRT-approximation-failure-P}. To bound $\bp\left(L_d([\ch_d],[\td\ch_d])\le e^{1/16}\theta^*\right)$, we apply~\eqref{eq:aux-LR-power} with $\theta=e^{1/16}\theta^*$ to the auxiliary test $\Phi$ in Proposition~\ref{prop:CorTest}. This gives
\begin{align}
    \bp\left(L_d([\ch_d],[\td\ch_d])\le e^{1/16}\theta^*\right)
    &\le\bp\bigl(\Phi([\ch_d],[\td\ch_d])=0\bigr)+e^{1/16}\theta^*\,\bq\bigl(\Phi([\ch_d],[\td\ch_d])=1\bigr)\nonumber\\
    &\le\frac{7}{\lambda}+e^{1/16}\theta^*n^{-\log\log n},
    \label{eq:LRT-power-aux}
\end{align}
where the last inequality follows from~\eqref{eq:CorTest-power} and~\eqref{eq:CorTest-typeI} in Proposition~\ref{prop:CorTest}. Substituting~\eqref{eq:LRT-power-aux} into~\eqref{eq:LRT-power-reduction} and using the definition of $\theta^*$, we obtain
\begin{align}
	\bp\bigl(\Psi([\ch_d],[\td\ch_d])=1\bigr)&\geq 1- \bp\left(L_d([\ch_d],[\td\ch_d])\le e^{1/16}\theta^*\right)-n^{-11}\nonumber\\
    &\ge1- \frac{7}{\lambda}-e^{1/16}\theta^*n^{-\log\log n}-n^{-11}\nonumber\\
    &=1-\frac{7}{\lambda}-e^{1/8}n^{-(\log\log n)/2}-n^{-11}.
    \label{eq:LRT-exact-lower-tail}
\end{align}
This proves~\eqref{eq:rank-LRT-power} and completes the proof of Theorem~\ref{thm:LRT}.
\end{proof}

\section{Analysis of \texttt{ApproxLR}}\label{appd:approxlr}

This appendix analyzes the algorithm \texttt{ApproxLR} defined in Section~\ref{sec:subset-dp-calculation}. In Appendix~\ref{appd:approxlr-lemmas}, we prove Lemmas~\ref{lem:approximate-L-recursion} and~\ref{lem:approximate-L-matching-permanent}. In Appendix~\ref{sec:approximate-L-guarantees}, we bound the accumulation of approximation errors and prove Lemma~\ref{lem:LR-approximation}. Finally, in Appendix~\ref{sec:approximate-L-complexity}, we prove Lemma~\ref{lem:approximate-L-complexity}, which bounds the computational complexity of \texttt{ApproxLR}.

\subsection{Proofs of Lemmas~\ref{lem:approximate-L-recursion} and~\ref{lem:approximate-L-matching-permanent}}
\label{appd:approxlr-lemmas}

\LRRecursion*

\begin{proof}
By~\cite[Lemma~2.1, Eqs.~(2.4)--(2.5)]{Luca24LRT}, the likelihood ratio satisfies
\begin{align}
    L_k(t,\td t)
    =\sum_{q=0}^{\min\{c,\td c\}}\frac{e^{\lambda s}s^q(1-s)^{c+\td c-2q}}{\lambda^q q!}\sum_{\substack{\sigma\in\mathrm{Inj}([1:q],[1:c])\\\td\sigma\in\mathrm{Inj}([1:q],[1:\td c])}}\prod_{a=1}^{q}L_{k-1}(t_{\sigma(a)},\td t_{\td\sigma(a)}),
    \label{eq:approximate-L-source-recursion}
\end{align}
where for $q=0$, the inner sum and product evaluate to $1$. 
For $q \ge 1$ and each pair $(\sigma,\td\sigma)$ in the inner sum, the definition of $\mathbf Z$ gives
\begin{align}
    \prod_{a=1}^{q}Z_{\sigma(a),\td\sigma(a)}
    =\left(\frac{s}{\lambda}\right)^q\prod_{a=1}^{q}L_{k-1}(t_{\sigma(a)},\td t_{\td\sigma(a)}).
    \label{eq:prod_LR_iter}
\end{align}
Equation~\eqref{eq:prod_LR_iter} also holds for $q=0$ since both sides equal $1$. Substituting~\eqref{eq:prod_LR_iter} into~\eqref{eq:approximate-L-source-recursion} and applying the definitions of $S_q(\mathbf Z)$ and $S_0(\mathbf Z)=1$ yields~\eqref{eq:approximate-L-exact-recursion}.

When $s=1$, the exponent $c+\td c-2q$ is strictly positive unless $q=c=\td c$. Thus, $(1-s)^{c+\td c-2q}=0$ for $q < c=\td c$. Under the convention $0^0=1$, the term for $q=c=\td c$ equals to $1$, verifying that~\eqref{eq:approximate-L-exact-recursion} holds for all $s\in(0,1]$. At depth $0$, both tree laws assign probability one to the singleton tree, yielding $L_0(t,\td t)=1$.
\end{proof}

The following proof is adapted from~\cite[Theorem~2.1]{Friedland06Matchings}, and we include it for completeness.

\PermanentIdentity*

\begin{proof}
By definition, the permanent of $\mathbf A_q(\mathbf Z)$ is a sum over all permutations of its row and column indices. The bottom-right block of $\mathbf A_q(\mathbf Z)$ is a $(\td c-q)\times(c-q)$ zero matrix, and any permutation selecting an entry from this block contributes zero to the permanent. 

For $q \ge 1$, permutations that avoid this zero block must map the bottom $\td c-q$ rows to distinct columns among the first $\td c$ columns. This leaves exactly $\td c - (\td c-q) = q$ columns among the first $\td c$ columns available for the top $c$ rows. Because the top-right block $\mathbf 1_{c\times(c-q)}$ has $c-q$ columns, the top $c$ rows must select exactly $q$ entries from the top-left block $\mathbf Z$, and the remaining $c-q$ entries from the top-right block $\mathbf 1_{c\times(c-q)}$. 

A choice of $q$ entries in $\mathbf Z$ (with no two in the same row or column) corresponds to a pair of injections $(\sigma, \td\sigma) \in \mathrm{Inj}([1:q],[1:c]) \times \mathrm{Inj}([1:q],[1:\td c])$. Summing over all valid pairs of injections counts each such choice of $q$ entries exactly $q!$ times, corresponding to the orderings of the index set $[1:q]$. For a fixed pair $(\sigma, \td\sigma)$, assigning the remaining $c-q$ top rows to the rightmost $c-q$ columns yields $(c-q)!$ bijections, and assigning the bottom $\td c-q$ rows to the remaining $\td c-q$ columns among the first $\td c$ columns yields $(\td c-q)!$ bijections. Since all entries in the $\mathbf 1$ blocks equal $1$, each such permutation evaluates to $\prod_{a=1}^{q}Z_{\sigma(a),\td\sigma(a)}$. Summing over all choices gives
\begin{align*}
    q!\operatorname{per}(\mathbf A_q(\mathbf Z))
    =(c-q)!(\td c-q)!\sum_{\substack{\sigma\in\mathrm{Inj}([1:q],[1:c])\\\td\sigma\in\mathrm{Inj}([1:q],[1:\td c])}}\prod_{a=1}^{q}Z_{\sigma(a),\td\sigma(a)}.
\end{align*}
Applying the definition of $S_q(\mathbf Z)$ from~\eqref{eq:approximate-L-matching-sum} and dividing by $q!(c-q)!(\td c-q)!$ yields~\eqref{eq:approximate-L-permanent-identity}.

For $q=0$, permutations avoiding the bottom-right zero block must map the bottom $\td c$ rows to the first $\td c$ columns, and the top $c$ rows to the rightmost $c$ columns. There are $\td c! c!$ such permutations. All selected entries lie in the $\mathbf 1$ blocks, yielding $\operatorname{per}(\mathbf A_0(\mathbf Z)) = c!\td c!$. Since $S_0(\mathbf Z)=1$, Equation~\eqref{eq:approximate-L-permanent-identity} holds. For $c=\td c=q=0$, the matrix $\mathbf A_0(\mathbf Z)$ has dimension $0\times 0$, and~\eqref{eq:approximate-L-permanent-identity} holds since $\operatorname{per}(\mathbf A_0(\mathbf Z))=1$ and $S_0(\mathbf Z)=1$.
\end{proof}

\subsection{Approximation Guarantee}
\label{sec:approximate-L-guarantees}

We first bound the error accumulated through the recursive approximations of the matching sums.

\begin{lem}[Approximation error]
\label{lem:approximate-L-error-propagation}
Let $\lambda>0$, $s\in(0,1]$, $\xi\in(0,1/2)$, and let $d\ge0$ be an integer. For each fixed pair $(t,\td t)\in\cz_d\times\cz_d$, let $\widehat L_d(t,\td t)$ be the output of Algorithm~\ref{algo:approximate-LR} with parameters $\lambda, s, \xi$ and $d$. If for every $h\in[1:d]$ and every pair of vertices $(u,\td v)\in\cs_{d-h}(t)\times\cs_{d-h}(\td t)$, each value $\widehat S_q(u,\td v)$ computed via~\eqref{eq:approximate-L-approximate-matching-sum} satisfies
\begin{align}
	    (1-\delta)S_q(\widehat{\mathbf Z}(u,\td v)) \le\widehat S_q(u,\td v) \le(1+\delta)S_q(\widehat{\mathbf Z}(u,\td v)), \label{eq:approx_error_cond}
\end{align}
where $\delta$ is the fixed parameter computed in Step~1, then we have
\begin{align}
    e^{-1/16}L_d(t,\td t)\le\widehat L_d(t,\td t)\le e^{1/16}L_d(t,\td t).
    \label{eq:approximate-L-root-error}
\end{align}
\end{lem}

\begin{proof}
Let $D$ and $\delta$ be the parameters computed in Step~1 of Algorithm~\ref{algo:approximate-LR}. Define the sequence $(b_h)_{h=0}^d$ by setting
\begin{align}
    b_0:=0,\qquad b_h:=D b_{h-1}+2\delta \quad \text{for } h\in[1:d].
    \label{eq:approximate-L-error-recursion}
\end{align}
Since $D\ge2$ and $\delta>0$, we have $b_h\ge0$ for every $h\in[0:d]$. We proceed by induction on $h$ to prove that for every $h\in[0:d]$ and every pair $(u,\td v)\in\cs_{d-h}(t)\times\cs_{d-h}(\td t)$,
\begin{align}
    e^{-b_h}L_h(t_u,\td t_{\td v})\le\widehat\ell(u,\td v)\le e^{b_h}L_h(t_u,\td t_{\td v}).
    \label{eq:approximate-L-level-error}
\end{align}

For the base case $h=0$, Lemma~\ref{lem:approximate-L-recursion} gives $L_0(t_u,\td t_{\td v})=1$, and~\eqref{eq:approximate-L-initialization} sets $\widehat\ell(u,\td v)=1$. Hence, \eqref{eq:approximate-L-level-error} holds with $b_0=0$.

Fix $h\in[1:d]$, and assume that~\eqref{eq:approximate-L-level-error} holds for every vertex pair at depth $d-h+1$. Consider an arbitrary pair $(u,\td v)\in\cs_{d-h}(t)\times\cs_{d-h}(\td t)$. Following Step~3, we enumerate the child sets as $\sfch_t(u)=\{y_1,\ldots,y_c\}$ and $\sfch_{\td t}(\td v)=\{\td y_1,\ldots,\td y_{\td c}\}$. Define the matrix $\mathbf Z(u,\td v)$ by setting
\[
    \mathbf Z(u,\td v)_{ij}:=\frac{s}{\lambda}L_{h-1}(t_{y_i},\td t_{\td y_j}),\qquad (i,j)\in[1:c]\times[1:\td c].
\]
Since every child pair $(y_i,\td y_j)$ resides at depth $d-h+1$, applying the induction hypothesis to these pairs and using~\eqref{eq:approximate-L-approximate-entries} yields
\begin{align}
    e^{-b_{h-1}}\mathbf Z(u,\td v)_{ij}\le\widehat{\mathbf Z}(u,\td v)_{ij}\le e^{b_{h-1}}\mathbf Z(u,\td v)_{ij}.
    \label{eq:approximate-L-entry-error}
\end{align}

Fix $q\in[1:\min\{c,\td c\}]$. For each pair of injections $\sigma\in\mathrm{Inj}([1:q],[1:c])$ and $\td\sigma\in\mathrm{Inj}([1:q],[1:\td c])$, all terms in~\eqref{eq:approximate-L-entry-error} are nonnegative. Multiplying these inequalities over the $q$ pairs $(\sigma(a),\td\sigma(a))$ gives
\begin{align*}
    e^{-q b_{h-1}}\prod_{a=1}^{q}\mathbf Z(u,\td v)_{\sigma(a),\td\sigma(a)}
    \le\prod_{a=1}^{q}\widehat{\mathbf Z}(u,\td v)_{\sigma(a),\td\sigma(a)}\le e^{q b_{h-1}}\prod_{a=1}^{q}\mathbf Z(u,\td v)_{\sigma(a),\td\sigma(a)}.
\end{align*}
By the definition of $S_q$ in~\eqref{eq:approximate-L-matching-sum}, summing these inequalities over all pairs $(\sigma,\td\sigma)$ and dividing by $q!$ gives
\begin{align}
    e^{-q b_{h-1}}S_q(\mathbf Z(u,\td v))\le S_q(\widehat{\mathbf Z}(u,\td v))\le e^{q b_{h-1}}S_q(\mathbf Z(u,\td v)).
    \label{eq:approximate-L-matching-input-error}
\end{align}
For $q=0$, since $S_0(\mathbf Z(u,\td v))=S_0(\widehat{\mathbf Z}(u,\td v))=\widehat S_0(u,\td v)=1$,~\eqref{eq:approximate-L-matching-input-error} also holds. Combining~\eqref{eq:approximate-L-matching-input-error} with the assumption~\eqref{eq:approx_error_cond}, we have for every $q\in[0:\min\{c,\td c\}]$,
\begin{align}
    (1-\delta)e^{-q b_{h-1}}S_q(\mathbf Z(u,\td v)) \le\widehat S_q(u,\td v) \le(1+\delta)e^{q b_{h-1}}S_q(\mathbf Z(u,\td v)).\label{eq:approx_error_recur3}
\end{align}
Since $\delta\in(0,1/2)$, the inequalities $e^{-2\delta}\le1-\delta$ and $1+\delta\le e^{2\delta}$ hold. Furthermore, $q \le \max\{c,\td c\} \le D$ and $b_{h-1}\ge0$, which implies $q b_{h-1}+2\delta\le D b_{h-1}+2\delta = b_h$. Applying these bounds to~\eqref{eq:approx_error_recur3}, we have
\begin{align}
    e^{-b_h}S_q(\mathbf Z(u,\td v))\le\widehat S_q(u,\td v)\le e^{b_h}S_q(\mathbf Z(u,\td v)).
    \label{eq:approximate-L-matching-total-error}
\end{align}

For $s\in(0,1]$ and under the convention $0^0=1$, the coefficients $e^{\lambda s}(1-s)^{c+\td c-2q}$ are nonnegative. Multiplying~\eqref{eq:approximate-L-matching-total-error} by these coefficients and summing over $q$ gives
\begin{align*}
    e^{-b_h}L_h(t_u,\td t_{\td v})\le e^{\lambda s}\sum_{q=0}^{\min\{c,\td c\}}(1-s)^{c+\td c-2q}\widehat S_q(u,\td v)=\widehat\ell(u,\td v)\le e^{b_h}L_h(t_u,\td t_{\td v}),
\end{align*}
where the inequalities follow from the exact recursion~\eqref{eq:approximate-L-exact-recursion} in Lemma~\ref{lem:approximate-L-recursion}, and the equality follows from the update rule~\eqref{eq:approximate-L-update}. This completes the induction step and proves~\eqref{eq:approximate-L-level-error} for all $h\in[0:d]$.

It remains to bound the final error exponent $b_d$. For $d\ge1$, the recurrence~\eqref{eq:approximate-L-error-recursion} gives
\[
    b_d=2\delta\sum_{j=0}^{d-1}D^j=2\delta\frac{D^d-1}{D-1}\le 2\delta D^d \le \frac{1}{16},
\]
where we used $D\ge2$ and $\delta=1/(64D^d)$ from~\eqref{eq:approximate-L-parameters}. For $d=0$, we have $b_d=b_0=0 \le 1/16$. Applying~\eqref{eq:approximate-L-level-error} at $h=d$ to the root pair $\rho$ and $\td\rho$, and using $\widehat L_d(t,\td t)=\widehat\ell(\rho,\td\rho)$ from~\eqref{eq:approximate-L-output}, we have~\eqref{eq:approximate-L-root-error}, which completes the proof.
\end{proof}

We now bound the probability that the condition in~\eqref{eq:approx_error_cond} fails. Combined with Lemma~\ref{lem:approximate-L-error-propagation}, this establishes the following approximation guarantee.

\begin{lem}[Likelihood-ratio approximation]
\label{lem:LR-approximation}
Let $\lambda>0$, $s\in(0,1]$, $\xi\in(0,1/2)$, and let $d\ge0$ be an integer. 
For a fixed pair $(t,\td t)\in\cz_d\times\cz_d$, let $\widehat L_d(t,\td t)$ be the output of Algorithm~\ref{algo:approximate-LR} with parameters $\lambda, s, \xi$, and $d$. Then
\begin{align}
    \P\left(e^{-1/16}L_d(t,\td t)\le\widehat L_d(t,\td t)\le e^{1/16}L_d(t,\td t)\right)\ge1-\xi,
    \label{eq:LR-approximation-uniform}
\end{align}
\end{lem}

\begin{proof}
Fix $(t,\td t)\in\cz_d\times\cz_d$, and let $D$, $\delta$, and $\eta$ be the parameters computed in Step~1 of Algorithm~\ref{algo:approximate-LR}. By~\eqref{eq:approximate-L-parameters}, we have $D\ge2$ and $\delta,\eta\in(0,1/2)$. For $d=0$, Step~2 and Lemma~\ref{lem:approximate-L-recursion} give $\widehat L_0(t,\td t)=L_0(t,\td t)=1$, so~\eqref{eq:LR-approximation-uniform} holds with probability $1$. We now consider $d\ge1$.

We first bound the total number of invocations of $\texttt{ApproxPer}$ in Step~3. Since every vertex in either tree has at most $D$ children, we have $|\cs_{d-h}(t)|\le D^{d-h}$ and $|\cs_{d-h}(\td t)|\le D^{d-h}$ for every $h\in[1:d]$. At each vertex pair, Step~3 invokes $\texttt{ApproxPer}$ at most $D$ times, as the number of invocations equals to the smaller number of their children. Thus, the total number of invocations is at most
\begin{align}
    D\sum_{h=1}^{d}|\cs_{d-h}(t)|\,|\cs_{d-h}(\td t)|
    &\le D\sum_{h=1}^{d}D^{2(d-h)}
    \le (D+1)\sum_{j=0}^{d}D^{2j}.
    \label{eq:approximate-L-call-count}
\end{align}

Fix $h\in[1:d]$ and $(u,\td v)\in\cs_{d-h}(t)\times\cs_{d-h}(\td t)$, and let $c=|\sfch_t(u)|$ and $\td c=|\sfch_{\td t}(\td v)|$. We next bound the probability that a value $\widehat S_q(u,\td v)$ fails to satisfy~\eqref{eq:approx_error_cond} for a fixed $q\in[1:\min\{c,\td c\}]$. The matrix $\widehat{\mathbf Z}(u,\td v)$ is completely determined by values computed at depths strictly larger than $d-h$. Conditioned on $\widehat{\mathbf Z}(u,\td v)=\mathbf Z$ for a fixed realization $\mathbf Z$, the value $\widehat S_q(u,\td v)$ is computed by $\texttt{ApproxPer}$ on the fixed input $\mathbf A_q(\mathbf Z)$. 

For any fixed realization $\mathbf Z$ of $\widehat{\mathbf Z}(u,\td v)$, the matrix $\mathbf A_q(\mathbf Z)$ is nonnegative. Conditioned on $\widehat{\mathbf Z}(u,\td v)=\mathbf Z$, we apply~\eqref{eq:approximate-L-permanent-guarantee} in Lemma~\ref{lem:approximate-L-permanent-FPRAS} to $\mathbf A_q(\mathbf Z)$ and divide the approximation bounds by the positive factor $(c-q)!(\td c-q)!$ to obtain
\[
    \P\left(\widehat S_q(u,\td v)\notin\bigl[(1-\delta)S_q(\mathbf Z),(1+\delta)S_q(\mathbf Z)\bigr]\cond\widehat{\mathbf Z}(u,\td v)=\mathbf Z\right)\le\eta,
\]
where we used the definition of $\widehat S_q(u,\td v)$ in~\eqref{eq:approximate-L-approximate-matching-sum} and the identity~\eqref{eq:approximate-L-permanent-identity} in Lemma~\ref{lem:approximate-L-matching-permanent}. Since this conditional bound holds uniformly for all possible realizations $\mathbf Z$, taking the expectation over $\widehat{\mathbf Z}(u,\td v)$ via the law of total probability yields the unconditional bound
\begin{align}
    \P\left(\widehat S_q(u,\td v)\notin\bigl[(1-\delta)S_q(\widehat{\mathbf Z}(u,\td v)),(1+\delta)S_q(\widehat{\mathbf Z}(u,\td v))\bigr]\right)\le\eta.
\end{align}

Applying the union bound over all values $\widehat S_q(u,\td v)$ computed via~\eqref{eq:approximate-L-approximate-matching-sum}, we obtain
\begin{align}
    \P\left(\text{\eqref{eq:approx_error_cond} fails for at least one }\widehat S_q(u,\td v)\right)
    &\le \eta D\sum_{h=1}^{d}|\cs_{d-h}(t)|\,|\cs_{d-h}(\td t)| \nonumber\\
    &\le \eta(D+1)\sum_{j=0}^{d}D^{2j}=\xi,
    \label{eq:approximate-L-all-calls-failure}
\end{align}
where the second inequality follows from~\eqref{eq:approximate-L-call-count}, and the equality follows from the choice of $\eta$ in~\eqref{eq:approximate-L-parameters}. Therefore, with probability at least $1-\xi$, all inequalities required by Lemma~\ref{lem:approximate-L-error-propagation} hold simultaneously. Applying~\eqref{eq:approximate-L-root-error} from Lemma~\ref{lem:approximate-L-error-propagation} establishes~\eqref{eq:LR-approximation-uniform}, which completes the proof.

\end{proof}

\subsection{Computational Complexity}
\label{sec:approximate-L-complexity}

In this subsection, we bound the computational complexity of $ \texttt{ApproxLR}$, as established in the following lemma.

\ApproxLRComplexity*

\begin{proof}
Fix $(t,\td t)\in\cz_d\times\cz_d$, and let $D$, $\delta$, $\eta$ be the parameters computed in Step~1 of Algorithm~\ref{algo:approximate-LR}. We bound the computational complexity of each step of this algorithm.

First, each input tree has at most $D^j$ vertices at depth $j$ for every $j\in[0:d]$. Hence,
\[
    |\cv(t)|+|\cv(\td t)|\le2\sum_{j=0}^{d}D^j\le4D^d,
\]
where the second inequality follows from $D\ge2$. Computing $D$ in Step~1 requires examining the children of each vertex, which has complexity $O(D^d)$. Once $D$ is known, computing $\delta$ and $\eta$ via~\eqref{eq:approximate-L-parameters} has complexity $O(d)$. Since $d\le D^d$, Step~1 has complexity $O(D^d)$. Step~2 initializes value $1$ for each pair in $\cs_d(t)\times\cs_d(\td t)$,and therefore has complexity $O(D^{2d})$.

Fix an iteration $h\in[1:d]$ and a pair $(u,\td v)\in\cs_{d-h}(t)\times\cs_{d-h}(\td t)$ in Step~3, and let $c=|\sfch_t(u)|$ and $\td c=|\sfch_{\td t}(\td v)|$. Since $c,\td c\le D$, enumerating the children and forming $\widehat{\mathbf Z}(u,\td v)$ via~\eqref{eq:approximate-L-approximate-entries} has complexity $O(D^2)$. For each $q\in[1:\min\{c,\td c\}]$, the nonnegative matrix $\mathbf A_q(\widehat{\mathbf Z}(u,\td v))$ is a square matrix with dimension $c+\td c-q\le2D$ and can be constructed in $O(D^2)$ time. By Lemma~\ref{lem:approximate-L-permanent-FPRAS}, the complexity of applying $\texttt{ApproxPer}$ to this matrix is bounded by $\operatorname{Poly}(2D,\delta^{-1},\log(\eta^{-1}))$. By~\eqref{eq:approximate-L-parameters}, we have
\begin{align*}
    \delta^{-1}=64D^d, &\quad \quad \quad\log(\eta^{-1})\le\log(\xi^{-1})+2(d+1)\log D.
\end{align*}
Therefore, computing each $\widehat S_q(u,\td v)$ via~\eqref{eq:approximate-L-approximate-matching-sum} has complexity $\operatorname{Poly}(D,D^d,\log(\xi^{-1})+(d+1)\log D)$. There are at most $D+1$ choices for $q$. The update in~\eqref{eq:approximate-L-update} combines these terms and requires $O(D)$ operations to calculate the coefficients $(1-s)^{c+\td c-2q}$. Thus, the total complexity of calculating $\widehat\ell(u,\td v)$ for the fixed pair $(u,\td v)$ is
\begin{align*}
	(D+1)\operatorname{Poly}\left(D,D^d,\log(\xi^{-1})+(d+1)\log D\right).
\end{align*}
  
Moreover, the total number of vertex pairs processed in Step~3 satisfies
\begin{align*}
	\sum_{h=1}^{d}|\cs_{d-h}(t)|\,|\cs_{d-h}(\td t)|\le\sum_{h=1}^{d}D^{2(d-h)}\le\sum_{j=0}^{d}D^{2j}.
\end{align*}

Finally, Step~4 returns the root value and has complexity $O(1)$. Summing up the costs from Steps~1 through~4, the overall computational complexity is
 \begin{align}
    &O(D^{d})+O(D^{2d})+(D+1)\sum_{j=0}^{d}D^{2j}\operatorname{Poly}\left(D,D^d,\log(\xi^{-1})+(d+1)\log D\right)+O(1)\nonumber\\
    &=(D+1)\sum_{j=0}^{d}D^{2j}\operatorname{Poly}\left(D,D^d,\log(\xi^{-1})+(d+1)\log D\right),\nonumber
 \end{align}
which proves~\eqref{eq:approximate-L-complexity-general}.  

For the last assertion, let $\xi=n^{-11}$ and assume $D^d=n^{o(1)}$. Since $D\ge2$ and $d\ge1$, we have
\begin{align*}
	D\le D^d=n^{o(1)},\qquad (D+1)\sum_{j=0}^{d}D^{2j}\le D^{2(d+1)}\le D^{4d}=n^{o(1)},
\end{align*}
and
\begin{align*}
	\log(\xi^{-1})+(d+1)\log D \le 11\log n+2\log(D^d)=O(\log n).
\end{align*}
Therefore, all arguments of the polynomial in~\eqref{eq:approximate-L-complexity-general} are $n^{o(1)}$, yielding a total complexity of $n^{o(1)}$. This completes the proof.
\end{proof}

\section{Seeded Graph Matching}\label{appd:seeded}
This appendix analyzes the seeded completion in Step~4 of Algorithm~\ref{algo:TBAlign}. We first state Algorithm~\ref{algo:seeded} and its guarantee in~\cite[Proposition~6]{Cheng26Chandelier}, and then give its greedy version executed in Step~4 of \texttt{TBAlign} with guarantees in Lemma~\ref{lem:greedy-seeded-completion}.
Recall that for a subset $\cs\subseteq \cv$ and a mapping $\hat\pi:\cs\rightarrow \td \cv$, we use $W(u,\td v)$ to denote the number of vertices $i\in \cv'$ such that $i$ is adjacent to $u$ in $G$ and $\hat\pi(i)$ is adjacent to $\td v$ in $\td G$.

\begin{algorithm}[H]
\caption{Seeded graph matching}
\begin{algorithmic}[1]\label{algo:seeded}
\REQUIRE Two graphs $G$ and $\td G$, a mapping $\pi':\cs\rightarrow \td\cv$, and parameter $\gamma$.
\ENSURE A mapping $\hat\pi$.
\STATE $\ci\gets \cs$ and $\hat{\pi}\gets \pi'$.
\WHILE{there exists $u\in \cv\setminus \ci$ and $\td v\in \td\cv\setminus\hat{\pi}(\ci)$ such that $W(u,\td v)\ge \frac{\gamma(n-2)\lambda^2}{n^2}$}
    \STATE $\ci\gets \ci\cup\{u\}$.
    \STATE $\hat{\pi}(u)\gets\td v$.
    \STATE Update $W(u,\td v)$ based on the new $\hat{\pi}$.
\ENDWHILE
\RETURN $\hat \pi$.
\end{algorithmic}
\end{algorithm}

The guarantee of Algorithm~\ref{algo:seeded} is provided by the following proposition. As established in \cite[Proposition~6]{Cheng26Chandelier} and its proof, this result holds simultaneously for all seed mappings and for every choice of the pair added in each iteration.

\begin{prop}[{\cite[Proposition~6]{Cheng26Chandelier}}]\label{prop:seedmatch}
    Fix an arbitrarily small constant $\eps>0$. Suppose $(G,\td G)\sim \mathrm{CER}(n,\lambda,s)$ with $\frac{\lambda}{n}\le \frac12$, $\lambda s\ge(1+\eps)\log n$, and $\frac{ns-\lambda}{n-\lambda}\ge\eps$. Let $\pi'$ be an arbitrary injective mapping: $\cs\rightarrow \td\cv$ such that $\pi'(u)=\pi(u)$ for every $u\in \cs$, and $|\cs|\ge (1-\eps/16)n$. Let $\gamma$ denote the unique solution in $(1,+\infty)$ to $\gamma\log \gamma-\gamma+1=\frac{3n^2\log n}{(n-2)\lambda^2}$. Then with probability $1-O(n^{-\eps/8})$, Algorithm~\ref{algo:seeded} with inputs $\gamma$ and any such $\pi'$ outputs $\hat\pi=\pi$ in $O(n\lambda^2)$ time.
\end{prop}

The following algorithm is a greedy version of Algorithm~\ref{algo:seeded} that sequentially selects the pair with the maximum weight. This algorithm serves as the actual subroutine for seeded completion in Step~4 of Algorithm~\ref{algo:TBAlign}.

\begin{algorithm}[H]
\caption{Greedy seeded graph matching}
\label{algo:greedy-seeded-completion}
\begin{algorithmic}[1]
\REQUIRE Two graphs $G$ and $\td G$, and a partial mapping $\pi'$.
\ENSURE A mapping $\hat\pi$.
\STATE Initialize $\hat\pi\gets\pi'$.
\STATE Compute $W(u,\td v)$ for all currently unmatched pairs.
\WHILE{there exist unmatched vertices in $\cv$}
    \STATE Choose a currently unmatched pair $(u^*,\td v^*)$ maximizing
    $W(u,\td v)$, breaking ties arbitrarily.
    \STATE Set $\hat\pi(u^*)\gets\td v^*$.
    \STATE Update $W(u,\td v)$ for all currently unmatched pairs.
\ENDWHILE
\RETURN $\hat\pi$.
\end{algorithmic}
\end{algorithm}

\begin{lem}
\label{lem:greedy-seeded-completion}
Under the same assumptions on the input graphs $(G,\td G)$ and the initial matching $\pi'$ as in Proposition~\ref{prop:seedmatch}, Algorithm~\ref{algo:greedy-seeded-completion} has time complexity $O\bigl(n^2+|\ce|\,|\td\ce|\bigr)$, and with probability $1-O(n^{-\eps/8})$, it outputs the correct matching $\hat\pi=\pi$ for any such initial matching $\pi'$.
\end{lem}

\begin{proof}
We prove correctness and the complexity bound separately.

\textbf{\underline{Proof of Correctness}.}
We consider any realization $(g,\td g)$ of $(G,\td G)\sim \text{CER}(n,\lambda,s)$ such that the event established in Proposition~\ref{prop:seedmatch} holds, and let $\gamma>1$ be the value defined in Proposition~\ref{prop:seedmatch}. Suppose, for contradiction, that Algorithm~\ref{algo:greedy-seeded-completion} selects an incorrect pair $(u^*,\td v^*)$ for the first time at some iteration. Before this iteration, the current $\hat\pi$ is a correct injective mapping containing at least as many matched pairs as $\pi'$. By Proposition~\ref{prop:seedmatch}, the incorrect pair $(u^*,\td v^*)$ must satisfy
\[
W(u^*,\td v^*) < \frac{\gamma(n-2)\lambda^2}{n^2};
\]
otherwise, Algorithm~\ref{algo:seeded} initialized with the current $\hat\pi$ could choose this pair and return the incorrect matching $(u^*,\td v^*)$, contradicting Proposition~\ref{prop:seedmatch}.

On the other hand, since $(u^*,\td v^*)$ maximizes $W(u,\td v)$ among all currently unmatched pairs, it follows that no unmatched pair reaches the threshold at this iteration. In this case, Algorithm~\ref{algo:seeded} initialized with the current $\hat\pi$ would therefore terminate before obtaining a complete matching, which is again a contradiction to Proposition~\ref{prop:seedmatch}. Thus, on the event considered above, Algorithm~\ref{algo:greedy-seeded-completion} never selects an incorrect pair. Since Algorithm~\ref{algo:greedy-seeded-completion} always returns a complete matching, the output $\hat\pi$ must equal to $\pi$ under the event we considered, which has probability $1-O(n^{-\eps/8})$.

\textbf{\underline{Proof of Complexity}.}
Starting from the input partial mapping $\pi'$, we first set $W(u,\td v)=0$ for every unmatched pair $(u,\td v)$. For every vertex $i$ already matched by $\pi'$, and for every currently unmatched pair $(u,\td v)$ such that $u\in\cn(i)$ and $\td v\in\td\cn(\pi'(i))$, set
\[
W(u,\td v)\gets W(u,\td v)+1.
\]
For each fixed vertex $i$ matched by $\pi'$, these updates have time complexity $O\left(|\cn(i)|\,|\td\cn(\pi'(i))|\right)$. 

Since every weight is an integer between $0$ and $n$, we place all unmatched pairs into $n+1$ buckets according to their weights. For each $k\in[0:n]$, define
\begin{align}
\cb_k:=\left\{(u,\td v):(u,\td v)\text{ is currently unmatched and }W(u,\td v)=k\right\}.
\end{align}
Placing all unmatched pairs into these buckets takes $O(n^2)$ time. In each iteration, we scan $\cb_n,\cb_{n-1},\ldots,\cb_0$ to find the largest index $k$ such that $\cb_k$ is nonempty, choose an arbitrary pair $(u^*,\td v^*)$ from $\cb_k$, and set $\hat\pi(u^*)\gets\td v^*$. We then remove all remaining pairs containing either $u^*$ or $\td v^*$. The time complexity of finding the largest nonempty bucket and removing conflicts is $O(n)$. After adding $(u^*,\td v^*)$, for every currently unmatched pair $(w,\td x)$ such that $w\in\cn(u^*)$ and $\td x\in\td\cn(\td v^*)$, update
\[
W(w,\td x)\gets W(w,\td x)+1,
\]
and move $(w,\td x)$ to the corresponding bucket. Therefore, updating the weights in this iteration takes $O\left(|\cn(u^*)|\,|\td\cn(\td v^*)|\right)$ time. Each vertex contributes to the weight updates exactly once, either during the initialization or when it is newly matched. Hence, the total number of weight updates is at most
\begin{align}
\sum_{u\in\cv}|\cn(u)|\,|\td\cn(\hat\pi(u))|\le\left(\sum_{u\in\cv}|\cn(u)|\right)\left(\sum_{\td v\in\td\cv}|\td\cn(\td v)|\right)=O\left(|\ce|\,|\td\ce|\right).
\end{align}
Since the number of iterations is $O(n)$, the total time complexity of completing $\pi'$ to $\hat\pi$ is
\[
O\left(n^2+|\ce|\,|\td\ce|\right),
\]
which completes the proof.
\end{proof}

\section{\erdos--\renyi Graph Properties}
The proofs in this appendix largely follow those in Appendices~C and~D of~\cite{Cheng26RBAlign}, where the results are established under the assumption that $\lambda=(\log n)^{\alpha+o(1)}$ for some $\alpha\in(0,1)$. Here we extend these results to the broader regime $\lambda=\omega(1)$ and $\lambda=(\log n)^{O(1)}$. The key arguments remain unchanged, and we present the proofs here for completeness.

\subsection{Neighborhood Size}
\label{appd:neighbor-size}
\begin{lem}
\label{lem:neighbor_size}
    Consider an \erdos--\renyi random graph $G(\cv,\ce)\sim\mathrm{ER}(n,\frac{\lambda}{n})$, where $\lambda=\omega(1)$ and $\lambda=(\log n)^{O(1)}$. Let $d$ be a positive integer with $d\le(\log n)^\gamma$ for some constant $\gamma\in (0,1)$, and for each vertex $i$ in $G$, let $\cn(i,d)$ denote the set of vertices within distance $d$ from $i$ in $G$. Then, we have
    \begin{equation}
        \label{eq:neighbor_size}
        \P\left(\exists i\in \cv:|\cn(i,d)|\ge 8e^2\lambda^d\log n\right)= \td{O}(n^{-1/3}).
    \end{equation}
\end{lem}

\begin{proof}
    Fix a vertex $i\in\cv$.
    We will show that
    \begin{equation*}
        \P\left(|\cn(i,d)|\ge 8e^2\lambda^d\log n\right)= \td{O}(n^{-4/3}).
    \end{equation*} 
    Then~\eqref{eq:neighbor_size} then follows by applying a union bound on the $n$ vertices in $\cv$.
    
    For each $k\in \{0,\ldots,d\}$, let $\cs(i,k)$ denote the set of vertices exactly  $k$-hop away from vertex $i$. It follows from this definition that 
    \[
    |\cn (i,d)|=\sum_{k=0}^d|\cs (i,k)|. 
    \]
    Define event 
    \[
    \cb_k:=\left\{|\cs (i,k)|<4(\log n)\lambda^k\prod_{h=0}^k(1+\lambda^{-h/2})\right\}.
    \]
    Notice that 
    \begin{align*}
        \sum_{k=0}^d4(\log n)\lambda^k\prod_{h=0}^k(1+\lambda^{-h/2})&\le 4(\log n)\prod_{h=0}^\infty(1+\lambda^{-h/2})\sum_{k=0}^d\lambda^k\\
        &\stackrel{(a)}{\le}4(\log n)e^2\frac{(\lambda^{d+1}-1)}{\lambda-1}\\
        &\le 8e^2\lambda^d\log n,
    \end{align*}
    where (a) follows because 
    \begin{align*}
        \prod_{h=0}^\infty(1+\lambda^{-h/2})\le \exp\left(\sum_{h=0}^\infty\lambda^{-h/2}\right)\le e^2
    \end{align*}
    for all large enough $n$.
    Therefore, it suffices to show that $\P(\cap_{k=0}^d\cb_k)= 1-\td{O}(n^{-4/3})$. We rewrite the probability of this intersection event as
    \begin{align}
       \P(\cap_{k=0}^d\cb_k)&=\P(\cb_0)-\sum_{k=1}^{d}\P(\cb_k^c\cap(\cap_{h=0}^{k-1}\cb_{h}))\nonumber\\
       &\ge \P(\cb_0)-\sum_{k=1}^{d}\P(\cb_k^c|\cap_{h=0}^{k-1}\cb_{h}).\label{eq:chain}
    \end{align}
    Notice that $\cb_0$ holds almost surely, and it suffices to bound each term in the summation in~\eqref{eq:chain}. Fix some $k\in\{0,\ldots,d\}$. Conditioned on events $\sum_{l=0}^{k-1}|\cs (i,l)|=m$ and $|\cs (i,k-1)|=m'$, we have
    \[
    |\cs (i,k)|\sim \bin(n-m,1-(1-\lambda/n)^{m'}).
    \]
    Notice that $1-(1-\lambda/n)^{m'}\le \lambda m'/n$. Then conditioned on event $\cap_{l=0}^{k-1}\cb_{l}$, we have
    \[
    |\cs (i,k)|\stackrel{\mathrm{sto.}}{\le}\bin\left(n,\frac{4(\log n)\lambda^{k}\prod_{h=0}^{k-1}(1+\lambda^{-h/2})}{n}\right),
    \]
    where $\sto$ is the stochastic dominance operator. By the Bennett's inequality stated in Lemma~\ref{lem:bennett}, we have
    \begin{align}
        &\P(\cb_k^c|\cap_{h=0}^{k-1}\cb_{h})\nonumber\\
        \le{}& \P\left(\bin\left(n,\frac{4(\log n)\lambda^{k}\prod_{h=0}^{k-1}(1+\lambda^{-h/2})}{n}\right)\ge 4(\log n)\lambda^{k}\prod_{h=0}^{k}(1+\lambda^{-h/2})\right)\nonumber\\
        \le{}& \exp\left(-4(\log n)\lambda^{k}\left(\prod_{h=0}^{k-1}(1+\lambda^{-h/2})\right)\left(1-\frac{4(\log n)\lambda^{k}\prod_{h=0}^{k-1}(1+\lambda^{-h/2})}{n}\right) \phi(\lambda^{-k/2})\right),\label{eq:size_bennett}
    \end{align}
    where $\phi(x):=(1+x)\log(1+x)-x$.
    We now bound each term in this exponent. 
    Notice that $\phi(x)\ge \frac{x^2}{3}$ for all $x\in [0,1]$, and it follows that $$\phi(\lambda^{-k/2})\ge \frac{\lambda^{-k}}{3}.$$ 
    We also have 
    $$\prod_{h=0}^{k-1}(1+\lambda^{-h/2})\ge 1+\lambda^{0}=2.$$
    Under the assumptions $\lambda=(\log n)^{O(1)}$ and $d\le(\log n)^\gamma$, we have $\lambda^d=n^{o(1)}$. Therefore, 
    \begin{align*}
        1-\frac{4(\log n)\lambda^{k}\prod_{h=0}^{k-1}(1+\lambda^{-h/2})}{n}\ge 1-\frac{4e^2(\log n)\lambda^k}{n}\ge\frac12
    \end{align*}
    for all large enough $n$. Substituting these three inequalities into~\eqref{eq:size_bennett} yields
    \begin{align*}
        P(\cb_k^c|\cap_{h=0}^{k-1}\cb_{h})&\le\exp\left(-4(\log n)\lambda^k\times 2\times\frac12\times \frac{\lambda^{-k}}{3}\right)\\
        &\le \exp\left(-\frac43\log n\right)=n^{-4/3}.
    \end{align*}

    Notice that the above inequality holds for every $k\in [d]$. Therefore, we have 
    \[
    \P(\cap_{k=0}^d\cb_k)\ge 1-dn^{-4/3}=1-\td{O}(n^{-4/3}),
    \]
    which completes the proof.
\end{proof}

\subsection{Comparison with Independent Tree Pairs}
\label{appd:tree-pmf}
For a pair of vertices $(u,w)\in \cv\times\cv$, $\ct$ denotes the local tree rooted at $w$ in $G\setminus u$ with depth up to $d$. For a pair of vertices $(\td v,\td x)\in \td\cv\times\td\cv$, $\td\ct$ denotes the local tree rooted at $\td x$ in $\td G\setminus \td v$ with depth up to $d$.
\begin{lem}
    \label{lem:tree_pmf}
    Suppose $(G,\td G)\sim \mathrm{CER}(n,\lambda,s)$.
    For any pair of unlabeled rooted tree $(t,\td t)\in \cz\times\cz$, both with depth within $d$, we have
    \begin{align}
        &\frac{\P\big(\{[\ct]=t\}\cap\{[{\td\ct}]=\td t\}\cap \{\bar\cv(\ct)\cap \bar\cv(\td\ct)=\emptyset\}\big)}{\bq([\ch_d]=t,[\td\ch_d]=\td t)}\nonumber\\
        &\le \exp\left(\frac{\lambda}{n}\left((|\cv(t)|+1)|\cv(t)|+|\cv(\td t)|(1+|\cv(t)|+|\cv(\td t)|)\right)\right)\label{eq:tree_pmf}
    \end{align}
    
\end{lem}

\begin{proof}
    To begin with, we define a mapping $\ell$ that relabels a local tree in an \erdos--\renyi graph into a Galton--Watson tree that is labeled by the children orders. Consider a non-empty local tree $\ct$. The image $\ell(\ct)$ has the same underlying rooted tree structure as $\ct$, but its vertex labels are replaced by order labels as follows. First, the root of $\ell(\ct)$ is assigned the label $(1)$. For each vertex $v\in\cv(\ct)$, we order the children of $v$ according to their indices in $\ct$: if the children of $v$ have indices
    \[
    i_1<i_2<\cdots<i_{c_v},
    \]
    then the child with index $i_j$ is assigned order $j$ among the children of $v$. Given these sibling orders, the label of each vertex in $\ell(\ct)$ is obtained by concatenating the label of its parent and its order.

    With the definition of $\ell$, we claim that for each $(h,\td h)\in \cx\times\cx$, both with depth up to $d$,
    \begin{align}
        &\frac{\P\big(\{\ell(\ct)=h\}\cap\{\ell(\td\ct)=\td h\}\cap \{\bar\cv(\ct)\cap \bar\cv(\td\ct)=\emptyset\}\big)}{\bq(\ch_d=h,\td\ch_d=\td h)}\nonumber\\
        \le{}& \exp\left(\frac{\lambda}{n}\left((|\cv(h)|+1)|\cv(h)|+|\cv(\td h)|(1+|\cv(h)|+|\cv(\td h)|)\right)\right).\label{eq:cxd}
    \end{align}

    Since the mapping $\ell$ preserves the tree structure, we know that $[\ct]=t$ only if $\ell(\ct)=h$ for some $h\in\cx$ that satisfies $[h]=t$. Therefore, we have for each $(t,\td t)\in\cz\times\cz$,
    \begin{align*}
        &\frac{\P\big(\{[{\ct}]=t\}\cap\{[{\td\ct}]=\td t\}\cap \{\bar\cv(\ct)\cap \bar\cv(\td\ct)=\emptyset\}\big)}{\bq([\ch_d]=t,[\td\ch_d]=\td t)}\\
        ={}& \frac{\sum_{\substack{(h,\td h)\in \cx\times\cx:\\
        [h]=t,[\td h]=\td t
        }}\P\big(\{\ell(\ct)=h\}\cap\{\ell(\td\ct)=\td h\}\cap \{\bar\cv(\ct)\cap \bar\cv(\td\ct)=\emptyset\}\big)}{\sum_{\substack{(h,\td h)\in \cx\times\cx:\\
        [h]=t,[\td h]=\td t
        }}\bq(\ch_d=h,\td\ch_d=\td h)}\\
        \le{}& \exp\left(\frac{\lambda}{n}\left((|\cv(t)|+1)|\cv(t)|+|\cv(\td t)|(1+|\cv(t)|+|\cv(\td t)|)\right)\right),
    \end{align*}
    where the last inequality follows by~\eqref{eq:cxd} and the fact that $|\cv(t)|=|\cv(h)|$ given that $[h]=t$.
    This completes the proof of~\eqref{eq:tree_pmf}. In the rest of the proof, we focus on proving the claimed inequality~\eqref{eq:cxd}.

        By the definition of the independent tree distribution $\bq(\ch,\td \ch)$, we have
        \[
        \bq(\ch_d=h,\td\ch_d=\td h)=\gw(\ch_d=h)\gw(\td\ch_d=\td h).
        \]
        
        In the following, we will focus on proving
        \begin{equation}
            \label{eq:t}
            \frac{\P(\ell(\ct)=h)}{\gw(\ch_d=h)}\le \exp\left(\frac{\lambda}{n}(|\cv(h)|+1)|\cv(h)|\right),
        \end{equation}
        and 
        \begin{equation}
            \label{eq:tdt}
            \frac{\P(\{\ell(\td\ct)=\td h\}\cap \{\bar\cv(\ct)\cap \bar\cv(\td\ct)=\emptyset\}\,|\,\ell(\ct)=h)}{\gw(\td\ch_d=\td h)}\le \exp\left(\frac{\lambda}{n}|\cv(\td h)|(1+|\cv(h)|+|\cv(\td h)|)\right).
        \end{equation}
        The desired inequality~\eqref{eq:cxd} then follows by multiplying~\eqref{eq:t} and~\eqref{eq:tdt}.

        \underline{\textbf{Proof of~\eqref{eq:t}:}} 
        For each $k\in\{0,\ldots,d-1\}$, we define a total ordering for the vertices at depth $k$ in the order-labeled rooted tree $h$: a vertex is ranked higher if its parent is ranked higher among vertices at depth $k-1$; among vertices sharing the same parent, the order is determined by the ordering of siblings. Recall that $\cv(h,k)$ denotes the set of vertices at depth $k$ in graph $h$. For each $k$ and $r\in [|\cv(h,k)|]$, let $c_{k,r}$ denote the number of children of the $r$-th ordered vertex at depth $k$. Then by the definition of the $\gw$ distribution, we have

        \begin{equation}
            \label{eq:gw_t}
            \gw(\ch_d=h)=\prod_{k=0}^{d-1}\prod_{r=1}^{|\cv(h,k)|}\P(\poi(\lambda)=c_{k,r}),
        \end{equation}
        where we define the product $\prod_{r=1}^{|\cv(h,k)|}\P(\poi(\lambda)=c_{k,r})$ to be $1$ when $|\cv(h,k)|=0$.

        To bound the numerator $\P(\ell(\ct)=h)$, we describe a breadth-first exploration procedure for constructing $\ct$. The exploration is performed in the graph $G\setminus u$ and is rooted at $w$. We reveal the vertices of $\ct$ level by level, following the ordering of vertices in the target tree $h$ introduced above.  More precisely, among vertices at the same depth, a vertex is explored earlier if its parent is explored earlier, and among siblings, the order is determined by their vertex labels in $G$. 

        At the beginning of the exploration, the root $w$ is the only discovered vertex. Since the vertex $u$ is removed, the number of neighbors of $w$ in $G\setminus u$ has distribution
        \[
        \bin\left(n-2,\frac{\lambda}{n}\right).
        \]
        To obtain a local tree isomorphic to $h$, this number must be equal to $c_{0,1}$.

        More generally, suppose that the exploration is about to reveal the children of the $r$-th ordered vertex at depth $k$ in the breadth-first order, where $k\in\{0,\ldots,d-1\}$ and $r\in[|\cv(h,k)|]$.
        These children must satisfy:
        \begin{itemize}
            \item The total number of these children is exactly $c_{k,r}$. We also note that the number of these children is distributed as 
            \[
            \bin\left(n-1-x_{k,r},\frac{\lambda}{n}\right),
            \]
            where $x_{k,r}$ denotes the number of vertices that have already been discovered before this step in the breadth-first exploration.
            \item In addition to matching the prescribed offspring numbers, the exploration must not create any cycle. Equivalently, every newly discovered vertex must be connected to the previously explored part only through its parent, and there must be no edges among vertices discovered at the same step. If any newly discovered vertex has an additional edge to an already discovered vertex, or if there is an edge among two newly discovered vertices, then the explored $d$-neighborhood is not a tree and $\ct$ becomes empty.
        \end{itemize}
        
        Recall that in Definition~\ref{de:local_tree_new}, $\ct$ is non-empty only when the $d$-depth neighborhood of $w$ in $G$ is a tree. Here, we loosen this constraint, and only require the $d$-depth neighborhood of $w$ in $G\setminus u$ is a tree. This results in an upper bound on the probability of $\ell(\ct)=h$.
        Through this breadth-first exploration, we have
        \begin{align}
            &\P(\ell(\ct)=h)\nonumber\\
            \le{}&\prod_{k=0}^{d-1}\prod_{r=1}^{|\cv(h,k)|}\P(\text{the number of children is $c_{k,r}$ and the new children do not create any cycle})\nonumber\\
            \le{}& \prod_{k=0}^{d-1}\prod_{r=1}^{|\cv(h,k)|}\P\left(\bin\left(n-1-x_{k,r},\frac{\lambda}{n}\right)=c_{k,r}\right)\nonumber\\
            \stackrel{(a)}{\le}{}&\prod_{k=0}^{d-1}\prod_{r=1}^{|\cv(h,k)|}\exp\left(\frac{\lambda }{n}(1+x_{k,r}+c_{k,r})\right)\P(\poi(\lambda)=c_{k,r})\nonumber\\
            \stackrel{(b)}{\le}{}&\exp\left(\frac{\lambda}{n}|\cv(h)|(1+|\cv(h)|)\right)\prod_{k=0}^{d-1}\prod_{r=1}^{|\cv(h,k)|}\P(\poi(\lambda)=c_{k,r}).\label{eq:t_bfs}
        \end{align}
        Here (a) follows by Lemma~\ref{lem:binom-poi-gap}, and (b) follows because $c_{k,r}+x_{k,r}\le |\cv(h)|$ throughout the exploration process and there are at most $|\cv(h)|$ terms in the double product. This completes the proof of~\eqref{eq:t}.

        \underline{\textbf{Proof of~\eqref{eq:tdt}:}} For the unlabeled rooted tree $\td h$, we similarly define the notations $\td c_{k,r}$ and $\cv(\td h,k)$. Then we have
        \begin{equation}
            \label{eq:gw_tdt}
            \gw(\td\ch_d=\td h)=\prod_{k=0}^{d-1}\prod_{r=1}^{|\cv(\td h,k)|}\P(\poi(\lambda)=\td c_{k,r}).
        \end{equation}

        Now we move on to bound the numerator. Notice that
        \begin{align}
            &\P(\{\ell(\td\ct)=\td h\}\cap \{\bar\cv(\ct)\cap \bar\cv(\td\ct)=\emptyset\}\,|\,\ell(\ct)=h)\nonumber\\
            \le{}& \max_{\substack{\bar\cs\subseteq \bar\cv:\\
            \bar w\in \bar\cs,\\
            \bar x\notin\bar\cs,\\
            |\bar\cs|=|\cv(h)|}}\P(\{\ell(\td\ct)=\td h\}\cap \{\bar\cv(\ct)\cap \bar\cv(\td\ct)=\emptyset\}\,|\,\{\ell(\ct)=h\}\cap \{\bar{\cv}(\ct)=\bar\cs\})\label{eq:max_cond}
        \end{align}
        Fix an arbitrary subset $\bar\cs\subseteq \bar\cv$ that satisfies the criteria in~\eqref{eq:max_cond}. We focus on bounding the conditional probability 
        \[
        \P(\{\ell(\td\ct)=\td h\}\cap \{\bar\cv(\ct)\cap \bar\cv(\td\ct)=\emptyset\}\,|\,\{\ell(\ct)=h\}\cap \{\bar{\cv}(\ct)=\bar\cs\}).
        \]

        We again consider the breadth-first exploration procedure to construct $\td\ct$. To guarantee the desired event $\{\ell(\td\ct)=\td h\}\cap \{\bar\cv(\ct)\cap \bar\cv(\td\ct)=\emptyset\}$, when we explore the children of the $r$-th vertex at depth $k$, we have to ensure
        \begin{itemize}
            \item None of the children are corresponding to the vertices in $\bar\cs$. Otherwise, $\td\ct$ has a non-empty overlap with the tree $\ct$;
            \item The number of children is exactly $\td c_{k,r}$. Given the event that the vertex has no children corresponding to $\bar\cs$, the distribution of the number of children at this step is given by 
            \[
            \bin\left(n-1-|\cv(h)|-\td x_{k,r},\frac{\lambda}{n}\right),
            \]
            where the excluded set is $\bar\cs$, whose size is $|\cv(h)|$ by~\eqref{eq:max_cond}, and where $\td x_{k,r}$ is the number of vertices discovered before this step. This is because conditioning on the event $\{\ell(\ct)=h\}\cap \{\bar{\cv}(\ct)=\bar\cs\}$ does not reveal any information about any edge without an endpoint in $\bar{\cs}$.
            \item The newly discovered children are connected to the previously explored part only through its parent, and there are no edges among them.
        \end{itemize}
        Recall that in Definition~\ref{de:local_tree_new}, $\td\ct$ is non-empty only when the $d$-depth neighborhood of $\td x$ in $\td G$ is a tree. Here, we loosen this constraint, and only require the $d$-depth neighborhood of $\td x$ in $\td G\setminus \td v$ is a tree. This results in an upper bound on the probability of $\ell(\td\ct)=\td h$.
        By this exploration process, we have
        \begin{align*}
            &\P(\{\ell(\td\ct)=\td h\}\cap \{\bar\cv(\ct)\cap \bar\cv(\td\ct)=\emptyset\}\,|\,\{\ell(\ct)=h\}\cap \{\bar{\cv}(\ct)=\bar\cs\})\\
            \le{}&\prod_{k=0}^{d-1}\prod_{r=1}^{|\cv(\td h,k)|}\P(\text{no children corresponding to any vertex in }\bar\cs\text{ and}\\
            &\quad\quad\quad\quad\quad\quad\text{the number of new children is $\td c_{k,r}$ and}\\
            &\quad\quad\quad\quad\quad\quad\text{the new children do not create any cycle})\\
            \le{}& \prod_{k=0}^{d-1}\prod_{r=1}^{|\cv(\td h,k)|} \P\left(\bin\left(n-1-|\cv(h)|-\td x_{k,r},\frac{\lambda}{n}\right)=\td c_{k,r}\right)\\
            \stackrel{(c)}{\le}{}&\prod_{k=0}^{d-1}\prod_{r=1}^{|\cv(\td h,k)|}\exp\left(\frac{\lambda}{n}(1+\td x_{k,r}+\td c_{k,r}+|\cv(h)|)\right)\P(\poi(\lambda)=\td c_{k,r})\\
            \stackrel{(d)}{\le}{}&\exp\left(\frac{\lambda}{n}|\cv(\td h)|(1+|\cv(h)|+|\cv(\td h)|)\right)\prod_{k=0}^{d-1}\prod_{r=1}^{|\cv(\td h,k)|}\P(\poi(\lambda)=\td c_{k,r}).
        \end{align*}
        Here (c) and (d) follows by the similar reason as (a) and (b) in~\eqref{eq:t_bfs}. This completes the proof of~\eqref{eq:tdt}.
    \end{proof}

\subsection{Coupling with Correlated Tree Pairs}

    Let $d\ge 0$ be an integer, and $u$ and $\td u$ be two corresponding vertices in $G$ and $\td G$. Define $\ct$ to be the local tree rooted at $u$ in $G$ such that $\ct$ is the $d$-hop-neighborhood of $u$ in $G$ if it is a tree; otherwise, $\ct$ is an empty tree. Similarly, define $\td\ct$ as the local tree rooted at $\td u$ in $\td G$.
\begin{lem}
\label{lem:tree_tv_new}
    Suppose $(G,\td G)\sim \mathrm{CER}(n,\lambda,s)$, where $\lambda=\omega(1)$, $\lambda = (\log n)^{O(1)}$, and $s\in (\sqrt{C_\mathrm{Otter}},1]$ is a constant. Assume $d\le(\log n)^{\gamma}$ for some constant $\gamma\in (0,1)$.
 Then there exists a coupling  between $(\ct,\td \ct)$ and $(\ch,\td\ch)\sim \bp$ such that
    \begin{equation}
        \label{eq:dtv_new}
        \P([\ct]=[\ch_d],[\td\ct]=[\td\ch_d])= 1-O(n^{-1/2}).
    \end{equation}
    
\end{lem}

\begin{proof}
    We first define a mapping $\ell$ that relabels a pair of local trees to the format of a pair of Galton--Watson trees. We first describe the relabeling of $\ct$. If $\ct$ is an empty tree, then the output $\ell(\ct)$ is also empty. Now suppose $\ct$ is non-empty. The root of $\ell(\ct)$ is assigned the label $(1)$. For a vertex $v\in\cv(\ct)$, if the corresponding vertex $\td v$ also belongs to $\cv(\td\ct)$, then the children of $v$ are ordered in two blocks. First, we list the common children, namely those children $y$ of $v$ such that the corresponding vertex $\td y$ is a child of $\td v$ in $\td\ct$. These common children are ordered increasingly according to their indices. If there are $c^*_v$ such children, they receive orders $1,\ldots,c^*_v$. Next, we list the remaining children of $v$. These children are again ordered increasingly according to their indices and receive orders $c^*_v+1,\ldots,c^*_v+c_v$, where $c_v$ is the number of such remaining children.
    If, on the other hand, the corresponding vertex $\td v$ does not belong to $\cv(\td\ct)$, then all children of $v$ are ordered increasingly according to their indices and receive orders $1,\ldots,c_v$, where $c_v$ is the number of children of $v$. After this ordering of the children has been specified for every vertex in $\ell(\ct)$, the order labels are assigned recursively: if a vertex has label $(a_0,\ldots,a_k)$ and its $i$-th ordered child is considered, then this child is assigned the label $(a_0,\ldots,a_k,i)$. The relabeling of $\td\ct$ is defined analogously.

    In the rest of this proof, we will focus on constructing a coupling such that
    \begin{equation}
        \label{eq:couple_new}
        \P(\ell(\ct,\td \ct)=(\ch_d,\td\ch_d))=1-O(n^{-1/2}).
    \end{equation}
    This implies the lemma statement~\eqref{eq:dtv_new} because $\ell(\ct,\td\ct)=(\ch_d,\td\ch_d)$ implies $[\ct]=[\ch_d]$ and $[\td\ct]=[\td\ch_d]$. Therefore, we are left to construct a coupling so that~\eqref{eq:couple_new} holds.

    The coupling is constructed recursively over depths. For each $k\in [d]$, let $\ct_k$ denote the rooted tree obtained by truncating $\ct$ at depth $k$. More specifically, $\ct_k$ is the subgraph of $\ct$ induced by the set of vertices within depth $k$ from the root.
    We note that we have $\ct=\ct_d$ under this definition.
    The truncated trees $\td\ct_k$, $\ch_k$ and $\td\ch_k$ are defined similarly.  By these definitions, we have
    \[
    \{\ell(\ct,\td\ct)=(\ch_d,\td\ch_d)\}=\cap_{k=1}^d \{\ell(\ct_k,\td\ct_k)=(\ch_k,\td\ch_k)\}.
    \]

    \textbf{\underline{Step 1. Coupling $(\ct_1,\td\ct_1)$ with $(\ch_1,\td\ch_1)$:}} Recall that $u$ and $\td u$ are the roots of $\ct$ and $\td\ct$, respectively. Under the mapping $\ell$, both roots are relabeled by $(1)$, which agrees with the root labels of $\ch$ and $\td\ch$. 
    Let $\sfch_u^*$ denote the set of common children of $u$ and $\td u$ in $\ct$ and $\td\ct$. More precisely, $\sfch_u^*$ 
    consists of all indices $i\in[n]$ such that $i$ is a child of $u$ in $\ct$ and $\td i$ is a child of $\td u$ in $\td\ct$. Let $\sfch_u$ denote the set of children of $u$ in $\ct$ whose corresponding vertices are not children of $\td u$ in $\td\ct$, and let $\widetilde\sfch_{\td u}$ denote the analogous set of children of $\td u$ in $\td\ct$ that do not correspond to children of $u$ in $\ct$. We refer to these as the exclusive children of $u$ and $\td u$, respectively.
    We define the following three events:
    \begin{itemize}
        \item $\ca_1^*$ is the event that under the maximal coupling~\cite{levin2017markov} between random variables $|\sfch_u^*|$, $|\sfch_u|$, and  $|\wdtd\sfch_{\td u}|$, and independent Poisson random varables, $X_u^*\sim\poi(\lambda s)$, $X_u\sim\poi(\lambda(1-s))$, and $\td X_{\td u}\sim\poi(\lambda(1-s))$, we have $|\sfch_u^*|=X_u^*$, $|\sfch_u|=X_u$ and $|\wdtd\sfch_{\td u}|=\td X_{\td u}$.
        
        \item $\cb_1$ is the event that $|\sfch_u^*\cup\sfch_u\cup\wdtd\sfch_{\td u}|\le \lambda(2-s)\log n$.
        \item $\cc_1$ is the event that there are no edges in the union graph $\bar G$ between any pair of vertices in $\sfch_u^*\cup\sfch_u\cup\wdtd\sfch_{\td u}$.

    \end{itemize}
    We claim that $\ca_1^*\cap \cc_1$ implies $\ell(\ct_1,\td\ct_1)=(\ch_1,\td\ch_1)$. First, under event $\cc_1$, no cycles exist in the $1$-hop-neighborhood of $u$ and $\td u$. Therefore, both $\ct_1$ and $\td \ct_1$ are non-empty.
    Under the correlated tree-pair distribution $\bp$, the number of children of the root in the intersection tree has distribution $\poi(\lambda s)$, while the numbers of exclusive children added to the two trees have distribution $\poi(\lambda(1-s))$ independently. Hence, on $\ca_1^*$, the numbers of common and exclusive root children in $(\ct_1,\td\ct_1)$ agree with those in $(\ch_1,\td\ch_1)$. Moreover, the definition of $\ell$ orders common children first and exclusive children afterwards, with each block ordered increasingly by index. 
    Therefore, after applying $\ell$, the labels of all vertices up to depth one agree with those in $(\ch_1,\td\ch_1)$.
    
    The event $\cb_1$ is not needed for this depth-one agreement. It is used to control the size of the explored neighborhood when bounding $\P(\cc_1^c)$ and the other error events at deeper depths.
    
    We claim that
    \begin{equation}
    \label{eq:root_multi_new}
    (|\sfch_u^*|,|\sfch_u|,|\wdtd\sfch_{\td u}|,|\mathsf{NC}_u|)\sim \mathrm{Multi}\left(n-1,\frac{\lambda s}{n},\frac{\lambda (1-s)}{n},\frac{\lambda (1-s)}{n},1-\frac{\lambda(2-s)}{n}\right),
    \end{equation}
    where $|\mathsf{NC}_u|$ is the number of vertices that are not connected to $u$ in $G$ and not connected to $\td u$ in $\td G$. This follows by the definition of the correlated \erdos--\renyi graph pair model, in which an edge appears in both $G$ and $\td G$ with probability $\frac{\lambda s}{n}$, only in $G$ with probability $\frac{\lambda(1-s)}{n}$ and only in $\td G$ with probability $\frac{\lambda(1-s)}{n}$. Also, these edges are mutually independent.
    
    By Lemma~\ref{lem:multi-poi-new}, we can couple these random variables with $X_u^*$, $X_u$ and $\td X_{\td u}$ such that
    \begin{equation}
        \label{eq:a1_new}
        \P(\ca_1^{*c})=O\left(n^{-3/4}\right).
    \end{equation}
    By the Chernoff bound (equation~\eqref{eq:CB_Upper} in Lemma~\ref{lem:CB}), we have
    \begin{equation}
        \label{eq:b1_new}
        \P(\cb_1^c)\le \P\left(\bin\left(n-1,\frac{\lambda(2-s)}{n}\right)\ge \lambda(2-s)\log n\right)= n^{-\omega(1)}.
    \end{equation}
    By the union bound over all vertex pairs in $\sfch_u^*\cup\sfch_u\cup\wdtd\sfch_{\td u}$, we have 
    \begin{equation}
        \label{eq:c1_new}
        \P(\cc_1^c\,|\,\cb_1)=O\left(\frac{\lambda^3\log^2 n}{n}\right).
    \end{equation}
    Equations~\eqref{eq:a1_new},~\eqref{eq:b1_new} and~\eqref{eq:c1_new} together imply that 
    \begin{equation}
        \label{eq:depth1_new}
        \P(\ca^{*c}_1\cup\cb_1^c\cup\cc_1^c)=O\left(\frac{n^{1/4}+\lambda^3\log^2 n}{n}\right)=O(n^{-3/4}).
    \end{equation}

    \textbf{\underline{Step 2. Coupling $(\ct_k,\td\ct_k)$ with $(\ch_k,\td\ch_k)$ given $\ell(\ct_{k-1},\td\ct_{k-1})=(\ch_{k-1},\td\ch_{k-1})$:}} 
    For each depth $k\ge 2$, we recursively define three sets of vertices: $\com_k$, $\excl_k$, and $\wdtd{\excl}_k$. The set $\com_k$ consists of the vertices at depth $k$ that appear in both local trees under the correspondence, while $\excl_k$ and $\wdtd{\excl}_k$ consist of the vertices appearing exclusively in $\ct$ and $\td\ct$, respectively. At depth one, we initialize
    \[
        \com_1:=\sfch_u^*,\qquad
        \excl_1:=\sfch_u,\qquad
        \wdtd{\excl}_1:=\wdtd{\sfch}_{\td u}.
    \]

    Now we move on to consider a general $k\ge 2$, and recursively define the sets $\com_k$, $\excl_k$, and $\wdtd{\excl}_k$.
    For each $v\in\com_{k-1}$, let $\sfch_v^*$ denote the set of common children of $v$ and $\td v$ in $\ct$ and $\td\ct$; that is, $\sfch_v^*$ contains all children $y$ of $v$ in $\ct$ such that $\td y$ is a child of $\td v$ in $\td\ct$. Let $\sfch_v$ denote the set of children of $v$ in $\ct$ whose counterparts are not children of $\td v$ in $\td\ct$, and let $\wdtd{\sfch}_{\td v}$ denote the set of children of $\td v$ in $\td\ct$ whose counterparts are not children of $v$ in $\ct$. For each $v\in\excl_{k-1}$, let $\sfch_v$ denote the set of children of $v$ in $\ct$. Similarly, for each $\td v\in\wdtd{\excl}_{k-1}$, let $\wdtd{\sfch}_{\td v}$ denote the set of children of $\td v$ in $\td\ct$.

    We then define
    \[
    \com_{k}
    :=
    \bigcup_{v\in\com_{k-1}}\sfch_v^*,
    \]
    \[
    \excl_{k}
    :=
    \left(\bigcup_{v\in\com_{k-1}}\sfch_v\right)
    \cup
    \left(\bigcup_{v\in\excl_{k-1}}\sfch_v\right),
    \]
    and
    \[
    \wdtd{\excl}_{k}
    :=
    \left(\bigcup_{v\in\com_{k-1}}\wdtd{\sfch}_{\td v}\right)
    \cup
    \left(\bigcup_{\td v\in\wdtd{\excl}_{k-1}}\wdtd{\sfch}_{\td v}\right).
    \]

    For each depth $k\ge 2$, we define the following events that can be shown to imply the desired property $\ell(\ct_k,\td\ct_k)=(\ch_k,\td\ch_k)$.
    \begin{itemize}
        \item $\ca_k^*$ is the event that for each $v\in \com_{k-1}$, 
        under the maximal coupling between random variables $|\sfch_v^*|$, $|\sfch_v|$, and  $|\wdtd\sfch_{\td v}|$, and independent Poisson random varables, $X_v^*\sim\poi(\lambda s)$, $X_v\sim\poi(\lambda(1-s))$, and $\td X_{\td v}\sim\poi(\lambda(1-s))$, we have $|\sfch_v^*|=X_v^*$, $|\sfch_v|=X_v$ and $|\wdtd\sfch_{\td v}|=\td X_{\td v}$.

        \item $\ca_k$ is the event that for each $v\in \excl_{k-1}$, 
        under the maximal coupling between the random variable $|\sfch_v|$ and the Poisson random variable $X_v\sim\poi(\lambda)$, we have $|\sfch_v|=X_v$.

        \item $\td\ca_k$ is the event that for each $\td v\in \wdtd \excl_{k-1}$,  under the maximal coupling between the random variable $|\wdtd\sfch_{\td v}|$ and the Poisson random variable $\td X_{\td v}\sim\poi(\lambda)$, we have $|\wdtd\sfch_{\td v}|=\td X_{\td v}$.

        \item $\cb_k$ is the event that $|\com_k\cup\excl_k\cup\wdtd\excl_{k}|\le 2^{k-1}(\lambda(2-s))^k\log n$.
        
        \item $\cc_k$ is the event that there are no edges in the union graph $\bar G$ between any pair of vertices in $\com_k\cup\excl_k\cup\wdtd\excl_k$.

        \item $\cd_k$ is the event that in the union graph $\bar G$, there exists no vertex outside the two trees $\ct_{k-1}$ and $\td\ct_{k-1}$ that is adjacent to more than one vertex in $\com_{k-1}\cup\excl_{k-1}\cup\wdtd\excl_{k-1}$.

        \item $\ci_k$ is the event that there are no edges in $G$ that has an endpoint in $\cup_{i=1}^{k}(\excl_i\cup\com_i)$ and the other endpoint corresponding to any vertex in $\cup_{i=1}^{k}\wdtd\excl_i$.

        \item $\td\ci_k$ is the event that there are no edges in $\td G$ that has an endpoint in $\cup_{i=1}^{k}(\wdtd\excl_i\cup\com_i)$ and the other endpoint corresponding to any vertex in $\cup_{i=1}^{k}\excl_i$.
    \end{itemize}

    We claim that, for every $k\ge 1$,
    \[
        (\ca^*_1\cap\cb_1\cap\cc_1)
        \cap
        \bigcap_{i=2}^k
        (\ca_i^*\cap\ca_i\cap\wdtd{\ca}_i
        \cap \cb_i\cap\cc_i\cap\cd_i\cap\ci_i\cap\td\ci_i)
        \subseteq
        \{\ell(\ct_k,\td\ct_k)=(\ch_k,\td\ch_k)\}.
    \]
    We prove this claim by induction on $k$. The case $k=1$ has already been
    established in Step 1.

    Now suppose that the claim holds up to depth $k-1$, where $k\ge 2$. That is, assume
    \[
        (\ca^*_1\cap\cb_1\cap\cc_1)
        \cap
        \bigcap_{i=2}^{k-1}
        (\ca_i^*\cap\ca_i\cap\wdtd{\ca}_i
        \cap \cb_i\cap\cc_i\cap\cd_i\cap\ci_i\cap\td\ci_i)
        \subseteq
        \{\ell(\ct_{k-1},\td\ct_{k-1})=(\ch_{k-1},\td\ch_{k-1})\}.
    \]
    We show that, on the additional event
    \[
        \ca_k^*\cap\ca_k\cap\wdtd{\ca}_k
        \cap\cb_k\cap\cc_k\cap\cd_k\cap\ci_k\cap\td\ci_k,
    \]
    the equality extends to depth $k$.

    By the induction hypothesis, the order-labeled structures of $(\ct,\td\ct)$ and $(\ch,\td\ch)$ agree up to depth $k-1$. It remains to compare the children generated by vertices at depth $k-1$. For each common vertex $v\in\com_{k-1}$, the event $\ca_k^*$ couples the children numbers $|\sfch_v^*|$, $|\sfch_v|$ and $|\wdtd{\sfch}_{\td v}|$
    with independent Poisson random variables of means $\lambda s$, $\lambda(1-s)$, and $\lambda(1-s)$, respectively. These are exactly the offspring distributions used in the correlated tree-pair construction $\bp$ for vertices in the intersection tree and for the exclusive children added to the two trees. Similarly, for each exclusive vertex $v\in\excl_{k-1}$, the event $\ca_k$ couples $|\sfch_v|$ with a $\poi(\lambda)$ random variable, which is the offspring distribution of an exclusive Galton--Watson subtree attached to $\ch$. The event $\wdtd{\ca}_k$ gives the analogous coupling for every exclusive vertex $\td v\in\wdtd{\excl}_{k-1}$ in $\td\ct$. Therefore, the numbers of children generated from all vertices at depth $k-1$ agree with those in the coupled Galton--Watson tree pair. Moreover, by the definition of the relabeling map $\ell$, common children are ordered first, followed by exclusive children, and each block is ordered increasingly by index. This is precisely the deterministic ordering convention used in the construction of $(\ch,\td\ch)$. Hence, after applying $\ell$, the labels of all newly generated vertices at depth $k$ agree with those in $(\ch_k,\td\ch_k)$.

    Moreover, event $\cc_k$ rules out edges among vertices in $\com_k\cup\excl_k\cup\wdtd{\excl}_k,$ and event $\cd_k$ rules out any vertex outside the previously explored trees that is connected to more than one vertex in $\com_{k-1}\cup\excl_{k-1}\cup\wdtd{\excl}_{k-1}.$ Together, these events ensure that the newly revealed vertices do not create any cycles. Therefore, we have $\ell(\ct_k,\td\ct_k)=(\ch_k,\td\ch_k),$ which completes the induction.

    We note that the events $\cb_k$, $\ci_k$ and $\td\ci_k$ are not needed for the immediate implication $\ell(\ct_k,\td\ct_k)=(\ch_k,\td\ch_k)$. The event $\cb_k$ provides a size bound on the newly explored vertices, which is used to control the error events. Events $\ci_k$ and $\td\ci_k$ prevent overlaps of the exclusive part of the trees at later depths, thus ruling out undesired correlations that could affect the coupling.

    Next, we separately bound the probability of these events at depth $k$ conditioned on the event
    \[
        \cg_{k-1}:=(\ca^*_1\cap\cb_1\cap\cc_1)
        \cap
        \bigcap_{i=2}^{k-1}
        (\ca_i^*\cap\ca_i\cap\wdtd{\ca}_i
        \cap \cb_i\cap\cc_i\cap\cd_i\cap\ci_i\cap\td\ci_i).
    \]
    Notice that given $\cap_{i=1}^{k-1}\cb_i$ the total number of vertices within $k-1$ depth of the two trees satisfies
    \begin{align*}
            |\bar\cv(\ct_{k-1})\cup\bar\cv(\td\ct_{k-1})|=& 1+\sum_{i=1}^{k-1}|\com_i\cup\excl_i\cup\wdtd\excl_{i}|\\
            \le & 1+\sum_{i=1}^{k-1}2^{i-1}(\lambda(2-s))^i\log n\\
            \le & (2\lambda(2-s))^{k-1}\log n,
    \end{align*}
    where $\bar\cv(\ct_{k-1})$ (resp. $\bar\cv(\td\ct_{k-1})$) denotes the image of the nodes in $\ct_{k-1}$ (resp. $\td\ct_{k-1}$) in the union graph $\bar G$. 

    Given event $\cg_{k-1}$, for each $v\in \com_{k-1}$, the number of children has distribution  
    \begin{align}
    &(|\sfch_v^*|,|\sfch_v|,|\wdtd\sfch_{\td v}|,|\mathsf{NC}_v|)\nonumber\\& \quad \quad \quad\quad\quad\quad\sim \mathrm{Multi}\left(n-|\bar\cv(\ct_{k-1})\cup\bar\cv(\td\ct_{k-1})|,\frac{\lambda s}{n},\frac{\lambda (1-s)}{n},\frac{\lambda (1-s)}{n},1-\frac{\lambda(2-s)}{n}\right),\label{eq:general_multi_new}
    \end{align}
    where $|\mathsf{NC}_v|$ is the total number of vertices outside $|\bar\cv(\ct_{k-1})\cup\bar\cv(\td\ct_{k-1})|$ that are not connected to $v$ in $G$ and not connected to $\td v$ in $\td G$.
    The reasoning for~\eqref{eq:general_multi_new} is similar to that for~\eqref{eq:root_multi_new}. 
    The only difference is that we need to exclude all the already explored vertices in $|\bar\cv(\ct_{k-1})\cup\bar\cv(\td\ct_{k-1})|$, and hence there are $n-|\bar\cv(\ct_{k-1})\cup\bar\cv(\td\ct_{k-1})|$ vertices taken into consideration. 
    
    Applying Lemma~\ref{lem:multi-poi-new} with $n-m = |\bar\cv(\ct_{k-1})\cup\bar\cv(\td\ct_{k-1})|$, we can couple $(|\sfch_v^*|,|\sfch_v|,|\wdtd\sfch_{\td v}|)$ with $(X^*_v,X_v,\td X_{\td v})$ such that
    \[
    \P(|\sfch_v^*|=X_v^*,|\sfch_v|=X_v,|\wdtd\sfch_{\td v}|=\td X_{\td v})=1-O\left(\frac{(2\lambda(2-s))^k\log n+n^{1/4}}{n}\right).
    \]
    Taking a union bound over the set $\com_{k-1}$, whose size is bounded by $(2\lambda(2-s))^{k-1}\log n$ conditioned on event $\cap_{i=1}^{k-1}\cb_i$, yields 
    \begin{equation}
        \label{eq:a*k_new}
        \P(\ca^{*c}_k\,|\,\cg_{k-1})=O\left(\frac{(2\lambda(2-s))^{2k}\log^2 n+(2\lambda(2-s))^{k}n^{1/4}\log n}{n}\right).
    \end{equation}

    For each $v\in \excl_{k-1}$, we have
    \[
    |\sfch_v|\sim \bin\left(n-|\bar\cv(\ct_{k-1})\cup\bar\cv(\td\ct_{k-1})|,\frac{\lambda}{n}\right).
    \]
    By Lemma~\ref{lem:binom-poi-tv}, we can couple $|\sfch_v|$ with $X_v$ such that
    \[
    \P(|\sfch_v|=X_v)\ge 1-\frac{\lambda^2+(2\lambda(2-s))^{k-1}\lambda\log n}{n},
    \]
    and a union bound over $\excl_{k-1}$,  whose size is bounded by $(2\lambda(2-s))^{k-1}\log n$ conditioned on event $\cap_{i=1}^{k-1}\cb_i$ yields 
    \begin{equation}
        \label{eq:ak_new}
        \P(\ca^{c}_k\,|\,\cg_{k-1})=O\left(\frac{(2\lambda(2-s))^{2k}\log^2 n+(2\lambda(2-s))^{k+1}\log n}{n}\right).
    \end{equation}
    By an analogous argument, we can also get
    \begin{equation}
        \label{eq:tdak_new}
        \P(\td \ca^{c}_k\,|\,\cg_{k-1})=O\left(\frac{(2\lambda(2-s))^{2k}\log^2 n+(2\lambda(2-s))^{k+1}\log n}{n}\right).
    \end{equation}

    Because we have $|\com_{k-1}\cup\excl_{k-1}\cup\wdtd\excl_{k-1}|\le 2^{k-2}(\lambda(2-s))^{k-1}\log n$ on the event $\cap_{i=1}^{k-1}\cb_i$,
    there are at most $n2^{2k-4}(\lambda(2-s))^{2k-2}\log^2 n$ ways to pick two distinct vertices from $\com_{k-1}\cup\excl_{k-1}\cup\wdtd\excl_{k-1}$ and one vertex from the unexplored part of the graph. In the union graph $\bar G$, the probability that the two picked vertices both have an edge with the vertex in the unexplored part is $\left(\frac{\lambda(2-s)}{n}\right)^2$. 
    Then, we can apply the union bound to obtain
    \begin{equation}
        \label{eq:dk_new}
        \P(\cd_k^c\,|\,\cg_{k-1})=O\left(\frac{\lambda^2(2-s)^2n2^{2k-4}(\lambda(2-s))^{2k-2}\log^2 n}{n^2}\right)=O\left(\frac{(2\lambda(2-s))^{2k}\log^2 n}{n}\right).
    \end{equation}

    Notice that 
    \[
    |\com_{k}\cup\excl_{k}\cup\wdtd\excl_{k}|\sto\bin\left(n,\frac{\lambda(2-s)}{n}|\com_{k-1}\cup\excl_{k-1}\cup\wdtd\excl_{k-1}|\right).
    \]
    We can apply the Chernoff bound (equation~\eqref{eq:CB_upper2} in Lemma~\ref{lem:CB}) to obtain
    \begin{align}
        \P(\cb^c_k\,|\, \cg_{k-1})&\le \P\left(\bin\left(n,\frac{\lambda(2-s)}{n}2^{k-2}(\lambda(2-s))^{k-1}\log n\right)\ge 2^{k-1}(\lambda(2-s))^k\log n\right)\nonumber\\
        &\le \exp\left(-\frac{2^{k-2}(\lambda(2-s))^k\log n}{3}\right)=n^{-\omega(1)}.
        \label{eq:bk_new}
    \end{align}

    Now further condition on event $\cb_k$, which implies there are at most $2^{2k-2}(\lambda(2-s))^{2k}\log^2 n$ vertex pairs in $\com_{k}\cup\excl_{k}\cup\wdtd\excl_{k}$. we can apply a union bound over these vertex pairs to obtain
    \begin{equation}
        \label{eq:ck_new}
        \P(\cc_k^c\,|\,\cg_{k-1}\cap\cb_k)\le 2^{2k-2}(\lambda(2-s))^{2k}\frac{\lambda(2-s)}{n}\log^2 n=O\left(\frac{(2\lambda(2-s))^{2k+1}\log^2 n}{n}\right).
    \end{equation}

    By a union bound over the vertex pairs in $(\cup_{i=1}^{k}(\excl_i\cup\com_i))\times (\cup_{i=1}^{k}\wdtd\excl_i)$,  whose size is bounded by $$\left(\sum_{i=1}^{k} 2^{i-1}(\lambda(2-s))^i\log n\right)^2$$ conditioned on event $\cap_{i=1}^k\cb_i$, we get 
    \begin{align}
        \P(\ci_k^c\,|\,\cg_{k-1}\cap\cb_k)&\le \left(\sum_{i=1}^{k} 2^{i-1}(\lambda(2-s))^i\log n\right)^2\frac{\lambda}{n}\nonumber\\
        &\le (2\lambda(2-s))^{2k}\frac{\lambda}{n}\log^2 n \nonumber\\
        &=O\left(\frac{(2\lambda(2-s))^{2k+1}\log^2 n}{n}\right).\label{eq:fk_new}
    \end{align}
    Similarly, we can bound 
    \begin{equation}
        \label{eq:tdfk_new}
        \P(\td\ci_k^c\,|\,\cg_{k-1}\cap\cb_k)=O\left(\frac{(2\lambda(2-s))^{2k+1}\log^2 n}{n}\right).
    \end{equation}
    By equations~\eqref{eq:a*k_new} -~\eqref{eq:tdfk_new}, we have
    \begin{align}
        \label{eq:depthk_new}
        &\P(\ca^{*c}_k\cup\ca^c_k\cup\td\ca_k^c\cup\cb_k^c\cup\cc_k^c\cup\cd^c_k\cup\ci^c_k\cup\td\ci^c_k\,|\,\cg_{k-1})\nonumber\\
        ={}&O\left(\frac{(2\lambda(2-s))^{2k+1}\log^2 n+(2\lambda(2-s))^{k}n^{1/4}\log n}{n}\right)=O(n^{-2/3}),
    \end{align}
    where the last inequality follows because $(2\lambda(2-s))^{2k+1}=n^{o(1)}$ and $(2\lambda(2-s))^{k}=n^{o(1)}$ under the assumptions $\lambda = (\log n)^{O(1)}$ and $k\le d\le (\log n)^\gamma$ for some $\gamma\in (0,1)$.

    Finally, by equations~\eqref{eq:depth1_new} and~\eqref{eq:depthk_new},
    under the coupling we constructed
    \begin{align*}
        &\P(\ell(\ct_d,\td\ct_d)=(\ch_d,\td\ch_d))\\
        \ge{}& 1-\P(\ca^{*c}_1\cup\cb_1^c\cup\cc_1^c)-\sum_{k=2}^d \P(\ca^{*c}_k\cup\ca^c_k\cup\td\ca_k^c\cup\cb_k^c\cup\cc_k^c\cup\cd^c_k\cup\ci^c_k\cup\td\ci^c_k\,|\,\cg_{k-1})\\
        ={}&1-O\left(n^{-1/2}\right),
    \end{align*}
    which completes the proof.

 \end{proof}

\subsection{Degree Cutoff}
\label{appd:bad-nodes}
Recall the parameters $d$ and $\tau$ defined  in~\eqref{eq:TBAlign-parameters}, where
\begin{align}
    d=\left\lceil(\log n)^{1/3} \right\rceil,\quad
    \tau=\left\lceil\max\left\{\sqrt{\log n}, \lambda\left(1+\frac{1}{\log\log n}\right)\right\}\right\rceil.
    \label{eq:unified-degree-cutoff}
\end{align}

We define $\cb$ and $\td\cb$ as the sets of vertices whose $(d+1)$-hop
neighborhoods contain a vertex with degree greater than $\tau$:
\begin{align}
    \cb &:={} \left\{ u\in\cv:
    \exists v\in\cv\text{ such that } \dist_G(u,v)\le d+1
    \text{ and } \deg_G(v)>\tau \right\}, \label{eq:Bset}\\
    \td\cb &:={}
    \left\{ \td u\in\td\cv: \exists\td v\in\td\cv\text{ such that } \dist_{\td G}(\td u,\td v)\le d+1 \text{ and } \deg_{\td G}(\td v)>\tau \right\}. \label{eq:tBset}
\end{align}

\begin{lem}\label{lem:TBAlignBadNodes}
Consider two \erdos--\renyi graph $G,\td G\sim \mathrm{ER}(n,\frac{\lambda}{n})$ such that $\lambda=\omega(1)$ and $\lambda=(\log n)^{O(1)}$. Choose $d$ and $\tau$ as in~\eqref{eq:TBAlign-parameters} when constructing the local trees. Then for the sets $\cb$ and $\td\cb$ defined in \eqref{eq:Bset} and \eqref{eq:tBset}, we have
\begin{align}
    |\cb|+|\td\cb|  &\le
    2n\exp\left(-(\log n)^{2/5}\right) \label{eq:bad-node-bound}
\end{align}
with probability at least
\begin{align*}
    1-2\exp\left(-(\log n)^{2/5}\right)
\end{align*}
for all sufficiently large $n$.
\end{lem}

\begin{proof}
    To prove~\eqref{eq:bad-node-bound}, we will establish the following expecation bounds for $|\cb|$ and $|\td\cb|$:
    \begin{align}
    \be[|\cb|]&\le n\exp\left(-2(\log n)^{2/5}\right).
    \label{eq:eb}\\
    \be[|\td\cb|] &\le n\exp\left(-2(\log n)^{2/5}\right).
    \label{eq:etdb}
\end{align}
Notice that
\[
\left\{|\cb|+|\td\cb|> 2n\exp\left(-(\log n)^{2/5}\right)\right\}\subset\left\{|\cb|> n\exp\left(-(\log n)^{2/5}\right)\right\}\bigcup \left\{|\td\cb|> n\exp\left(-(\log n)^{2/5}\right)\right\}.
\]
We have
\begin{align*}
    &\P\left(
    |\cb|+|\td\cb|> 2n\exp\left(-(\log n)^{2/5}\right) \right)\\
    &\qquad\le \P\left( |\cb|> n\exp\left(-(\log n)^{2/5}\right) \right) + \P\left( |\td\cb|> n\exp\left(-(\log n)^{2/5}\right) \right) \\
    &\qquad\le 2\exp\left(-(\log n)^{2/5}\right),
\end{align*}
where the last inequality follows by equations~\eqref{eq:eb} and~\eqref{eq:etdb} and Markov's inequality. In the rest of this proof, we focus on proving~\eqref{eq:eb}, and ~\eqref{eq:etdb} follows by symmetry.

    \textbf{\underline{Proof of~\eqref{eq:eb}:}}
    Fix a vertex $u\in \cv$. Define event
    \[
    \mathsf{BAD}_u:=\{\exists v\in \cv\text{ such that }
    \dist(u,v)\le d+1
    \text{ and }
    \deg(v)> \tau\}.
    \]
    By~\eqref{eq:Bset}, $u\in\cb$ if and only if $\mathsf{BAD}_u$ occurs. 
Equivalently, either $\deg_G(u)>\tau$, or there exist an integer $l\in[d+1]$ and distinct vertices $v_1,\ldots,v_l\in\cv\setminus\{u\}$ such that
$(u,v_1,\ldots,v_l)$ is a path in $G$ and $\deg_G(v_l)>\tau$. Since
$\deg_G(u)\sim\bin(n-1,\lambda/n)$, we have 
    \begin{align}
        \P(\deg_G(u)>\tau)&= \P\left(\bin\left(n-1,\frac{\lambda}{n}\right)>\tau\right)\nonumber\\
        &\le \P\left(\bin\left(n,\frac{\lambda}{n}\right)\ge\tau-1\right)\nonumber\\
        &\le
    \exp\left(-4(\log n)^{2/5}\right). \label{eq:degu}
    \end{align}
    Here, the last inequality is obtained by applying the Chernoff bound. We defer the detailed reasoning to the end of this proof.
    Fix distinct vertices $v_1,\ldots,v_l$ in $\cv$. Because each edge appears in the graph $G$ with probability $\lambda/n$, the path $(u,v_1,v_2,\ldots,v_l)$ appears in $G$ with probability $\frac{\lambda^l}{n^l}$. Moreover, conditioned on the path $(u,v_1,v_2,\ldots,v_l)$, we have
    \[
    \deg(v_l)\sim 1+\bin(n-2,\lambda/n).
    \]
    This is because the path implies that $v_{l-1}$ is a neighbor of $v_l$, while the edge between $v_l$ and any other vertex in the graph appears independently with probability $\lambda/n$. From this argument, we know that
    \begin{align}
        &\P(\text{the path }(u,v_1,\ldots,v_l) \text{ appears in $G$ and }\deg(v_l)>\tau)\nonumber\\
        &=\left(\frac{\lambda}{n}\right)^l \P\left( \bin\left(n-2,\frac{\lambda}{n}\right)>\tau -1 \right) \nonumber\\
        &\le \left(\frac{\lambda}{n}\right)^l \P\left(\bin\left(n,\frac{\lambda}{n}\right)\ge\tau-1\right) \nonumber\\
        &\le \left(\frac{\lambda}{n}\right)^l \exp\left(-4(\log n)^{2/5}\right),\label{eq:degv}
    \end{align}
    Here the last inequality follows by the same argument as in~\eqref{eq:degu}.
    
    For each fixed $l$, there are at most $n^l$ ways to choose the vertices $v_1,\ldots,v_l$ from the graph. Therefore, we can apply the union bound with~\eqref{eq:degu} and~\eqref{eq:degv} to obtain
    \begin{align}
        \P(\mathsf{BAD}_u)&\le
        \exp\left(-4(\log n)^{2/5}\right)+
        \sum_{l=1}^{d+1}n^l\, \left(\frac{\lambda}{n}\right)^l\,\exp\left(-4(\log n)^{2/5}\right)\nonumber\\
        &= \sum_{l=0}^{d+1}\lambda^l \exp\left(-4(\log n)^{2/5}\right)\nonumber\\
        &\le \lambda^{d+2} \exp\left(-4(\log n)^{2/5}\right)\nonumber\\
        & \stackrel{(a)}{\le} \exp\left( ((\log n)^{1/3}+3)\log \lambda-4(\log n)^{2/5}\right)\nonumber\\
        &\stackrel{(b)}{\le} \exp\left(-2(\log n)^{2/5}\right)\label{eq:badu}
    \end{align}
    for all sufficiently large $n$, where (a) holds because $d+2=\lceil(\log n)^{1/3}\rceil+2\leq (\log n)^{1/3}+3$, and (b) holds because $\log \lambda=O(\log\log n)$.
    
     Finally, using $|\cb|=\sum_{u\in\cv}\indi\{\mathsf{BAD}_u\}$ and linearity of
expectation,~\eqref{eq:badu} gives
\begin{align*}
    \be[|\cb|]
    =\sum_{u\in\cv}\P(\mathsf{BAD}_u)
    \le
    n\exp\left(-2(\log n)^{2/5}\right),
\end{align*}
which completes the proof of~\eqref{eq:eb}.

\textbf{\underline{Proof of the last inequality in~\eqref{eq:degu}:}}
By~\eqref{eq:TBAlign-parameters}, we have $\tau\ge\sqrt{\log n}$ and $\tau\ge\lambda(1+1/\log\log n)$.
Since $\tau\ge\sqrt{\log n}\ge 4\log\log n$ for all sufficiently large $n$, we have

\begin{align*}
    (\tau-1)\left(1+\frac{1}{\log\log n}\right)&\ge\tau\left(1+\frac{1}{\log\log n}-\frac{2}{\tau}\right)\ge
    \tau\left(1+\frac{1}{2\log\log n}\right).
\end{align*}
Then it follows that
\begin{align*}
    \lambda
    \le \frac{\tau}{1+1/\log\log n}
    \le \frac{\tau-1}{1+1/(2\log\log n)}.
\end{align*}
Hence, by the
Chernoff bound~\eqref{eq:CB_upper2} in Lemma~\ref{lem:CB},
\begin{align*}
    \P\left(\bin\left(n,\frac{\lambda}{n}\right)\ge\tau-1\right)
    &\le\P\left(\bin\left(n,\frac{\tau-1}{n(1+1/(2\log\log n))}\right)\ge\tau-1\right)\\
    &\le\exp\left(-\frac{\tau-1}{12(1+1/(2\log\log n))(\log\log n)^2}\right)\\
    &\stackrel{(c)}{\le} \exp\left(-\frac{\tau}{48(\log\log n)^2}\right)\\
    &\stackrel{(d)}{\le} \exp\left(-4(\log n)^{2/5}\right),
\end{align*}
where (c) holds because $\tau-1>\tau/2$ and $1+1/(2\log\log n)<2$ for large enough $n$, and (d) holds because
$\tau\ge\sqrt{\log n}$.
\end{proof}

\section{Concentration Inequalities and Poisson Approximations}
This appendix collects the concentration inequalities and Poisson approximations used in the preceding appendices.

\subsection{Concentration Inequalities}
\begin{lem}[Bennett's inequality~\cite{bennett1962}]
\label{lem:bennett}
    
    Suppose $X_1,\ldots,X_n$ are independent random variables with finite second moments. Assume $X_i-\E[X_i]\le B$ a.s. for each $i\in [n]$ and some constant $B>0$. Let $V=\sum_{i=1}^n\E[(X_i-\E[X_i])^2]$. Then, for every $x\ge 0$, 
    \[
    \P\left(\sum_{i=1}^n(X_i-\E[X_i])\ge x\right)\le \exp\left(-\frac{V}{B^2}\cdot \phi(xB/V)\right),
    \]
    where $\phi(t):=(1+t)\log (1+t)-t$ for $t\ge 0$.
\end{lem}

\begin{lem}[Chernoff bound for Binomial distribution~\texorpdfstring{\cite[Theorems~20~and~21]{mitzenmacher2017probability}}{}]\label{lem:CB}
    Let $X\sim \mathrm{Bin}(n,p)$ be a Binomial distribution with mean $\mu:=np$. Then the following bounds hold.
    \begin{enumerate}[label=\textbf{(\roman*)}]
        \item \textbf{(Upper Tails).} For any $\tau> \mu$, 
        \begin{align}
            \P(X\ge \tau)< \Bigl(\frac{e\mu}{\tau}\Bigr)^{ \tau}=\exp \Bigl(\tau\bigl(1+\log\mu-\log\tau\bigr)\Bigr).\label{eq:CB_Upper}
        \end{align}
        For any $\delta\in(0,1]$,
        \begin{equation}
            \P(X\ge (1+\delta)\mu)<\exp\Bigl(-\frac{\mu\delta^2}{3}\Bigr).
            \label{eq:CB_upper2}
        \end{equation}
        \item \textbf{(Lower Tail).} For any $\delta\in(0,1)$, 
        \begin{align}
            \P(X\le (1-\delta)\mu)\le \exp\Bigl(-\frac{\mu\delta^2}{2}\Bigr).\label{eq:CB_Lower}
        \end{align}
    \end{enumerate}
\end{lem}

\subsection{Concentration Bounds on Tree Generation Size}
\label{appd:gen-size-rev}
\begin{lem}\label{lem:gen-size-rev}
Let $\mu\ge 2$, and let $\ch$ be a Poisson Galton--Watson tree with offspring mean $\mu$. Let $\cs_l(\ch)$ be the set of nodes in the $l$-th generation of $\ch$ for $l\in\mathbb N$. Then for any constant $C_2>1$,
\begin{align}
    \P\bigl(\lvert \cs_{l}(\ch) \rvert \ge (C_2 \mu)^l \bigr)
    &\le \exp\left(-\tfrac14 C_2^l + \tfrac12\right),\label{eq:gen_LB} \\
    \P\bigl(\lvert \cs_{l}(\ch) \rvert\le C_2^{-1}\mu^l \bigr)
    &\le \frac{C_2^2}{(C_2-1)^2}\cdot\frac{1}{\mu-1}.\label{eq:gen_UB}
\end{align}
\end{lem}
\begin{proof}
We first set up the log-moment generating function of the generation sizes, and then prove the upper-tail bound~\eqref{eq:gen_LB} and the lower-tail bound~\eqref{eq:gen_UB} in turn.

For any fixed $k\in[l]$, each node in $\cs_{k-1}(\ch)$ has an independent $\mathrm{Poi}(\mu)$ number of children. Hence, conditioned on $\lvert \cs_{k-1}(\ch) \rvert$, by the Poisson summation property,
\begin{align}
    \lvert \cs_{k}(\ch) \rvert\sim \mathrm{Poi}\bigl(\lvert \cs_{k-1}(\ch) \rvert\mu\bigr).\label{eq:gen_size_distribution}
\end{align}
Define the log-moment generating function $
\varphi_k(t) := \log \E\left[e^{t \lvert \cs_{k}(\ch) \rvert}\right]$ for all $k\in[l]$.
Since $|\cs_{0}(\ch)|=1$ and $|\cs_{1}(\ch)|\sim \poi(\mu)$, we have
\begin{align}
    \varphi_0(t) =& \log \E[e^{t |\cs_{0}(\ch)|}] = t, \nonumber\\ \varphi_1(t) =& \log \E[e^{t |\cs_{1}(\ch)|}]
= \mu(e^t - 1)
= \mu\bigl(e^{\varphi_0(t)} - 1\bigr),\nonumber
\end{align}
and for general $1 \le k\le l$, the tower property and \eqref{eq:gen_size_distribution} give
\begin{align}
\varphi_k(t)
&= \log \E\left[e^{t \lvert \cs_{k}(\ch) \rvert}\right]
= \log \E\left[\E\left[e^{t \lvert \cs_{k}(\ch) \rvert}\,\middle|\, \lvert \cs_{k-1}(\ch) \rvert\right]\right] \nonumber\\
&= \log \E\left[\exp\bigl(\mu\cdot \lvert \cs_{k-1}(\ch) \rvert\cdot(e^t-1)\bigr)\right] \nonumber\\
&= \varphi_{k-1}\bigl(\mu(e^t-1)\bigr).\label{eq:phi_rec_rev}
\end{align}

Iterating~\eqref{eq:phi_rec_rev} yields $\varphi_k=\underbrace{\varphi_1\circ\cdots\circ\varphi_1}_{k\ \text{times}}$, where $\circ$ denotes composition. Separating the leftmost factor then gives
\begin{align}
    \varphi_k(t)=\varphi_1\bigl(\varphi_{k-1}(t)\bigr)=\mu\bigl(e^{\varphi_{k-1}(t)}-1\bigr),\qquad 1\le k\le l.\label{eq:phi_last_rev}
\end{align}

\underline{\textbf{Proof of~\eqref{eq:gen_LB}.}} By the Chernoff bound, for any $t>0,$
\begin{align}
    \P\left(\lvert \cs_{l}(\ch) \rvert \ge (C_2 \mu)^l\right)
\le \exp\left(-t (C_2 \mu)^l + \varphi_l(t)\right).\label{eq:child_size_Chernoff}
\end{align}
By choosing  $t=\frac{1}{4\mu^l}>0$, we have 
\begin{align}
    \P\left(\lvert \cs_{l}(\ch) \rvert \ge (C_2 \mu)^l\right)
\le & \exp\left(-\frac{1}{4\mu^l} (C_2 \mu)^l +\varphi_l\left(\frac{1}{4\mu^l}\right)\right)= \exp\left(-\tfrac14 C_2^l + \varphi_l\left(\frac{1}{4\mu^l}\right)\right).\label{mgb}
\end{align} 

To prove \eqref{eq:gen_LB}, it suffices to show that
\begin{align}
    \varphi_l\left(\tfrac{1}{4\mu^l}\right)\le \tfrac12,\label{eq:logMGF_rev}
\end{align}
since substituting~\eqref{eq:logMGF_rev} into~\eqref{mgb} then proves~\eqref{eq:gen_LB}.

\underline{\textbf{Proof of~\eqref{eq:logMGF_rev}:}} Since $\mu\ge2$,
\begin{align}
    \prod_{j=1}^{\infty}\Bigl(1+\frac{1}{2\mu^j}\Bigr)
    \le \exp\Bigl(\sum_{j=1}^{\infty}\frac{1}{2\mu^j}\Bigr)
    = \exp\Bigl(\frac{1}{2(\mu-1)}\Bigr)
    \le e^{1/2},\label{eq:prod_bound_rev}
\end{align}
where the first inequality uses $1+x\le e^x$ and the last uses $\mu\ge2$. We now prove by induction on $k=0,1,\ldots,l$ that
\begin{align}
    \varphi_k\left(\frac{1}{4\mu^l}\right)\le \frac{1}{4\mu^{l-k}}\prod_{j=l-k+1}^{l}\Bigl(1+\frac{1}{2\mu^j}\Bigr).\label{eq:phi_ind_rev}
\end{align}
For the base case $k=0$, the product is empty and~\eqref{eq:phi_ind_rev} holds since $\varphi_0\left(\frac{1}{4\mu^l}\right)=\frac{1}{4\mu^l}$. Now assume~\eqref{eq:phi_ind_rev} holds for $k-1$. Bounding the partial product by~\eqref{eq:prod_bound_rev}, the induction hypothesis gives
\begin{align}
    \varphi_{k-1}\left(\frac{1}{4\mu^l}\right)\le \frac{1}{\mu^{l-k+1}}\cdot\frac{e^{1/2}}{4}\le \frac{1}{2\mu^{l-k+1}}\le \frac12,\label{eq:phi_prev_rev}
\end{align}
where the second inequality uses $\frac{e^{1/2}}{4}\le\frac12$ and the third uses $\mu^{l-k+1}\ge\mu\ge2$. Moreover, since $\lvert\cs_{k-1}(\ch)\rvert\ge0$, we have 
\[\varphi_{k-1}\left(\frac{1}{4\mu^l}\right)=\log\E\left[e^{\frac{1}{4\mu^l}\lvert\cs_{k-1}(\ch)\rvert}\right]\ge0,\]
and thus $\varphi_{k-1}\left(\frac{1}{4\mu^l}\right)\in[0,\frac12]$. By~\eqref{eq:phi_last_rev}, we have
\begin{align}
    \varphi_k\left(\frac{1}{4\mu^l}\right)=\mu\bigl(e^{\varphi_{k-1}\left(\frac{1}{4\mu^l}\right)}-1\bigr)
    &\le \mu\varphi_{k-1}\left(\frac{1}{4\mu^l}\right)\left(1+\varphi_{k-1}\left(\frac{1}{4\mu^l}\right)\right)\nonumber\\
    &\le \mu\varphi_{k-1}\left(\frac{1}{4\mu^l}\right)\Bigl(1+\frac{1}{2\mu^{l-k+1}}\Bigr),\nonumber
\end{align}
where the first inequality holds by applying $x=\varphi_{k-1}\left(\frac{1}{4\mu^l}\right)$ to the inequality $e^{x}-1\le x(1+x)$ for $x\in[0,\frac12]$, and the last inequality uses $\varphi_{k-1}\left(\frac{1}{4\mu^l}\right)\le\frac{1}{2\mu^{l-k+1}}$ in~\eqref{eq:phi_prev_rev}. Substituting the induction hypothesis~\eqref{eq:phi_ind_rev} for the remaining factor $\varphi_{k-1}\left(\frac{1}{4\mu^{l}}\right)$ gives
\begin{align}
    \varphi_k\left(\frac{1}{4\mu^{l}}\right)
    \le \mu\cdot\frac{1}{\mu^{l-k+1}}\cdot\frac14\prod_{j=l-k+2}^{l}\Bigl(1+\frac{1}{2\mu^j}\Bigr)\cdot\Bigl(1+\frac{1}{2\mu^{l-k+1}}\Bigr)
    = \frac{1}{\mu^{l-k}}\cdot\frac14\prod_{j=l-k+1}^{l}\Bigl(1+\frac{1}{2\mu^j}\Bigr),\nonumber
\end{align}
which completes the induction on $k$ for~\eqref{eq:phi_ind_rev}. Taking $k=l$ in~\eqref{eq:phi_ind_rev} and applying~\eqref{eq:prod_bound_rev} yields
\[
    \varphi_l\left(\frac{1}{4\mu^{l}}\right)\le \frac14\prod_{j=1}^{l}\Bigl(1+\frac{1}{2\mu^j}\Bigr)\le \frac{e^{1/2}}{4}\le \frac12,
\]
which proves~\eqref{eq:logMGF_rev}.

\underline{\textbf{Proof of~\eqref{eq:gen_UB}.}} Since $\ch$ is a Poisson Galton--Watson tree with offspring mean $\mu$, equation~\eqref{eq:gen_size_distribution} gives $\E[\lvert\cs_k(\ch)\rvert]=\mu\E[\lvert\cs_{k-1}(\ch)\rvert]$ for each $k\in[l]$, so $\E[\lvert \cs_{l}(\ch) \rvert] = \mu^l$ since $\lvert\cs_0(\ch)\rvert=1$. Also $\mathrm{Var}(\lvert\cs_0(\ch)\rvert)=0$, and for each $k\in[l]$, by the law of total variance and~\eqref{eq:gen_size_distribution},
\begin{align*}
\mathrm{Var}(\lvert \cs_{k}(\ch) \rvert)
&= \E\bigl[\mathrm{Var}(\lvert\cs_k(\ch)\rvert \mid \lvert \cs_{k-1}(\ch) \rvert)\bigr]
   + \mathrm{Var}\bigl(\E[\lvert \cs_{k}(\ch) \rvert\mid \lvert \cs_{k-1}(\ch) \rvert]\bigr) \\
&= \mu \E[\lvert \cs_{k-1}(\ch) \rvert] + \mu^2 \mathrm{Var}(\lvert \cs_{k-1}(\ch) \rvert),
\end{align*}
where the second equality uses~\eqref{eq:gen_size_distribution}, under which the conditional variance and mean of $\lvert\cs_k(\ch)\rvert$ both equal $\mu\lvert\cs_{k-1}(\ch)\rvert$. Induction on $k\in[l]$ then gives $\mathrm{Var}(\lvert \cs_{l}(\ch) \rvert)= \mu^l \cdot \frac{\mu^l-1}{\mu-1}$. Hence, by Chebyshev's inequality,
\begin{align}
\P\bigl(\lvert \cs_{l}(\ch) \rvert \le C_2^{-1}\mu^l\bigr)
&= \P\bigl(\lvert \cs_{l}(\ch) \rvert \le C_2^{-1}\E[\lvert \cs_{l}(\ch) \rvert]\bigr)\nonumber \\
&\le \P\bigl(\bigl|\lvert \cs_{l}(\ch) \rvert-\E[\lvert \cs_{l}(\ch) \rvert]\bigr|
\ge (1-C_2^{-1})\E[\lvert \cs_{l}(\ch) \rvert]\bigr)\nonumber \\
&\le \frac{\mathrm{Var}(\lvert \cs_{l}(\ch) \rvert)}
{(1-C_2^{-1})^2(\E[\lvert \cs_{l}(\ch) \rvert])^2}
\le  \frac{C_2^2}{(C_2-1)^2}\cdot\frac{1}{\mu-1},\nonumber
\end{align}
where the last inequality uses $\frac{\mu^l-1}{\mu^l}\le1$. This proves~\eqref{eq:gen_UB}.
\end{proof}

\subsection{Poisson Approximation}
\begin{lem}[Lemma 22 in~\cite{wang2026}]
    \label{lem:binom-poi-gap}
Let $n\in \mathbb{N}^+$, $\mu\in\mathbb{R}^+$ and $x,y\in\mathbb{N}$ be such that $x<n$ and $y\le n-x$.
    Then we have
    \[
    \frac{\P(\bin(n-x,\mu/n)=y)}{\P(\poi(\mu)=y)}\le \exp\left(\frac{\mu}{n}(x+y)\right).
    \]
\end{lem}

\begin{restatable}{lem}{lembinompoi}
    \label{lem:binom-poi-tv}
        If $m$ and $n$ are positive integers such that $m\le n$ and $\mu$ is a positive real number, then
        \[
        \dtv\left(\bin\left(m,\frac{\mu}{n}\right),\poi(\mu)\right)\le \frac{\mu^2+(n-m)\mu}{n}.
        \]
    \end{restatable}

    \begin{proof}
    By the triangle inequality, we have
    \begin{align*}
        &\dtv\left(\bin\left(m,\frac{\mu}{n}\right),\poi\left(\mu\right)\right)\\
        \le{}&  \dtv\left(\bin\left(m,\frac{\mu}{n}\right),\poi\left(\frac{m\mu}{n}\right)\right)+\dtv\left(\poi(\mu),\poi\left(\frac{m\mu}{n}\right)\right),
    \end{align*}
    and it suffices to bound these two total variation distances separately. 
    
    It follows by the classic Poisson approximation of binomial random variables (see for example~\cite{hodges1960}) that
    \begin{equation}
        \label{eq:poi-approx}
        \dtv\left(\bin\left(m,\frac{\mu}{n}\right),\poi\left(\frac{m\mu}{n}\right)\right)\le \frac{m\mu^2}{n^2}\le \frac{\mu^2}{n}.
    \end{equation}

    On the other hand, for a random variable $X\sim \poi(\mu)$, we can write $X=Y+Z$, where $Y\sim \poi(m\mu/n)$ and $Z\sim\poi(\mu-m\mu/n)$ are two independent Poisson random variables. We then have
    \begin{equation}
        \dtv\left(\poi(\mu),\poi\left(\frac{m\mu}{n}\right)\right)\le \P(X\neq Y)=\P(Z\neq 0)= 1-\exp\left(-\mu+\frac{m\mu}{n}\right)\le \frac{(n-m)\mu}{n}.\label{eq:tv_poi}
    \end{equation}
    Equations~\eqref{eq:poi-approx} and~\eqref{eq:tv_poi} together imply the desired inequality.
\end{proof}
\begin{restatable}{lem}{lemmultipoi_new}
        \label{lem:multi-poi-new}
        Let $m$ and $n$ be two positive integers satisfying $m\le n$ and $n-m=o(\frac{n}{\lambda})$. Let $s\in (0,1]$ be a constant.
        Suppose $ \lambda = (\log n)^{O(1)}$.
        Consider random variables $$(X_1,X_2,X_3,X_4)\sim\mathrm{Multi}(m,\lambda s/n,\lambda(1-s)/n,\lambda(1-s)/n,1-(2-s)\lambda/n),$$ and three independent Poisson random variables $X_1'\sim \poi(\lambda s)$, $X_2'\sim \poi(\lambda(1-s))$ and $X_3'\sim\poi(\lambda(1-s))$. Then we have 
        \begin{equation}
        \label{eq:multi_dtv_new}
        \dtv(\cl(X_1,X_2,X_3),\cl(X_1',X_2',X_3'))=O\left(\frac{(n-m)\lambda+n^{1/4}}{n}\right).
        \end{equation}
        
    \end{restatable}
    \begin{proof}
        First assume $s=1$. In this case,~\eqref{eq:multi_dtv_new} simply follows by Lemma~\ref{lem:binom-poi-tv}. We assume $s<1$ in the rest of the proof.

        By the definition of the multinomial distribution, random variables $X_1,X_2,X_3$ can be sampled in a sequential manner:
        \begin{enumerate}
            \item Sample $X_1\sim\bin(m,\lambda s/n)$.
            \item Given $X_1$, sample $X_2\sim\bin\left(m-X_1,\frac{\lambda(1-s)}{n-\lambda s}\right)$.
            \item Given $X_1$ and $X_2$, sample $X_3\sim\bin\left(m-X_1-X_2,\frac{\lambda(1-s)}{n-\lambda}\right)$.
        \end{enumerate}
    By Lemma~\ref{lem:binom-poi-tv}, we can couple $X_1$ and $X_1'$ such that under the coupling, 
        \begin{equation}
        \label{eq:X1_new}
        \P(X_1\neq X_1')\le \frac{\lambda^2 s^2+(n-m)\lambda s}{n}=O\left(\frac{(n-m)\lambda+n^{1/4}}{n}\right),
        \end{equation}
    because $\lambda=O(n^{1/8})$.
    By the Chernoff bound (equation~\eqref{eq:CB_Upper} in Lemma~\ref{lem:CB}), we know that 
        \begin{equation}
        \label{eq:X1_bound_new}
        \P(X_1> n^{1/8})<\exp\left(n^{1/8}\left(1+\log \frac{\lambda sm}{n}-\frac18\log n\right)\right)=n^{-\omega(1)},
        \end{equation}
    where the last equality follows because $\log\frac{\lambda sm}{n}=O(\log \lambda)=O(\log\log n)$.

    Conditioned on the event $\{X_1=X_1'=x\}$, 
    we have 
    \[
    X_2\sim \bin\left(m-x,\frac{\lambda(1-s)}{n-\lambda s}\right),
    \]
    and 
    \[
    X_2'\sim\poi(\lambda (1-s)).
    \]
    We can couple these two random variables such that
    \begin{align}
        \P(X_2\neq X_2'\cond X_1=X_1'=x)
        &\le \dtv\left(\bin\left(m-x,\frac{\lambda(1-s)}{n-\lambda s}\right),\poi(\lambda (1-s))\right)\nonumber\\
        &\le \dtv\left(\bin\left(m-x,\frac{\lambda(1-s)}{n-\lambda s}\right),\poi\left(\frac{\lambda(1-s)(m-x)}{n-\lambda s}\right)\right)\nonumber\\
        &\;\;+\dtv\left(\poi(\lambda (1-s)),\poi\left(\frac{\lambda(1-s)(m-x)}{n-\lambda s}\right)\right)\nonumber\\
        &\le \frac{(m-x)\lambda^2(1-s)^2}{(n-\lambda s)^2}+\Bigg|\lambda(1-s)-\frac{\lambda(1-s)(m-x)}{n-\lambda s}\Bigg|,\label{eq:littlex_new}
    \end{align}
    where the last inequality follows by~\eqref{eq:poi-approx} and~\eqref{eq:tv_poi} in the proof of Lemma~\ref{lem:binom-poi-tv}. 
    
    Under the assumption that $x\le n^{1/8}$, we further have
    \[
    \frac{(m-x)\lambda^2(1-s)^2}{(n-\lambda s)^2}=O\left(\frac{n^{1/4}}{n}\right)
    \]
    and
    \[
    \Bigg|\lambda(1-s)-\frac{\lambda(1-s)(m-x)}{n-\lambda s}\Bigg|=\Bigg|\frac{\lambda(1-s)(n-m-\lambda s+x)}{n-\lambda s}\Bigg|=O\left(\frac{(n-m)\lambda+n^{1/4}}{n}\right).
    \]
    So the right-hand side of~\eqref{eq:littlex_new} is equal to
    \[
     O\left(\frac{(n-m)\lambda+n^{1/4}}{n}\right).
    \]
    This implies that conditioned on events $\{X_1=X_1'\}$ and $\{X_1\le n^{1/8}\}$, we can couple $X_2$ and $X_2'$ so that
    \begin{equation}
        \P(X_2\neq X_2'\cond X_1=X_1',X_1\le n^{1/8})=O\left(\frac{(n-m)\lambda+n^{1/4}}{n}\right)\label{eq:x2_new}.
    \end{equation}

    We can apply the Chernoff bound (equation~\eqref{eq:CB_Upper} in Lemma~\ref{lem:CB}) to get
        \begin{equation}
        \label{eq:X2_bound_new}
        \P(X_2> n^{1/8}\cond X_1=X_1',X_1\le n^{1/8})= n^{-\omega(1)}.
        \end{equation}
     Conditioned on events $\{X_1=X_1'=x\}$ and $\{X_2=X_2'=y\}$, we have
     \[
    X_3\sim \bin\left(m-x-y,\frac{\lambda(1-s)}{n-\lambda }\right),
    \]
    and 
    \[
    X_3'\sim\poi(\lambda (1-s)).
    \]
    Similarly to~\eqref{eq:littlex_new}, we can couple these two random variables so that
    \begin{equation}
        \label{eq:littlexy_new}
        \P(X_3\neq X_3'\cond X_1=X_1'=x,X_2=X_2'=y)\le\frac{(m-x-y)\lambda^2(1-s)^2}{(n-\lambda )^2}+\Bigg|\lambda(1-s)-\frac{\lambda(1-s)(m-x-y)}{n-\lambda }\Bigg|.
    \end{equation}
    Under the assumptions $x\le n^{1/8}$ and $y\le n^{1/8}$, the right-hand side of~\eqref{eq:littlexy_new} is
    \[
    O\left(\frac{(n-m)\lambda+n^{1/4}}{n}\right).
    \]
    This implies that conditioned on events $\{X_1=X_1'\}$, $\{X_1\le n^{1/8}\}$, $\{X_2=X_2'\}$ and $\{X_2\le n^{1/8}\}$, we can further couple $X_3$ with $X_3'$ such that
    \begin{equation}
        \label{eq:x3_new}
        \P(X_3\neq X_3'|X_1=X_1',X_2=X_2',X_1\le n^{1/8},X_2\le n^{1/8})=O\left(\frac{(n-m)\lambda+n^{1/4}}{n}\right).
    \end{equation}

    Finally, putting~\eqref{eq:X1_new},~\eqref{eq:X1_bound_new},~\eqref{eq:x2_new},~\eqref{eq:X2_bound_new} and~\eqref{eq:x3_new} yields that we can couple $(X_1,X_2,X_3)$ and $(X_1',X_2',X_3')$ such that
    \begin{align*}
    &\P(X_1=X_1',X_2=X_2',X_3=X_3')\\
    \ge{}& \P(X_1=X_1',X_2=X_2',X_3=X_3',X_1\le n^{1/8},X_2\le n^{1/8})\\
    \ge{}& 1-\P(X_1\neq X_1')-\P(X_1> n^{1/8})-\P(X_2\neq X_2'\cond X_1=X_1',X_1\le n^{1/8})\\
    &\;\;\;\;-\P(X_2> n^{1/8}\cond X_1=X_1',X_1\le n^{1/8})-\P(X_3\neq X_3'|X_1=X_1',X_2=X_2',X_1\le n^{1/8},X_2\le n^{1/8})\\
    ={}&1-O\left(\frac{(n-m)\lambda+n^{1/4}}{n}\right),
    \end{align*}
    which completes the proof.
    
    \end{proof}
\end{document}